\documentclass[10pt,prd,preprintnumbers,floatfix,aps,nofootinbib,showpacs,twocolumn,superscriptaddress,noshowpacs]{revtex4}

\usepackage{epsfig}
\usepackage{url}
\usepackage{hyperref}
\usepackage[normalem]{ulem} 

\usepackage{latexsym}
\usepackage{epsfig}
\usepackage{amsmath}
\usepackage{amssymb}
\usepackage{wasysym}
\usepackage{graphicx}
\usepackage{dcolumn}
\usepackage{verbatim}
\usepackage{enumerate,mdwlist}
\usepackage[titletoc]{appendix}
\usepackage{amsfonts}
\usepackage{fancyvrb} 
\usepackage{tikz} 
\usepackage{braket}
\usetikzlibrary{calc}
\usepackage[export]{adjustbox}
\usepackage{bm}

\usepackage[normalem]{ulem}

\newcommand{\ba}{\begin{eqnarray}}
\newcommand{\ea}{\end{eqnarray}}
\newcommand{\be}{\begin{equation}}
\newcommand{\ee}{\end{equation}}

\newcommand{\nn}{\nonumber}

\newcommand{\de}{{\rm d}}
\newcommand{\vect}[1]{\boldsymbol{#1}}
\newcommand*{\Cc}{\mathcal}%

\newcommand{\Mvir}{M_{\rm vir}} 
\newcommand{\rvir}{r_{\rm vir}} 
\newcommand{\MLz}{M_{Lz}}       
\newcommand{\Mbbh}{M_{\rm BBH}} 

\definecolor{grey}{rgb}{0.4,0.4,0.4}
\definecolor{dullmagenta}{rgb}{0.4,0,0.4}
\definecolor{darkblue}{rgb}{0,0,0.4}
\definecolor{midblue}{rgb}{0,0,0.5}
\definecolor{midred}{rgb}{0.5,0,0}
\definecolor{orange}{rgb}{1,0.5,0}
\definecolor{lightbrown}{rgb}{0.75,0.5,0.25}
\definecolor{tan}{cmyk}{0.14,0.42,0.56,0}
\definecolor{djunglegreen}{cmyk}{0.99,0,0.52,0}
\definecolor{lightgreen}{rgb}{0,1,0}
\definecolor{olivegreen}{cmyk}{0.64,0,0.95,0.40}
\definecolor{midgreen}{rgb}{0.0,0.675,0.0}
\definecolor{darkgreen}{rgb}{0,0.5,0}
\definecolor{ceruleanblue}{rgb}{0.0, 0.2, 0.7}
\definecolor{burgundy}{rgb}{0.5, 0.0, 0.13}
\definecolor{hvred}{RGB}{186,12,47}

\hypersetup{
    colorlinks=true,
    linkcolor=ceruleanblue,
    filecolor=ceruleanblue,      
    urlcolor=midblue,
    citecolor=burgundy,
}

\newcommand{\wowsfull}{Wave-Optics Features}
\newcommand{\wow}{WOF}
\newcommand{\wows}{WOFs}

\usepackage[all]{xy} 
\usepackage{amsfonts}

\makeatletter

\def\l@subsubsection#1#2{}
\makeatother

\begin{document} 

\title{Weakly Lensed Gravitational Waves: \\ Probing Cosmic Structures with Wave-Optics Features}

\author{Stefano Savastano}
\email{stefano.savastano@aei.mpg.de}
\affiliation{Max Planck Institute for Gravitational Physics (Albert Einstein Institute) \\
Am Mühlenberg 1, D-14476 Potsdam-Golm, Germany}

\author{Giovanni Tambalo}
\email{giovanni.tambalo@aei.mpg.de}
\affiliation{Max Planck Institute for Gravitational Physics (Albert Einstein Institute) \\
Am Mühlenberg 1, D-14476 Potsdam-Golm, Germany}

\author{Hector Villarrubia-Rojo}
\email{hector.villarrubia-rojo@aei.mpg.de}
\affiliation{Max Planck Institute for Gravitational Physics (Albert Einstein Institute) \\
Am Mühlenberg 1, D-14476 Potsdam-Golm, Germany}
\affiliation{Instituto de Física de Partículas y del Cosmos (IPARCOS-UCM),
Universidad Complutense de Madrid, 28040 Madrid, Spain}

\author{Miguel Zumalac\'arregui}
\email{miguel.zumalacarregui@aei.mpg.de}
\affiliation{Max Planck Institute for Gravitational Physics (Albert Einstein Institute) \\
Am Mühlenberg 1, D-14476 Potsdam-Golm, Germany}

\begin{abstract}
Every signal propagating through the universe is at least weakly lensed by the intervening gravitational field. 
In some situations, wave-optics phenomena (diffraction, interference) can be observed as frequency-dependent modulations of the waveform of gravitational waves (GWs). We will denote these signatures as \textit{\wowsfull{}} (\wows) and analyze them in detail. 
Our framework can efficiently and accurately compute \wow{} in the single-image regime, of which weak lensing is a limit.
The phenomenology of \wow{} is rich and offers valuable information: the dense cusps of individual halos appear as peaks in Green's function for lensing.
If resolved, these features probe the number, effective masses, spatial distribution and inner profiles of substructures.
High signal-to-noise GW signals reveal \wows{} well beyond the Einstein radius, leading to a fair probability of observation by upcoming detectors such as LISA. Potential applications of \wow{} include reconstruction of the lens' projected density, delensing standard sirens and inferring large-scale structure morphology and the halo mass function. Because \wow{} are sourced by light halos with negligible baryonic content, their detection (or lack thereof) holds promise to test dark matter scenarios. 
\end{abstract}

\date{\today}

\maketitle

{
  \hypersetup{hidelinks}
  \tableofcontents
}

\section{Introduction}

Gravitational lensing, the effect of gravitational fields on the propagation of signals through the universe, predicts a plethora of observable effects \cite{Schneider:1992,Bartelmann:2010fz}. Many gravitational lensing phenomena have been observed using light and other electromagnetic (EM) signals, leading to a wide-range of applications in astrophysics, cosmology and fundamental physics. Now, the advent of gravitational-wave (GW) astronomy provides the prospect of observing novel lensing phenomena.

Lensed GW signals stand out as highly complementary to EM observations. GWs can be detected at very high redshift and are free from many of the systematic uncertainties present in EM probes. For this reason, GWs have been proposed as an alternative tool to test the cosmological model, e.g. studying the cross-correlation of GW observations and galaxy surveys \cite{Mukherjee:2019wfw, Mukherjee:2019wcg, Balaudo:2022znx}. Moreover, the low frequency and phase-coherence of observable GW sources make them ideal ground to probe the wave-optics (WO) regime \cite{Leung:2023lmq}. WO encompasses phenomena such as diffraction and interference, which imprint frequency-dependent signatures on GW waveforms. They can hence be used to identify a signal as lensed \cite{Dai:2018enj,Diego:2019lcd,Yeung:2021roe,Meena:2022unp,Shan:2023ngi,Meena:2023qdq} and even infer the lens properties accurately \cite{Takahashi:2003ix,Caliskan:2022hbu,Tambalo:2022wlm,Lin:2023ccz}, at least in the strong-lensing regime. In contrast, identifying lensing in the geometric-optics (GO) limit, i.e.~the high-frequency limit, requires associating multiple images from the same event, a method prone to false alarm \cite{Caliskan:2022wbh}, or identifying subtle waveform differences \cite{Dai:2017huk}, which requires sources with large mass ratios \cite{Ezquiaga:2020gdt,Vijaykumar:2022dlp}.

WO signatures require lenses in a restricted mass range, set by the frequency range of the observed signal. For observable GWs, this limitation implies that only relatively light structures can be detected. This dramatically reduces the probability of detecting WO features, at least for strongly-lensed signals in which the most likely lenses are massive galaxies \cite{Shajib:2022con}. This led to a pessimistic prospect to detect WO imprints in strongly lensed signals, e.g.~by LISA \cite{Sereno:2010dr}.

WO features can also be searched for in weakly-lensed signals, which do not require a close alignment of source-lens and observer and thus have a higher probability of occurring. It was estimated that LISA may detect WO effects at $\sim 50\times$ the strong-lensing impact parameter, corresponding to $\mathcal{O}(1\%)$ of massive BH binaries \cite{Gao:2021sxw}. While the above study was based purely on a mismatch analysis between lensed and unlensed waveforms, a more detailed estimate (accounting for waveform and lens parameter correlations) yields comparable values \cite{Caliskan:2022hbu}.
The trade-off between strong and weak lensing is that of rare and dramatic versus frequent but subtle signatures.

The lens distribution can be probed in a rather transparent way via weakly-lensed GWs that contain WO features.
Reference \cite{Choi:2021jqn} showed that the frequency-dependent amplification factor is determined by the shear of the Fermat potential at distances from the source given by the Fresnel radius $\propto 1/\sqrt{f}$. In this way, weakly-lensed GWs probe entire regions of a lens as the source inspirals towards merger. In contrast, signals in the GO regime are sensitive to a very small portion of the lens plane where the image forms. Thus, WO provides a unique opportunity to test the structure of gravitational lenses. 

The purpose of this paper is two-fold: first, we address the problem of computing WO signatures. Focusing on the single-image regime, we develop a framework to compute lensed waveforms of arbitrary lenses, efficiently and accurately. Second, we use these tools to explore the phenomenology of weakly lensed gravitational waves, the possibility of inferring structural features of lenses and the prospect of detection.
The paper is organized as follows. Section \ref{sec:lensing} summarizes the WO regime of gravitational lensing and introduces two frameworks for the single-image regime: a general numerical computation and an expansion suitable for the weak-lensing (WL) limit.
In Section \ref{sec:phenomenology} we discuss WO phenomenology, using a Green's function approach to analyze symmetric lenses, before addressing the case of a lens with substructure.
Section \ref{sec:observation} discusses the prospects of observation by
future detectors. Possible future applications of our formalism are presented in Sec.~\ref{sec:applications}. We conclude by  discussing our results in Sec.~\ref{sec:conclusions}.

\section{Gravitational lensing in the single-image regime} \label{sec:lensing}
In this section, we develop a framework to compute diffraction effects in the single-image regime. We will first summarize the WO formalism in the frequency and time domain (Sec.~\ref{sec:lensing_wo}). 
We will then present a method to numerically evaluate single-image signals in the time domain (Sec.~\ref{sec:lensing_nonperturbative}). Finally, we will develop a perturbative WL expansion in the time domain (Sec.~\ref{sec:lensing_perturbative}).

\subsection{Wave optics formalism} \label{sec:lensing_wo}

In the frequency domain, the effect of lensing is characterized by a multiplicative factor $F(f)$, called \emph{amplification factor}:
\begin{equation}
    F(f) \equiv \frac{\tilde h(f)}{\tilde h_0(f)}
    \,,
\end{equation}
where $\tilde h_0(f)$ and $\tilde h(f)$ are respectively the Fourier transforms of the unlensed and lensed strain amplitude. 
The frequency-domain amplification factor is obtained as
\begin{equation}\label{eq:amplification_F}
        F(w) = 
        \frac{w}{2\pi i}
        \int \de^2 \vect x 
        \exp\left(i w \phi(\vect x, \vect y)\right)
        \,,
\end{equation}
(see Ref.~\cite{Schneider:1992} for a derivation assuming the weak-field limit, the thin-lens approximation and a static configuration). The integration is over the lens plane, with the coordinates rescaled by an arbitrary dimensionful scale $\xi_0$ (e.g.~a characteristic scale of the lens), so $\vect x$ is dimensionless. The impact parameter $\vect y$ is rescaled by $\eta_0 \equiv D_S \xi_0 / D_L$, where $D_S,\, D_L$ are the angular diameter distances to the lens and the source, respectively.

Here we introduced the \emph{dimensionless frequency}
\begin{equation} \label{eq:lensing_freq_dimensionless}
 w \equiv 8\pi G \MLz f \,,
\end{equation}
which is given in terms of a \emph{redshifted effective lens mass}:
\begin{equation}\label{eq:def_MLz}
    \MLz 
    = 
    \frac{\xi_0^2 }{4 G d_{\rm eff}}
    \,.
\end{equation}
The factor $d_{\rm eff}\equiv \frac{D_L D_{LS}}{(1+z_L)D_S}$ also depends on the angular diameter distance between the observer and the source $D_{LS}$. For a point lens, $\MLz$ is equal to the total mass of the lens times $(1+z_L)$ if $\xi_0$ is set to the Einstein radius. However, this is not true for extended lenses (e.g.~Eq.~\eqref{eq:virial_mass} below). 

The integral in Eq.~\eqref{eq:amplification_F} depends on the \emph{Fermat potential}:
\begin{equation}\label{eq:fermat}
    \phi(\vect x, \vect y) 
    =
    \frac{1}{2}|\vect x - \vect y|^2- \psi(\vect x) - \phi_m(\vect y)\;,
\end{equation}
which is a dimensionless version of the time delay.
Here $\psi(\vect x)$ is the lensing potential, which depends on the matter distribution projected on the lens plane and whose derivative gives the deflection angle. 
In particular, it is obtained as the solution of $\nabla^2_{\vect x} \psi(\vect x) = 2 \Sigma(\xi_0 \vect x) / \Sigma_{\rm cr}$, with $\nabla^2_{\vect x}$ being the 2D Laplacian, $\Sigma(\xi_0 \vect x)$ the projected matter density of the lens, and $\Sigma_{\rm cr} \equiv (4 \pi G (1+z_L)d_{\rm eff})^{-1}$ the critical density.
We shift the Fermat potential by a constant $\phi_m(\vect y)$, defined in such a way as to make the minimum time delay equal to zero. 

An important case of WO lensing is the GO limit, corresponding to the $w\to \infty$ limit of the diffraction integral Eq.~\eqref{eq:amplification_F}:
\begin{equation}\label{eq:lensing_geometric_optics}
    F(w) 
    =
    \sum_J \sqrt{|\mu_J|} \, e^{i w \phi_J- i\pi n_J}
    \,.
\end{equation}
Here the index $J$ labels the GO images, located at stationary points of the Fermat potential $\vect x_J$  such that $\phi_{,i}(\vect x_J,\vect y)=0$, where a comma subscript indicates derivative with respect to lens-plane coordinates. The \emph{magnification} $ \mu^{-1} \equiv  \det\left(\phi_{,ij}(\vect x_J)\right)$ and the \emph{time delay} $\phi_J \equiv  \phi(\vect x_J,\vect y)$ are evaluated on the image positions. The Morse phase is $n_J = 0$, $\pi/2$ or $\pi$ depending on whether $\vect x_J$ corresponds to a minimum, saddle point or maximum of $\phi$, respectively.
In the single-image regime, GO is simply a rescaling of the waveform,  $F(w)=\sqrt{|\mu|}$.

\label{sec:time_domain_lensing}

We will now compute the amplification factor in time domain. We define the time-domain signal as the Fourier transform of $i F(w) / w$:
\begin{eqnarray}
 \label{eq:lensing_contour_time_integral}
  \Cc I(\tau) 
  &\equiv&
  \int_{-\infty}^{+\infty}  \de w \frac{i F(w)}{w} e^{-i w \tau}
  \nn \\
  &=& 
  \int \de^2 \vect x \int_{-\infty}^{+\infty} \frac{\de w}{2\pi} \exp\left({i w\left(\phi(\vect x, \vect y)-\tau\right)}\right)  
  \nn \\
  &=&  
  \int \de^2 \vect x \, \delta\left( \phi(\vect x,\vect y) - \tau\right) \,,
\end{eqnarray}
where $\delta(x)$ is the Dirac-delta function. The expression above reduces the computation of $\Cc I(\tau)$ to a one-dimensional integral over contours of equal time delay $\phi(\vect x, \vect y) = \tau$, see \cite{Ulmer:1994ij}. 
The amplification factor \eqref{eq:amplification_F} follows from Fourier-transforming back to the frequency domain.

By choosing coordinates that follow the contours, the equation above reduces to
\begin{equation}\label{eq:contour_decomposition}
        \Cc I(\tau) 
        = 
        \sum_k \int \frac{\de t \, \de s}{|\vect \nabla \phi|}
        \delta\left(t-\tau\right)
        \;,
\end{equation}
where the coordinate $t \equiv \phi(\vect x, \vect y)$ and $s$ is the arc-length distance along contours of equal time delay. The summation is over distinct contours with same time delay. We give a detailed derivation of this expression in App.~\ref{sec:appendix_ders}.

We will consider the \textit{Green's function}, defined as
    \begin{equation}\label{eq:green_function_def}
        G(\tau) 
        \equiv 
        \frac{1}{2 \pi}
        \frac{\de}{\de \tau}\Cc I(\tau)
        \,,
    \end{equation}
(also the Fourier transform of the amplification factor).
The time-domain lensed waveform is given as a convolution of the unlensed waveform $h_0$ with Green's function
    \begin{equation}
        h(t) 
        = 
        \int_{-\infty}^{+\infty} \de t^{\prime} 
        \, G(t^\prime-t) h_0(t^\prime)
        \,,
    \end{equation}
where $t \equiv 4 G \MLz \tau$.
The GO image in $G(\tau)$ appears as a singular contribution, stemming form the discontinuity of $\Cc I$ at $\tau_I$ (a Dirac delta function in the single-image regime). 
We will hereafter use the term Green's function when referring to the regular part, defined as
\begin{equation}\label{eq:green_function_split}
    \Cc G(\tau) 
    =    
    G(\tau) - \sqrt{|\mu|} \, \delta(\tau-\tau_I)
    \,,
\end{equation}
When necessary, we will refer to $G(\tau)$ as the full Green's function.

\begin{figure*}
    \centering
    \includegraphics[width=0.99\textwidth]{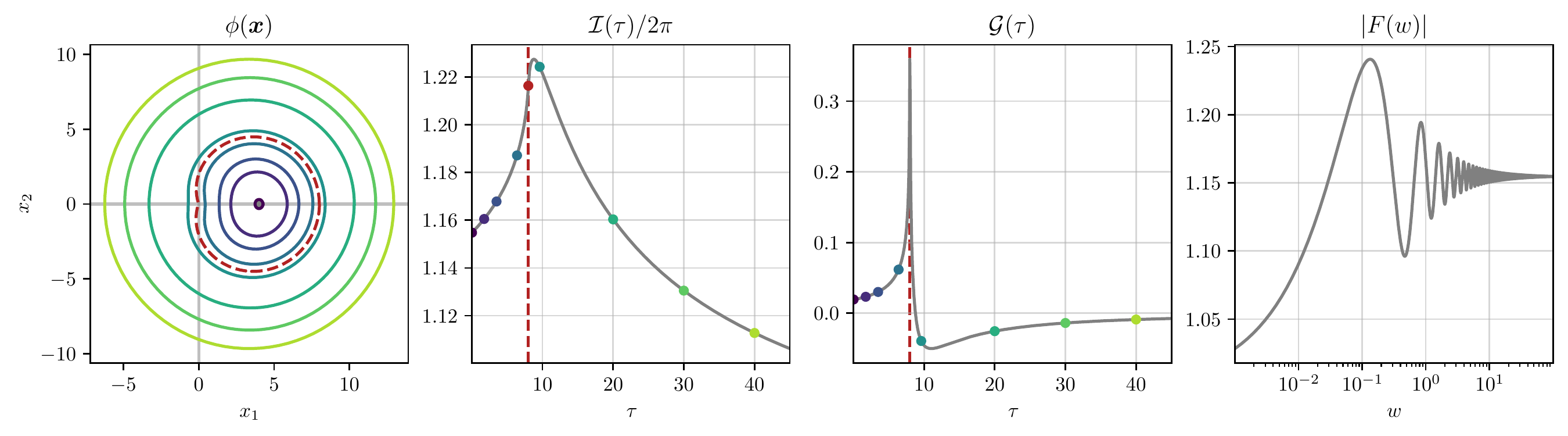}
    \caption{
    Computation of the 
    amplification factor in the single-image regime for an SIS with impact parameter $y=3$. 
    Each contour of constant Fermat potential, $\phi(\vect x)=\tau$, contributes to a point in
    $\mathcal{I}(\tau)$, see Eq.~\eqref{eq:It_non_pert}. Green's function $\mathcal{G}(\tau)$ 
    is then computed as the derivative of $\mathcal{I(\tau)}$, see Eqs.~\eqref{eq:green_function_def}
    and \eqref{eq:green_function_split}.
    Finally, the amplification factor $F(w)$ is the (inverse) Fourier transform of the full Green's 
    function $G(\tau)$. The sharp peak in Green's function is associated with the center of the 
    lens, which features a cusp in the SIS. In our case, the lens is located at $(0, 0)$ and the 
    corresponding contour is represented with a dotted line. 
    }
    \label{fig:It_contours}
\end{figure*}

Figure \ref{fig:It_contours} shows the procedure to compute WO predictions in the single-image regime. The lens is a Singular Isothermal Sphere (SIS), see Appendix \ref{sec:lens_models}.
Panels show the countours of constant Fermat potential for an example lens, the time-domain integral \eqref{eq:contour_decomposition}, Green's function \eqref{eq:green_function_split} and the amplification factor \eqref{eq:amplification_F}. Colored points in $\Cc I(\tau)$ and $\mathcal{G}(\tau)$ correspond to the contours in $\phi(\vect x)$. 
Most discussions on WO lensing have focused on the amplification factor $F(w)$. However, Green's function offers complementary insights into WO phenomena and their relation to lens properties, making certain features particularly transparent. Unlike $F(w)$, $\mathcal{G}(\tau)$ is real-valued and thus easier to display.

\subsection{Non-perturbative single-image framework}\label{sec:lensing_nonperturbative}

In this work, we focus on the single-image regime, in which only one GO image forms. The WL limit emerges as a particular case, in which deflections are small.
While the WL regime is amenable to a perturbative treatment, which we will develop in Sec.~\ref{sec:lensing_perturbative}, in this section we will first present the full framework needed to
compute the amplification factor, in the single-image regime, without additional approximations.
     
The starting point for the full computation of the 
time-domain amplification factor will be Eq.~\eqref{eq:lensing_contour_time_integral}, 
but expressed in polar coordinates
\begin{subequations}
\begin{align}
    x_1 &= x_{m\,,1} + R\cos \theta\ ,\\
    x_2 &= x_{m\,,2} + R\sin \theta\ ,
\end{align}
\end{subequations}
where $\vect{x}_m$ is the location of the minimum time delay, i.e.~$\phi(\vect{x}_m)=0$. 
With this change of coordinates we get
\begin{equation}\label{eq:It_non_pert}
    \mathcal{I}(\tau) = \int \de R \, \de \theta\, R \, \delta\left(\phi(R,\theta)-\tau\right)\ .
\end{equation}
Our main assumption to solve this integral will be that we are in the 
single-image regime, so that the global minimum is the only critical point of the
Fermat potential. Furthermore, if $\partial_R\phi\neq 0$, we can invert
\begin{equation}\label{eq:phi_R_theta}
    \phi(R, \theta) = \tau\ ,
\end{equation}
to obtain $R(\theta, \tau)$. Once this solution is found, we can plug it back and 
compute the integral as
\begin{equation}
    \mathcal{I}(\tau) 
    = 
    \int^{2\pi}_0 \de\theta \, \frac{R(\theta, \tau)}{|\partial_R\phi|}\ .
\end{equation}
In practice, what we will do is to solve the system of differential equations
\begin{align}
    \frac{\de \mathcal{I}}{\de\theta} &= \frac{R}{|\partial_R\phi|}\ ,\\
    \frac{\de R}{\de \theta} &= -\frac{\partial_\theta\phi}{\partial_R\phi}\ .
\end{align}
In this way, the curve will be sampled with the precision needed to achieve a given
tolerance in $\mathcal{I}$. The system is integrated from $\theta=0$ to $2\pi$ and
with initial conditions $\mathcal{I}(\theta=0, \tau)=0$ and $R(\theta=0, \tau)$ chosen such
that $\phi(R(0, \tau), 0) = \tau$. The previous derivation relied on the fact that
$\partial_R\phi\neq 0$, which is always the case for axisymmetric lenses when there
are no critical points. Even though we will not need it in this work, 
the previous framework only needs to be slightly modified if this is not the case
and $\partial_R\phi=0$. 
The main change to be made is that, instead of parameterizing 
the curve \eqref{eq:phi_R_theta} as $R(\theta,\tau)$, one must use a parametric 
representation $R(\sigma, \tau)$ and $\theta(\sigma, \tau)$. In this case, one 
should also keep track of the values of $R$ and $\theta$ and finish the integration 
once the contour closes. 

\subsection{Perturbative weak lensing expansion}\label{sec:lensing_perturbative}
   
\begin{figure}[t!]
        \centering
        \includegraphics[width=0.7\columnwidth]{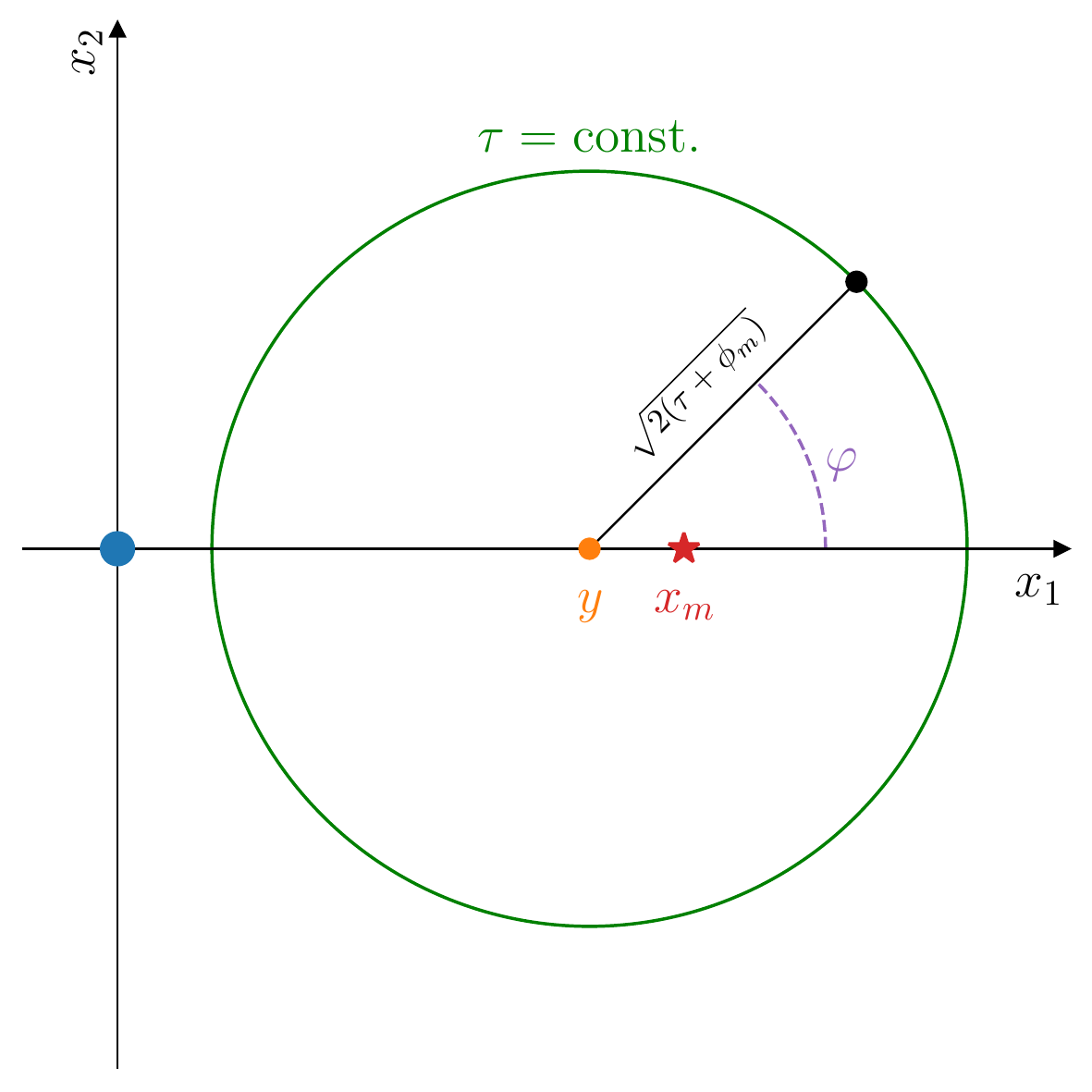}
            \caption{
                Diagram for computation $\Cc I(\tau)$ in the weak-lensing approximation.
            In the large-time delays region, contour of approximately equal $\tau$ are centered at $y$, have radius $\sqrt{2(\tau + \phi_m)}$ and angle $\varphi$ with respect to the $x_1$ axis. The lens is indicated by the blue dot, at the origin. The red star corresponds to the source, appearing at $x_m$. 
        }
        \label{fig:diag_It}
\end{figure}

We can understand how to set up a perturbative calculation in WL in the following way. Let us consider the image to be at $\vect x_m$ in the lens plane, as in Fig.~\ref{fig:diag_It}, and let us assume that $\psi(\vect x)$ grows less than the quadratic part of $\phi(\vect x, \vect y)$ at large $\vect x$.
Then, at sufficiently large $x$s from the image, the contours of constant $\phi(\vect x, \vect y) = \tau$ approach circles centered at $\vect y$ and are weakly influenced by $\psi(\vect x)$. One can then take into account the effect of $\psi(\vect x)$ perturbatively.
On the other hand, at radii comparable with the distance $|\vect y - \vect x_m|$ or smaller, the contours are still weakly affected by $\psi(\vect x)$, but cannot be parametrized at the lowest order as circles centered in $\vect y$. Indeed, the correct parametrization here is with ellipses centered at $\vect x_m$. This can be understood as the GO limit for $\Cc I(\tau)$ since regions of small time delay correspond to the high-frequency limit for $F(w)$ (see \cite{Ulmer:1994ij,Tambalo:2022plm}).
The two calculations for small and intermediate/large time delays can then be matched in an intermediate region. 
    
\subsubsection{Large time delays}\label{sec:wl_expansion}
The first step of the perturbative approach is to split the Fermat potential into a lens
contribution and a ``free'' part, $\phi(\vect x) = \phi_0(\vect x) - \psi(\vect x)$, where the
free piece
\begin{equation}
    \phi_0(\vect{x})\equiv \frac{1}{2}|\vect{x}-\vect{y}| - \phi_m\ ,
\end{equation}
still contains non-perturbative information about the lens in the minimum time delay $\phi_m$. 
After plugging this result into \eqref{eq:lensing_contour_time_integral}, we then expand in 
powers of the lensing potential $\psi$
\begin{align}
    \mathcal{I}(\tau) 
    &= 
    \int\de^2 \vect x \,
    \delta 
    \big(
    \phi_0(\vect{x})-\psi(\vect{x})
    -\tau
    \big)
    \nonumber\\
        &= \int\de^2 \vect x 
           \sum_{n \geq 0} \frac{(-1)^n}{n!} \psi^n(\vect{x}) \,
           \delta^{(n)}\big(\phi_0(\vect{x})-\tau\big)\nonumber\\
        &= \sum_{n \geq 0} \frac{1}{n!}\frac{\de^n}{\de\tau^n}
           \int\de^2 \vect x\, 
           \psi^n(\vect{x})\delta\big(\phi_0(\vect{x})-\tau\big)
           \ .
           \label{eq:It_sum_n}
\end{align}
In the second line, $\delta^{(n)}$ stands for the derivative of the Dirac delta with respect to its argument.
Without loss of generality, the impact parameter $\vect y$ can be taken to be parallel to the $x_1$ axis and with magnitude $y$.
We can use polar coordinates again, centered at the minimum of the time delay of the free case $(y, 0)$,
\begin{subequations}
\begin{align}
    x_1 &= y + r\cos \varphi\ ,\\
    x_2 &= r\sin \varphi\ ,
\end{align}
\end{subequations}
to evaluate the integrals in \eqref{eq:It_sum_n}. In general, for a generic
function $f(\vect{x})$, we can simplify the integral as
\begin{align}
    \int\de^2 \vect x \,&f(\vect{x}) \, \delta\big(\phi_0(\vect{x})-\tau\big)\nonumber\\
        &= 
        \int^\infty_0 
        r \, \de r
        \int^{2\pi}_0 
        \de\varphi \,
        f\big(\vect{x}(r,\varphi)\big) \,
            \delta\left(\frac{1}{2}r^2-\tau'\right)
            \nonumber\\
        &= 
        \Theta(t)
        \int^{2\pi}_0 \de\varphi\,
            f
            \left(\vect{x}(\sqrt{2\tau'}, \varphi)\right)\ ,
\end{align}
where $\tau' \equiv \tau + \phi_m$ and $\Theta$ is the Heaviside step function. Using this result, we can finally write the linear
approximation as
\begin{align}\label{eq:linearWL_final}
    \mathcal{I}(\tau) 
    \simeq 
    2\pi 
    + \sum_{n=1} \mathcal{I}_{(n)}(\tau)
    \ ,\qquad
    \tau > -\phi_m > 0
    \ ,
\end{align}
where
\begin{subequations}
\begin{align}
    \mathcal{I}_{(n)}(\tau) 
    &= 
    \frac{1}{n!}\frac{\de^n}{\de\tau^n}
    \int^{2\pi}_0 \de\varphi \,
        \psi^n\big(\vect{x}(\tau, \varphi)\big)\ ,
        \label{eq:time_domain_perturbative_expansion}
        \\
    x_1 &= y + \sqrt{2(\tau+\phi_m)}\cos \varphi
    \label{eq:time_domain_perturbative_expansion_x1}
    \ ,
    \\
    x_2 &= \sqrt{2(\tau+\phi_m)}\sin \varphi
    \label{eq:time_domain_perturbative_expansion_x2}
    \ .
\end{align}
\end{subequations}
This formula already captures all the essential diffraction features of the amplification 
factor with a very good accuracy that improves as $y$ increases. In the next subsection, we
will study the region of small time delays, $-\phi_m > \tau \geq 0$, where the linear 
formalism cannot be applied anymore, but analytic results from the GO expansion are available.

We now discuss the frequency-domain version of our approximation for WL signals. This will serve to illustrate how the \wow{} appears in the amplification factor. However, for applications we will evaluate lensed signals starting from the time domain.
Also, in the frequency domain, WL effects are given by an expansion in powers of $\psi(\vect x)$. We will show that at leading order in the lensing potential and for large $y$, the amplification factor is obtained by Fourier-transforming the signal from the large-time delay region.
To obtain this result, let us write $F(w) \simeq F_{(0)}(w) + F_{(1)}(w)$, where $F_{(0)}(w)$ and $F_{(1)}(w)$ are the Fourier transforms of $\Cc I_{(0)}(\tau)$ and $\Cc I_{(1)}(\tau)$, respectively.
Then, by using the expressions in Eq.~\eqref{eq:linearWL_final} and by performing the $\de \tau$ integration using the delta function, we have
\begin{align}
    F_{(0)}(w) 
    &=
    \frac{w}{2 \pi i}
    \int_{-\infty}^{+\infty} \de \tau \, e^{i w \tau} \, \Cc I_{(0)}(\tau)
    =
    e^{-i w \phi_m}
    \;,
    \label{eq:F_0}
    \\
    F_{(1)}(w) 
    &=
    \frac{w}{2 \pi i}
    \int_{-\infty}^{+\infty} \de \tau \, e^{i w \tau} \, \Cc I_{(1)}(\tau)
    \nn \\
    &=
    - \frac{w^2}{2 \pi}
    \int \de^2 \vect x
    \,
    e^{i w \phi_0(\vect x, \vect y)} \, \psi(\vect x)
    \;.
    \label{eq:F_1}
\end{align}
Notice that $\phi_m$ and $\phi_0(\vect x, \vect y)$ still depend on the lensing potential. However, we are interested in keeping only leading order terms in $\psi$. Hence, we can expand the exponents of Eqs.~\eqref{eq:F_0} and \eqref{eq:F_1} in powers of $\psi_m \equiv \psi(\vect x_m)$, truncating at linear order.
We also make use of the lensing equation at leading order: $\vect x_m \simeq \vect y + \vect \nabla_{\vect y}\psi(\vect y)$. This gives $\psi_m \simeq \psi(\vect y)$.
Then, expanding $F_{(0)}(w)$ up to first order in $\psi(\vect y)$ and adding the contribution from $F_{(1)}(w)$ leads to
\begin{equation}
    F(w) 
    \simeq 
    1 
    - 
    \frac{w^2}{2 \pi}
    \int \de^2 \vect x
    \,
    e^{i w |\vect x- \vect y|^2 / 2}
    \left(
     \psi(\vect x) - \psi(\vect y)
    \right)
    \;.
    \label{eq:F_wl}
\end{equation}
One can check that this expression correctly captures the WL features. Moreover, in the limit of large $y$ the GO result is approximately recovered. 
Indeed, in this limit the location of the image approaches $\vect x_m \simeq \vect y$.
Expanding the integrand in Eq.~\eqref{eq:F_wl} around this point and performing the Gaussian integral we obtain, at leading order in $w \gg 1$, $F(w) \simeq 1 +  \nabla^2_{\vect x} \psi(\vect y) / 2 \simeq \sqrt{|\mu|}$, as expected (see next subsection and the GO expressions in App.~\ref{sec:appendix_ders}).
Equation \eqref{eq:F_wl} can be applied to simple lenses to obtain analytic expressions in the WL regime. In App.~\ref{sec:appendix_an_res} we present the result for the SIS lens. 
An expression similar to Eq.~\eqref{eq:F_wl} is also derived in Ref.~\cite{Choi:2021jqn}.
\footnote{From our understanding, this reference subtracts $\psi(\vect x_m)$ instead of $\psi(\vect y)$ in Eq.~\eqref{eq:F_wl}. In this way, their $F(w)$ \emph{grows} at high $w$ and does not reduce to the GO limit. Therefore such an expression does not reproduce the \wow{} features we discuss in the next sections.}

\subsubsection{Small time delays}\label{subsubsec:small_t_exp}
For $\tau$ approaching the minimum time delay, the formalism above cannot be straightforwardly applied. The main effect of the lensing potential in this region is to shift the minimum (from the lens equation, at leading order in $\psi$ one has $\vect x_m \simeq \vect y + \psi(\vect y)$). At subleading order we also have a deformation of the contours contributing to a change in the magnification of the image.

The expansion of the contours near the minimum time delay leads to the GO expansion, which corresponds with the high-frequency limit of $F(w)$. One can systematically obtain this expansion without making assumptions about the size of $\psi$. Following \cite{Ulmer:1994ij,Tambalo:2022wlm}, we have in the time domain
\begin{equation}\label{eq:bgo_time_dom}
    \Cc I(\tau) 
    \simeq
    2 \pi \sqrt{|\mu|} \, \Theta(\tau)
    \left( 
    1 + \Delta_{(1)} \tau + \Delta_{(2)} \frac{\tau^2}{2}
    \right)
    \;.
\end{equation}
Here, $\mu$ is the magnification factor of the image, while $\Delta_{(1)}$ and $\Delta_{(2)}$ are the first two \emph{beyond geometric optics} (bGO) corrections. 
We give the explicit expression for these coefficients in App.~\ref{sec:appendix_ders}. Higher-order terms in $\tau$ can be obtained in a similar fashion.

In the frequency domain, Eq.~\eqref{eq:bgo_time_dom} becomes
\begin{equation}\label{eq:bgo_freq_dom}
    F(w) 
    \simeq 
    \sqrt{|\mu|} 
    \left(
    1 + \frac{i \Delta_{(1)}}{w} - \frac{\Delta_{(2)}}{w^2}
    \right)
    \;.
\end{equation}
Notice that all these expressions only require knowledge of the image location $x_m$ in order to be evaluated. Moreover, higher orders in the bGO expansion decay with higher powers of $w$ and are therefore subleading at high frequencies.
On top of the bGO terms associated with the image, other locations in the lens plane can contribute at subleading orders in $1 / w$. For instance, this is the case for points where $\psi(\vect x)$ develops cusps (typically at the lens' center). We will elaborate on this point later, see also \cite{Takahashi:2004mc,Tambalo:2022plm} for more details.

In order to connect with the expansion of Sec.~\ref{sec:wl_expansion} we need to pick a time delay $\tau_{\rm match}$ where to match the two expressions. In practice, in the large-$\tau$ expansion, it is convenient to use $\tau' = 0$ (or $\tau = -\phi_m$) as the matching point and it
is usually enough to keep only the leading order term in \eqref{eq:bgo_time_dom}. As shown in App.~\ref{app:Comparisons}, it is possible to achieve $\mathcal{O}(1\%)$ accuracy for $y>2$
by setting $\Delta_{(1)}=\Delta_{(2)}=0$ and interpolating between $\tau = 0$ and $\tau = -\phi_m$
computed using \eqref{eq:linearWL_final} with $n=1$.

\section{Analysis of lensing diffractive features}\label{sec:phenomenology}

The framework developed in the previous section allows us to compute wave-optics features (\wows) in the single-image regime. Let us now discuss the phenomenology of \wows, their dependence on the lens properties and the prospect of individually identifying and characterizing sub-lenses. We will first start with the analysis of symmetric lenses (Sec.~\ref{sec:pheno_symmetric}) before addressing models of composite ones (Sec.~\ref{sec:pheno_composite_model}) and their signatures (\ref{sec:pheno_composite_signatures}).

\subsection{Symmetric lenses} \label{sec:pheno_symmetric}

Let us start our discussion by considering simple, symmetric lenses (where $\psi$ only depends on $x \equiv |\vect x|$).
First, we will introduce the symmetric lens models (a detailed description of these lenses and their phenomenology is given in Ref.~\cite{Tambalo:2022wlm}). Then, we will discuss the \wow{} in the time domain and their dependence on the lens parameters.
Since GW analyses are often performed in the frequency domain, we will also discuss the \wow{} as a function of $w$.

\begin{figure*}
    \centering
    \includegraphics[width=\textwidth]{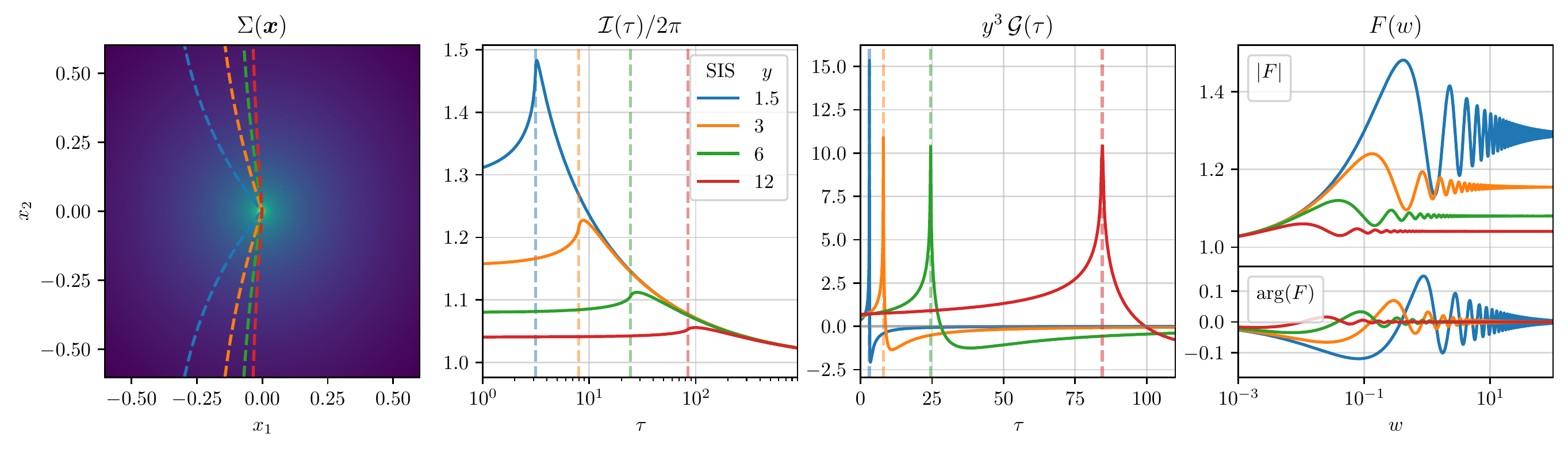}
    \caption{Role of the impact parameter $y$ on an SIS lens. The columns show the projected density, time-domain integral, Green's function and amplification factor. Green's function {omits the GO contribution and} has been rescaled by $y^3$ so the \wows{} can be appreciated at large impact parameter. {Dashed lines show the $\phi(\vect x,\vect y)=\tau_C$ contours passing through the lens' center and the associated values of $\tau$.}
    }
    \label{fig:SIS_vary_y}
\end{figure*}
\begin{figure*}
    \centering
    \includegraphics[width=\textwidth]{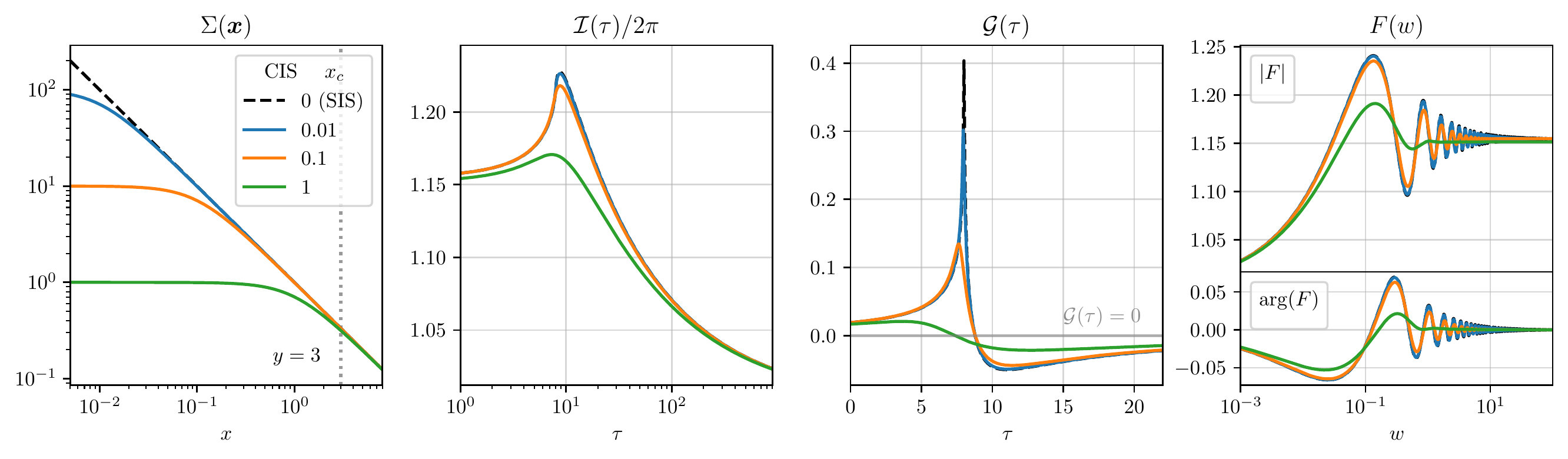}
    \caption{Role of the core size $x_c$ on an CIS lens. The columns show the projected density, time-domain integral, Green's function and amplification factor. These examples demonstrate the effect of the matter distribution on the shape of the lensing diffractive feature.
    }
    \label{fig:CIS_vary_xc}
\end{figure*}

Our discussion of symmetric lenses will focus on the well-known SIS and a one-parameter generalization, the Cored Isothermal Sphere (CIS).
The SIS is characterized by a central cusp with diverging density, $\Sigma(x) \propto x^{-1}$. 
In the GO limit, the SIS can have one or two images depending on whether the impact parameter is outside or inside the caustic $y_{\rm cr} = 1$, respectively.
The CIS has finite density with  $\Sigma(x) \propto \left(x^2+x_c^2\right)^{-1/2}$, where $x_c \equiv r_c/\xi_0$ is the projected size of the core.
Similarly to the SIS, multiple images form for sources within the caustic $y_{\rm rc}(x_c)\leq 1$ (smaller than for SIS, multiple images also require $x_c < 1/2$). 
More details about both lens models are given in App.~\ref{sec:lens_models}.

Let us now describe the WO phenomenology of these lenses in the single-image regime.
Figures \ref{fig:SIS_vary_y} and \ref{fig:CIS_vary_xc} show, respectively, the predictions for an SIS at varying $y$ and for a CIS at fixed $y$ but with different $x_c$s. 
For concreteness, we will first discuss Green's function $\Cc G(\tau)$, which makes the discussion especially transparent, and comment on the amplification factor $F(w)$ afterwards.
We will describe the overall structlensing gravitational waves o3ure of the \wow, then the role of lens parameters and discuss how they could be measured from GW observations. 

Single-image \wows{} begin at $\tau_{I}$, which corresponds to the minimum of the time delay (i.e.~the type I image, we set $\tau_{I}=0$ by convention). The GO image appears as a delta function in the full Green's function, Eq.~\eqref{eq:green_function_split}, while the WO piece $\Cc G(\tau)$ features a discontinuity, associated to the bGO correction at the position of the image (originating from the coefficient $\Delta_{(1)}$ in Eq.~\eqref{eq:bgo_time_dom}). $\Cc G(\tau)$ is initially positive and increases with $\tau$, as the contours approach the center of the lens. The location of the lens is associated to a peak in Green's function. At slightly higher $\tau$, $G(\tau)$ becomes negative and asymptotes towards zero as $\tau\to \infty$.

For symmetric lenses with a cusp (e.g.~SIS) the peak of the \wow{} is located at $\tau_C = \phi(\vect x=0, y)$. The peak is due to the high curvature of the constant-$\phi$ contours. The curvature, and hence the height of the peak, depends on the lens profile as well as the impact parameter: denser lenses and lower impact parameter produce taller peaks.

In non-differentiable lenses the peak is singular. For instance, in the linear weak-lensing approximation for the SIS lens, Green's function is found to have a logarithmic divergence (see App.~\ref{sec:appendix_an_res} for a derivation)
\begin{equation}
    G(\tau) 
    \simeq 
    - \frac{1}{\pi y^3} \log\left(\left| \tau -\tau_C \right| \right)
    \;.
\end{equation}

In the frequency domain, the peak in $\Cc G(\tau)$ is directly related to the damped oscillations seen in $|F(w)|$: sharper peaks have more pronounced features that decay more slowly with $w$, and can thus be observed at higher frequencies.
The singular contribution is related, in the frequency domain at large $w$, with subleading terms in the bGO expansion originating from regions of the lens plane close to the center of the lend. See discussion in App.~\ref{sec:appendix_an_res} around Eq.~\eqref{eq:go_sis_feature} and Refs.~\cite{Takahashi:2004mc,Tambalo:2022wlm} for more details.

Let us now discuss how the lens parameters affect the \wow, separating the peak and broad shape.
WO predictions are independent of $\MLz$ when expressed in terms of $y$, $\tau$ and $w$. However, $\MLz$ can be inferred when restoring the units for $t$ or $f$, given an observed waveform. We will discuss this at the end of this subsection, together with prospects for lens parameter recovery.

The impact parameter $y$ controls the position and amplitude of the \wow{} peak. This is shown in Fig.~\ref{fig:SIS_vary_y} for an SIS at $y \in (1.5, 12)$, always in the single-image regime. We find the scalings
\begin{equation}\label{eq:SIS_ldf_peak}
    \tau_C \propto y^2/2 \,, \quad \Cc G(\tau\sim \tau_C)\propto y^{-3}\,.
\end{equation}
The time-delay scaling is exact for the SIS, but Green's functions dependence is only valid in the WL limit, $y\gg 1$. 
Because the peak in $\Cc G(\tau)$ involves short timescales, in the frequency domain it corresponds to high frequencies, i.e.~the damped oscillatory pattern after the maximum of $F(w)$.

The broadband shape of the \wow{} is also sensitive to the impact parameter. This is determined by the behaviour of $\mathcal G(\tau)$ over large time delays and is therefore captured by low $w$ features of $F(w)$. 
We can thus use the position and height of the first peak, $w_0$ and $F(w_0)$ respectively, to characterize the broadband \wow. We find the scalings
\begin{equation}\label{eq:SIS_ldf_broadband}
w_0 \propto \frac{1}{y^2}\,,\quad |F(w_0)-1| \propto \frac{1}{y}\,.
\end{equation}
The scaling of $w_0$ is approximate, but the scaling of $F(w_0)$ is very accurate. The likely cause is that the height is dominated by the transition between the behaviour near the image (with a well-defined scaling with $y$) and the asymptotics of the lensing potential (independent of $y$). 
See Eq.~\eqref{eq:go_sis_feature} for an analytic estimate for the scalings in the frequency domain.

Measuring a \wow{} may allow one to infer the magnification of the image, which is not directly observable in the GO limit. This is because $y$ determines $\mu$, as can be seen from either the height of $\Cc I(\tau \to 0)$ or the asymptotic value of $|F(w\to \infty)|$. For the SIS in the single-image regime
\begin{equation}
    \mu = 1+\frac{1}{y}\,,
\end{equation}
(this also holds approximately true for lenses with the same large-$x$ behaviour, like the CIS). Therefore, Eqs.~\eqref{eq:SIS_ldf_peak}, \eqref{eq:SIS_ldf_broadband} imply the following scalings
\begin{align} 
\label{eq:delensing_simple_lens_relations_peak}
    & \mu-1 \propto  \tau_C^{-1/2}\,, \quad  \mu-1\propto \Cc G(\tau\sim \tau_C)^{1/3}\,, 
    \\  \label{eq:delensing_simple_lens_relations_broadband}
    & \mu-1\propto w_0^{1/2}\,, \quad \;  \mu-1 \propto  |F(w_0)| \,.
\end{align}
The correlation between the $\mu$ and the \wow{} properties opens the possibility of mitigating the uncertainty due to WL in standard sirens. We will comment on this potential application in Sec.~\ref{sec:applications_delensing}.

We explore the role of the lens compactness and shape by considering a CIS with variable core size $x_c$.
Figure \ref{fig:CIS_vary_xc} shows results at fixed $y=3$, but varying $x_c$ between $0$ (the SIS limit) and 1 (a sub-critical lens, unable to form multiple images even for $y\to 0$). The main effect of $x_c$ is on the amplitude and shape of the \wow{} peak: smaller cores produce narrower and taller peaks that persist at higher frequencies. Larger cores also shift the position of the peak slightly towards lower $\tau_C$: in this case the peak is associated to the edge of the core region, where contour curvature is maximum, rather than the lens' center, where the contours are much smoother.
In the sub-critical lens case ($x_c=1$) the peak in $\Cc G(\tau)$ is barely recognizable. The broadband \wow{} is still apparent in $F(w)$ by the onset of diffraction, which is caused by the overall transition rather than the peak.

Let us briefly discuss the prospect of constraining lens parameters from the observation of weakly-lensed GWs. Such an inference is possible in principle, at least if we assume a lens model. If we assume an SIS, we can infer the lens parameters from the \wow{} peak position and height via Eq.~\eqref{eq:SIS_ldf_peak}. 
The degeneracy between $\MLz$ and $y$ can be broken because the peak's position $\tau_C = t_c/4G \MLz$ depends on the effective lens mass, while its amplitude $\Cc G(\tau_C)$ depends only on $y$. 
Converting the projected mass $\MLz$ into the halo mass, $\Mvir$, requires knowledge of $\xi_0$, which depends on the redshift of the source and the lens. While $z_S$ can be constrained by the amplitude of the signal, $z_L\in (0,z_S)$ is generally unknown and only a lower bound on $\Mvir$ can be derived (corresponding to the largest $\xi_0$). Nonetheless, assuming a halo mass function enables a probabilistic inference of $\Mvir$, which is sharply peaked around the minimum possible value (see Ref.~\cite{Tambalo:2022wlm} Sec.~V\,A for details). Additional leverage on the lens parameters can be obtained from the broadband feature.

More general lens models will make parameter inference from the \wow{} more challenging. As we saw in the case of the CIS (Fig.~\ref{fig:CIS_vary_xc}), the lens' internal parameters affect the height of the peak, complicating the distinction between $y$ and $\MLz$ outlined above. These additional parameters can be constrained by the shape of the peak in the \wow{} and the broadband feature. Nonetheless, degeneracies with $\MLz$ and $y$ will affect the precision (see Sec.~IV in Ref.~\cite{Tambalo:2022wlm} for examples in strong lensing) and will lead to biases if the wrong lens model is assumed. 
Because the \wow{} depends on the entire lens, it is possible to reconstruct $\psi(x)$ given $\Cc I(\tau)$ under several assumptions (weak lensing, symmetric lens and known $y$). We discuss this possibility in Sec.~\ref{sec:applications_reconstruction}.

\subsection{Modelling composite lenses} \label{sec:pheno_composite_model}

Let us now address how GW observations may probe a lens with a non-symmetric profile with an internal structure.
We will consider a matter distribution composed of $N_{\rm sub}$ objects with a common projected profile, $\Sigma_{\rm sub}$, and mass
    \begin{equation}\label{eq:comp_lens}
        \Sigma_{\rm N}(\vect x) 
        = 
        \frac{1}{N_{\rm sub}}
        \sum_{i=0}^{N_{\rm sub}} 
        \Sigma_{\rm sub}(\vect x-\vect x_i)
        \,.
    \end{equation}
We consider equal-mass sublenses for simplicity, but our expressions are easy to generalize to a mass distribution.
The centers of each sub-lens, $\vect x_i$, will be drawn from a distribution $P(\vect x_i)$. Together with $N_{\rm sub}$, the functions $P(\vect x_i)$ and $\Sigma_{\rm sub}$ determine the statistical properties of the composite lens.
We stress that this composite lens model is not intended to be a realistic realization of a halo, but it provides insights on the \wow{} produced by substructures.

The average surface density for the composite lens is a convolution of the distribution $P(\vect x)$ with the sub-halo profile $\Sigma_{\rm sub}$
\begin{equation}\label{eq:comp_lens_avg}
    \braket{\Sigma_{\rm N}(\vect x)} 
    =
    \int \de^2 \vect x' 
    P(\vect x')\Sigma_{\rm sub}(\vect x-\vect x')
    \,.
\end{equation}
Because of linearity, an analogue expression can be derived for the lensing potential: $\psi_{\rm sub}(\vect x) =  \int \de^2 \vect x' P(\vect x') \psi_{\rm sub}(\vect x-\vect x')$. 
Equation \eqref{eq:comp_lens_avg} has some obvious limits: if $\Sigma_{\rm sub}$ is a delta function then the average profile is given by $P(\vect x)$ and vice-versa. However, considering two extended functions gives non-trivial profiles in general.
It is in principle possible to derive an expression for the variance,
higher order statistics and correlations at different points $\vect x_1,\vect x_2$, to further characterize the convergence towards the average lens as $N_{\rm sub}\to \infty$.

\begin{figure*}
    \centering
    \includegraphics[width=\textwidth]{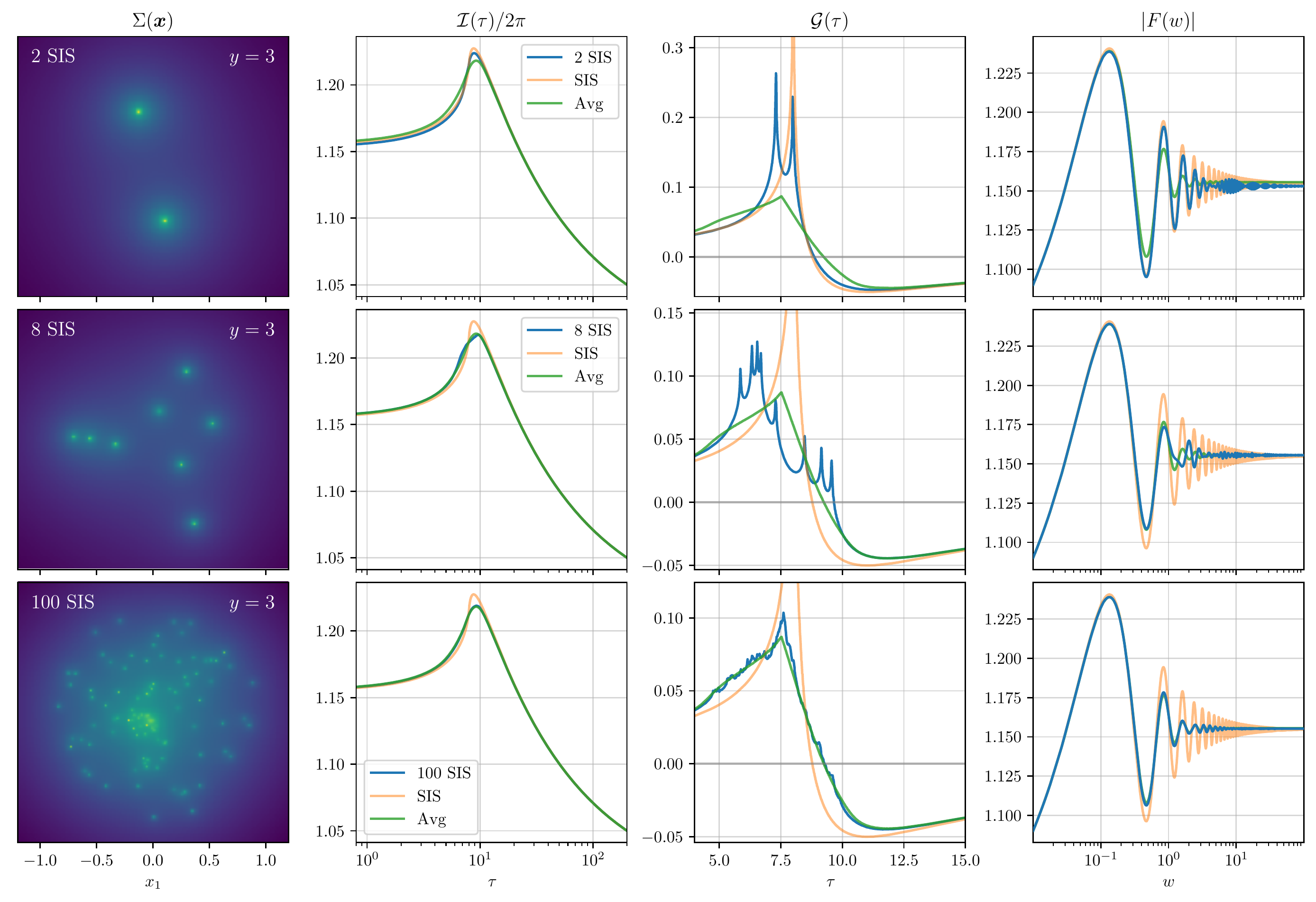}
    \caption{Convergence to the homogeneous lens when the number of sublenses is increased. The columns show the projected density, time-domain integral, Green's function and amplification factor for 2-100 SIS. Note that substructures in a realistic halo would appear far more scattered.
    }
    \label{fig:comp_lens_N}
\end{figure*}

For the composite lens, we will assume a distribution of sublenses that 
follows the SIS projected density profile, and model each of the sublenses as SISs
\begin{align}
    & \Sigma_{\rm sub}(\vect x)
    =
    \Sigma_{\rm SIS}(x) \, ,
    \\
    & P(x, \phi) 
    =
    \frac{1}{2\pi x} \Theta(x_{\rm max} - x)
    \,,
    \label{eq:composite_lens_Px_distrib}
\end{align}
where $\vect x$ is written in polar coordinates $x$ and $\varphi$. The truncation at $x_{\rm max}$ ensures a finite average profile \eqref{eq:comp_lens_avg} for the above $x$, $x^\prime$ dependence.
Writing $x' = \xi x$ and using the above definitions, we obtain the average projected profile and the lensing potential:
\begin{align}
    & \braket{\Sigma_{\rm N}} 
    = 
    \frac{1}{\pi}\int_0^{\pi} \de\varphi 
    \int_0^{\frac{x_{\rm max}}{x}}
    \frac{\de\xi}{\sqrt{1+\xi^2-2\xi\cos\varphi}}
    \,, \\
    & \braket{\psi_{\rm N}} 
    = 
    \frac{x^2}{\pi}\int_0^{\pi} \de\varphi 
    \int_0^{\frac{x_{\rm max}}{x}}
    \de\xi\sqrt{1+\xi^2-2\xi\cos\varphi}
    \,.
\end{align} 
These expressions can be solved numerically.
Let us now explore different realizations of this composite lens and their average limit.

\subsection{Signatures of composite lenses}\label{sec:pheno_composite_signatures}

As we saw, \wows{} are characterized by their broadband modulations and the peak's position/delay, amplitude and shape. Each of these characteristics depends on the lens parameters. 
An important property of a \wow{} is that it is approximately linear in the projected density $\Sigma(\vect x)$. Hence, Green's function $G(\tau)$ (or $\Cc I(\tau)$) of a composite lens is well described as the sum of the \wow{}s associated to each of the sublenses, appropriately time-shifted, at least in the WL limit.
Therefore, \wow{} peaks in GW data could be used to identify individual sub-lenses, infer their spatial distribution and constrain their properties. 
We will use the composite lens, Eq.~\eqref{eq:comp_lens}, and its average profile, \eqref{eq:comp_lens_avg}, to investigate how the number of sub-lenses and their distribution leave a characteristic imprint.

\begin{figure*}
    \centering
    \includegraphics[width=\textwidth]{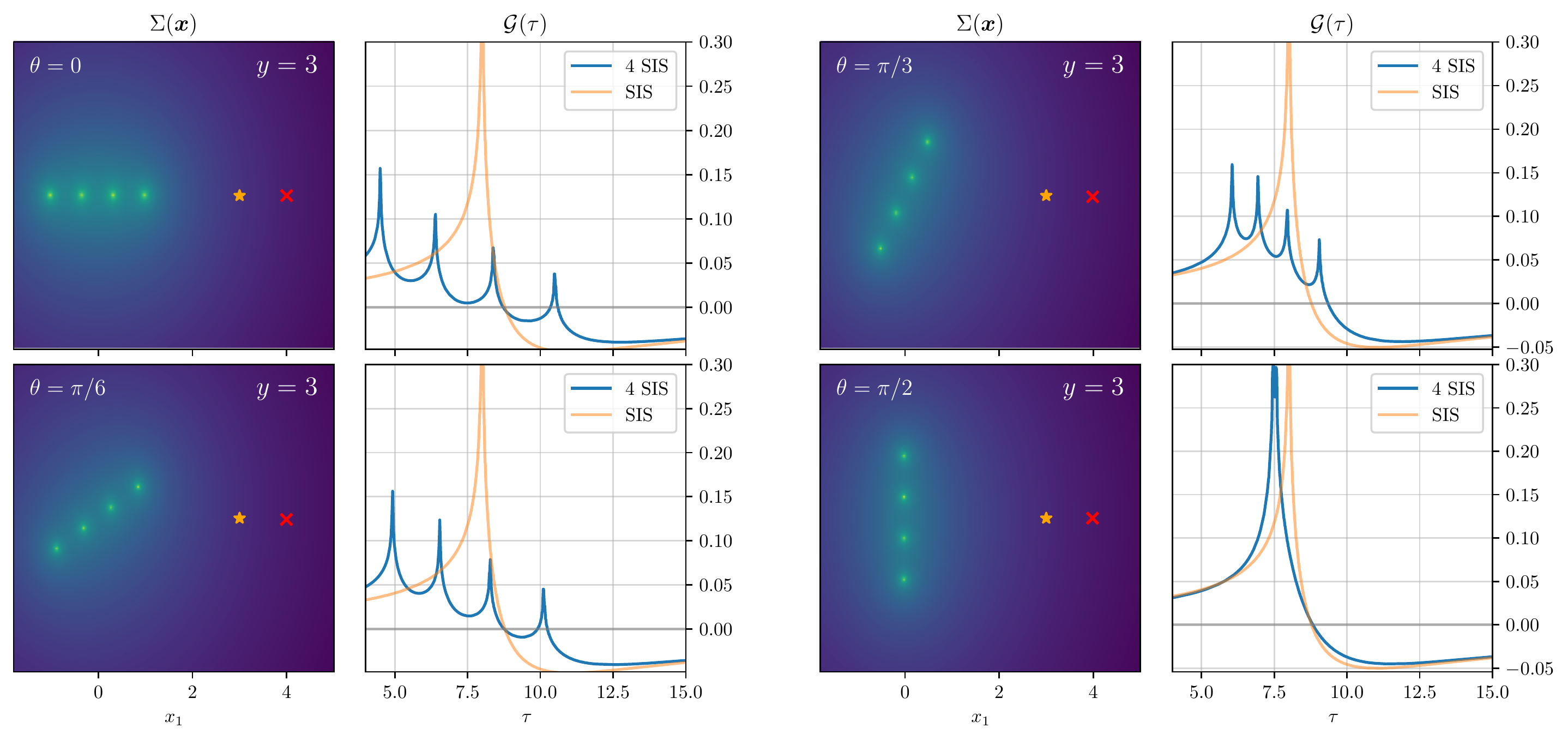}
    \caption{
    Predictions for a chain of 4 sub-lenses as a function of its orientation $\theta$ with respect to the optical axis aligned with the source (orange star). Only projected density and Green's function are shown. The sublenses appear as a series of peaks in $\Cc G(\tau)$, whose distribution over $\tau$ depends on the orientation. A chain perpendicular to the optical axis (lower right) produces closely overlapping peaks, which have a distinct shape from the SIS prediction.  The image position (red cross) at different orientations differ marginally.}
    \label{fig:comp_lens_line}
\end{figure*}

We will start exploring the effect of the number of sublenses.
Figure \ref{fig:comp_lens_N} shows $\Sigma(\vect x)$, $\Cc I(\tau)$, $\Cc G(\tau)$ and $F(w)$ for realizations of the composite lens \eqref{eq:comp_lens} with $N_{\rm sub}=2$, $4$, $100$ sublenses with the same total mass and at fixed $y=3$. Predictions for a single SIS and the averaged lensing profile, Eq.~\eqref{eq:comp_lens_avg}, are shown for comparison.
Information about the sublenses is most transparent in Green's function: each object forms a distinct peak, whose height and position are determined by its mass and separation from the GO image in the lens plane (following the trend seen for symmetric lenses in Fig.~\ref{fig:SIS_vary_y}). When the number of sublenses is large, $N_{\rm sub}=100$, $\Cc G(\tau)$ follows the averaged profile closely, with some stochastic ``cuspiness" added.
In the frequency domain, the existence of multiple peaks becomes an interference pattern in the damped oscillations. For $N_{\rm sub}=100$ the superposition is mostly incoherent, reproducing the prediction of the average profile.

The broadband shape of the \wow{} (as characterized by the position and height of the first peak) is independent of the number of sublenses. The relative difference in $F(w_0)$ between the models shown in Fig.~\ref{fig:comp_lens_N} is $\sim 10^{-3}$ between the SIS and the average lens and $\sim 10^{-4}$ between the average and composite lens $N_{\rm sub}$. 
This homogeneity is in stark contrast with the patterns of peaks/damped oscillations for different values of $N_{\rm sub}$. This consistency may reflect the broadband \wow{} only depending on the difference between the GO region and the asymptotic behaviour of the lensing potential, as argued above.

Let us now explore the effects of the mass distribution on the \wow. Figure \ref{fig:comp_lens_line} shows $\Sigma(\vect x)$ and $\Cc G(\tau)$ for composite lenses made of four equal-mass SIS profiles, equally spaced along a straight line. 
This type of sublens distribution is a crude approximation to filamentary structures in the cosmic web. 
In this case, the position of the four peaks satisfies $ \tau_n \sim \frac{1}{2}(y- \alpha n\cos\theta)^2+ (\alpha n \sin\theta)^2$, where $\theta$ is the angle between the line of sublenses and the optical axis and $\alpha$ controls the spacing. Therefore, the separation of \wow{} peaks in the time domain could be used to infer aspects of the spatial distribution. This shows that, despite all the information being compressed into a single dimension (e.g.~$G(\tau)$, $F(w)$), it might be possible to reconstruct the morphology of a 2-dimensional (projected) structure under certain circumstances. We will discuss this potential application in Sec.~\ref{sec:applications_LSS_morphology}.

Just as for individual lenses, identifying separate peaks in $\Cc G(\tau)$ can provide information on substructure mass and relative positions, cf.~Eq.~\ref{eq:SIS_ldf_peak} (although assumptions about the lens profile might be necessary, cf.~Sec.~\ref{sec:pheno_symmetric}). 
In the limit in which the linear approximation holds, $\Cc G(\tau)$ only depends on the projected matter distribution of the lens, with the amplitude determined by the overall impact parameter. A high-quality observation of a GW signal thus offers additional constraints on the inner structure of the lens.
Identifying the \wows{} allows the reconstruction of substructures, although certain degeneracies persist. First, since WO lensing encodes 1-dimensional information (through $\Cc I(\tau)$, $G(\tau)$ or $F(w)$) it is not possible to recover the 2-dimensional projected distribution $\Sigma(\vect x)$. This is obvious from the linear WL limit, in which the position and height of the peak only depend on the offset between the sub-lens and the GO image (for fixed sub-lens profile). 
Second, lack of knowledge of the lens redshift prevents us from accurately determining the virial mass of the lens (or sublenses), as converting $\MLz$ (observed) into a virial mass requires knowing $d_{\rm eff}$ and thus $z_S$ (constrained from the signal amplitude) and $z_L$ (unconstrained). As already discussed, this degeneracy can be partially broken using probabilistic information, see Sec.~VA in Ref.~\cite{Tambalo:2022wlm}. 

\section{Observational prospects} \label{sec:observation}

Let us now discuss the prospects of observing \wows{}. 
We will first derive the maximum impact parameter at which \wows{} can be detected for a given GW source (Sec.~\ref{sec:observation_ycrit}). We will then include information about the halo abundances to estimate the probability of observation (Sec.~\ref{sec:observation_rates}) and the prospect of constraining the halo mass function (Sec.~\ref{sec:observation_constraints}).
Our estimate of the probabilities includes only isolated halos: we conclude this section by discussing subhalos and their expected imprints (Sec.~\ref{sec:observation_subhalos}).

\subsection{Critical impact parameter} \label{sec:observation_ycrit}

Assessing whether \wows{} are detectable requires accounting for the details of GW sources and the instrument's sensitivity. We will assume sources to be equal-mass ratio, non-spinning compact binary coalescence and describe them with the \texttt{IMRPhenomD} \cite{Husa:2015iqa} model in the \texttt{PyCBC} package \cite{alex_nitz_2023_7885796}. 
We will focus on LISA \cite{LISA:2017pwj,LISACosmologyWorkingGroup:2022jok} and the Einstein Telescope (ET) \cite{Maggiore:2019uih,Kalogera:2021bya}. For each instrument, we will include the effect of sky inclination and polarization averaging over the antenna pattern functions \cite{Robson:2018ifk}.
Our results will then reflect typical sources: neither optimally aligned nor close to a blind spot of the instrument.
We further consider the detector to be static. This is a good approximation since there is a single image and the SNR is very concentrated around the merger (see Sec.~IVA in Ref.~\cite{Tambalo:2022plm} or Ref.~\cite{Marsat:2020rtl}). Finally, we will only obtain results for single detectors: a detector network will in general improve the prospects of detection by improving SNR and sky coverage.

We will assess the detectability of \wows{} based on the mismatch between the lensed and unlensed waveforms. For two generic waveforms, $h_1$ and $h_2$, the mismatch is defined as:
\begin{equation}
\label{eq:def_mismatch}
    \mathcal{M} 
    \equiv 
    1-\frac{( h_1|h_2 )}{\sqrt{( h_1|h_1)}\sqrt{( h_2|h_2 )}}
    \,.
\end{equation}
Here we introduced the noise-weighted inner product for two signals $h_1(t)$ and $h_2(t)$ with Fourier transforms $\tilde{h}_1(f)$ and $\tilde{h}_2(f)$:
\begin{equation}
    (h_1|h_2) \equiv 4\,\mathrm{Re} \int_{0}^{\infty} \frac{\de f}{S_n(f)} \tilde{h}_1(f)^* \tilde{h}_2(f)  \,,
\end{equation}
where $S_n(f)$ represents the sky-averaged one-sided detector power spectral density. In terms of this product, the signal-to-noise ratio is $\mathrm{SNR} \equiv \sqrt{(h|h)}$.

According to the Lindblom criterion~\cite{Lindblom}, two waveforms are considered indistinguishable if the condition $(\delta h|\delta h)<1$ is satisfied, where $\delta h \equiv h_1-h_2$. When considering signals with comparable SNR, as for a weakly lensed and an unlensed signal, this criterion requires $\mathcal{M} \times {\rm SNR}^2<1$. In general, the converse of the Lindblom criterion is not true. Other factors, such as correlations between waveform parameters and bias in the recovered source parameters, may restrict the detectability. 
Parameter degeneracies can be accounted for using the Fisher information matrix \cite{Vallisneri:2007ev}. However, Ref.~\cite{Tambalo:2022wlm} showed that the Fisher matrix can overestimate the precision in lensing parameters due to the breakdown of the linear signal approximation: this issue depends on the lens model and parameters and requires a case-by-case inspection. 
Detectability in the Fisher matrix approach is usually defined through the standard deviation of the marginalized posterior, leading to results that depend strongly on whether $\MLz$ or $y$ is used (see discussion below). 
Ultimately, Lindblom and Fisher analyses are answering different questions about the information gained from the signal.
Due to computational simplicity and the arguments discussed above, we will employ the \emph{flipped} Lindblom criterion, considering \wows{} to be detectable if $\mathcal{M} \times {\rm SNR}^2 > 1$. Henceforth, $\mathcal{M}$ is the lensed-to-unlensed mismatch, unless stated otherwise. At the end of the section we will compare our results to other methods.

Weak lensing on a single image can be detected at $y \gg 1$ via \wows{}. We stress that in GO this is not possible because a time delay and amplitude magnification on a single image are degenerate with source properties (luminosity distance, coalescence time). The condition $\mathcal{M}(y_{\rm cr}) \times {\rm SNR}^2 = 1$ determines the \emph{critical} impact parameter, $y_{\rm cr}$, which characterizes the minimum level of alignment between a lens and source required for detectability. We will omit the explicit dependence of $y_{\rm cr}$ on the source and lens properties. We will assume all lenses to be described by the SIS profile, regardless of their mass.

\begin{figure}[t!]
    \centering
    \includegraphics[width=\columnwidth]{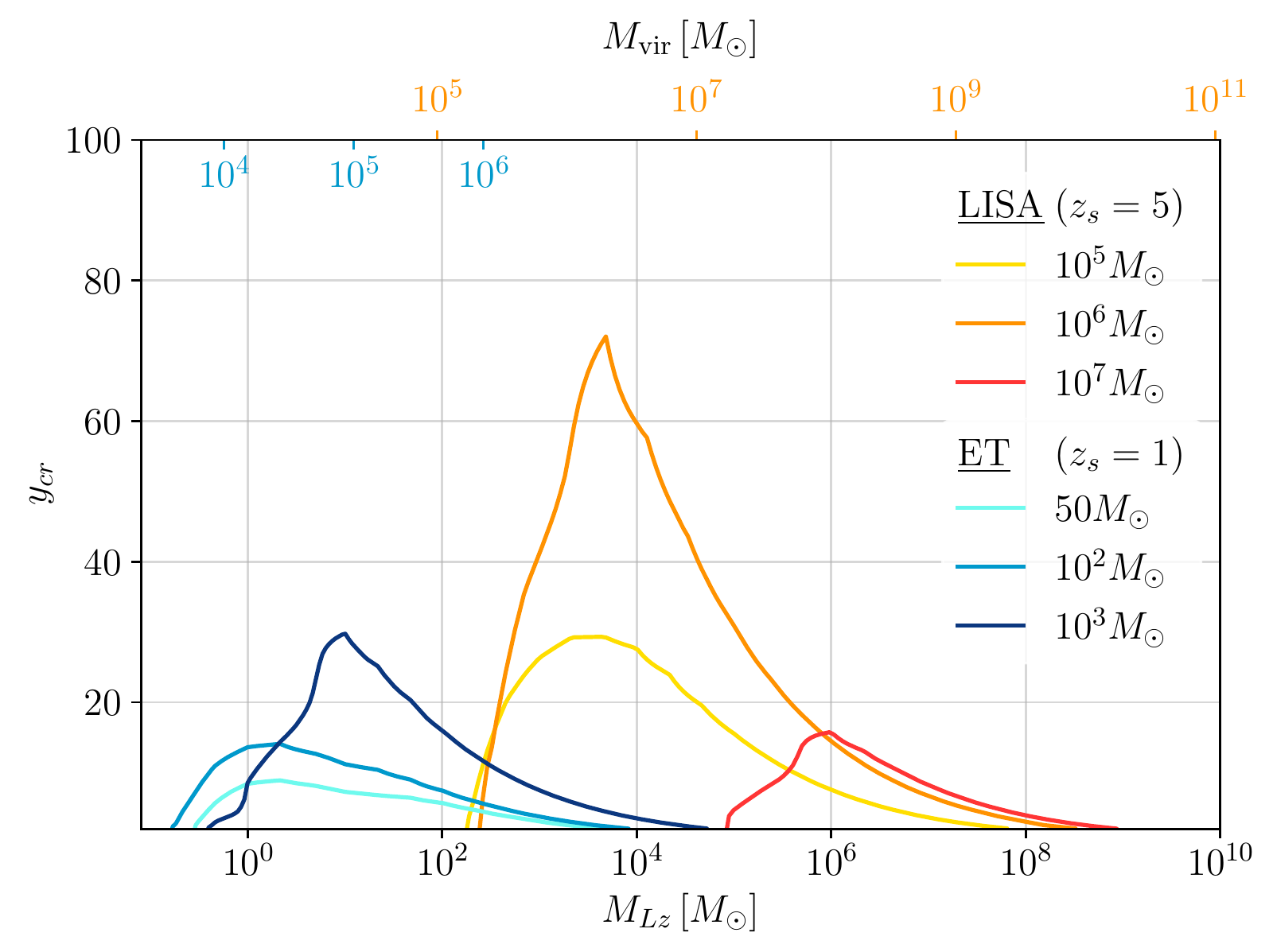}
    \caption{Critical impact parameter curves plotted against the lens mass $\MLz$ (bottom axis) and the virial mass $\Mvir$ (top axis), assuming $z_L=z_S/2$. The orange color scheme corresponds to lensed MBBH at $z_S$ observable by LISA, while lighter BBHs at $z_S=1$ observable by ET are in blue.}
    \label{fig:ycrit}
\end{figure}

\begin{figure}[t!]
    \centering
    \includegraphics[width=\columnwidth]{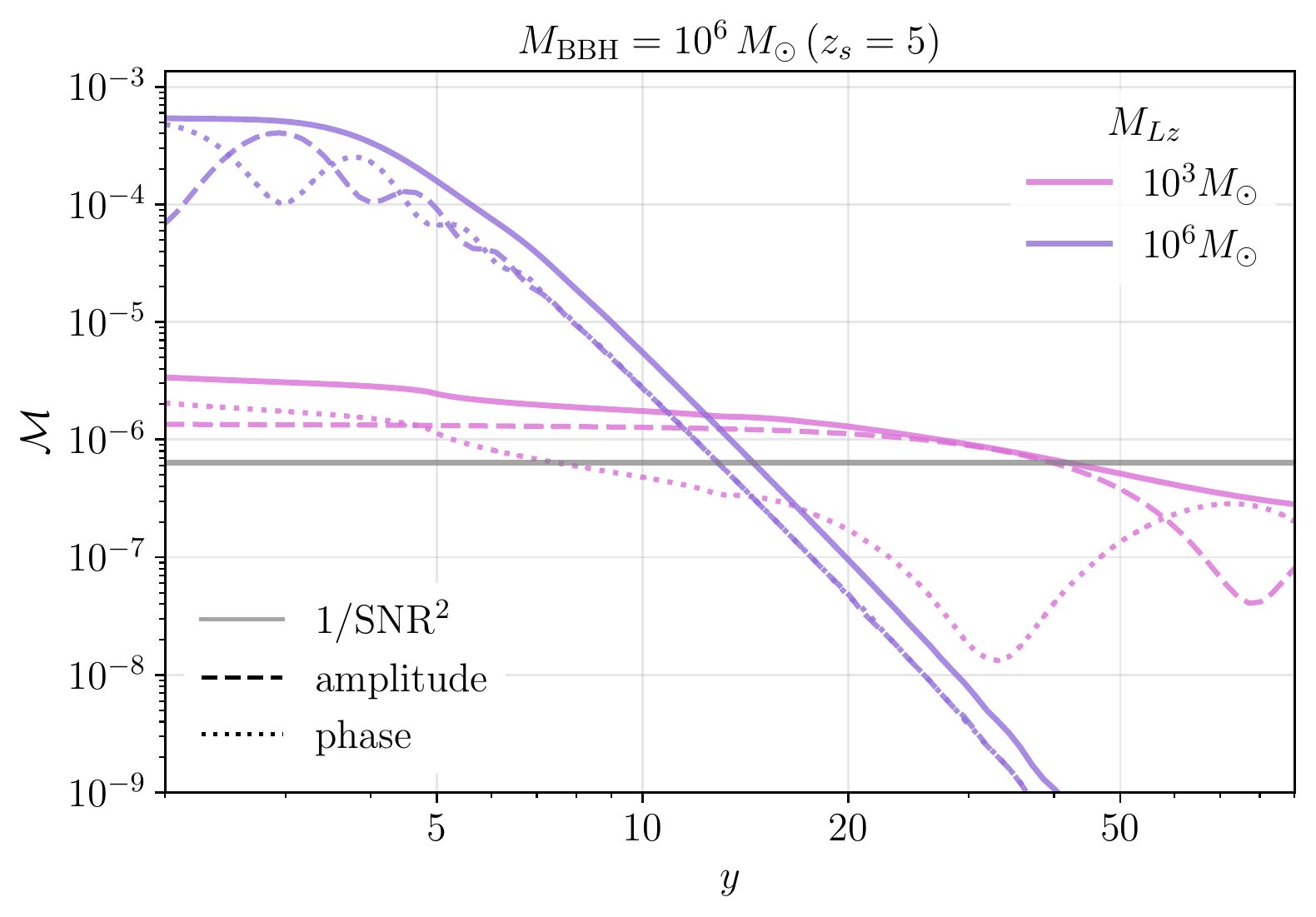}
    \caption{Lensed-to-unlensed (SIS) mismatch as a function of the impact parameter. The dashed and dotted lines correspond to modifications in the amplitude and phase in the lensed waveform, respectively. 
    }
    \label{fig:mismatch_amp_ph}
\end{figure}

The critical impact parameter depends on both the lens' and the binary's masses, as shown in Fig.~\ref{fig:ycrit}. Here, the virial mass, $\Mvir$, associated with $\MLz$ is shown on the upper scale, assuming a lens redshift of $z_L=z_S/2$ (the relationship between both masses is discussed in App.~\ref{sec:lens_models}).
The mass range covered by each curve depends on the frequency range spanned by the waveform and the corresponding detector response. 
Typically, since the merger frequency of a BBH dominates the SNR and scales as $1/M_{\rm BBH}$, the onset of WO is expected at larger lens masses for heavier BBH. This is evident when comparing LISA and ET systems in the Figure. However, for some signals, the dominant contribution to the SNR might come from a stage different from the merger. For instance, in LISA's MBBH with $M_{\rm BBH}<10^6 \, M_\odot$, the SNR is dominated by the contribution at the minimum of the detector's sensitivity curve, at $\sim 3 \times 10^{-3} \, {\rm Hz}$. The critical curves of such systems share a common range of lens masses.
From the Figure, we infer that lenses with $\MLz\simeq 5\times 10^3 \, M_\odot$ produce detectable \wows{} on the signals from MBBHs with $M_{\rm BBH}\simeq 10^6 \, M_\odot$ at $z_{S}=5$ even for misalignment as large as $y_{\rm cr}\simeq 71$. This is possible thanks to the large unlensed average SNR of such systems, ${\rm SNR}\sim 1.3 \times 10^3$, and to the onset of the WO effect at the merger frequency of the binary. Analogous conclusions can be drawn for binaries in the ET's band, such as BBHs with $M_{\rm BBH}\simeq 10^3 \, M_\odot$ at $z_{S}=1$ lensed by a $10 \, M_\odot$ object offset by $y_{\rm cr}\simeq 29$.

Figure \ref{fig:mismatch_amp_ph} shows the mismatch as a function of $y$ for fixed $M_{Lz}$ and source properties. The mismatch increases below $y<y_{\rm cr}$, but it saturates at low enough $y$. Interestingly, the mismatch at $y\ll y_{\rm cr}$ is lower for the cases with larger $y_{\rm cr}$: the signals with the more clear signatures (higher $\mathcal{M}$) are not the ones for which the detection probability is most enhanced (high $y_{\rm crit}$).
The Figure also shows the effect on both the amplitude and the phase of the GW signal. The two contributions oscillate as a function of the impact parameter, as when approaching the WO regime.


The behaviour of the $y_{\rm cr}$ curves presented above can be understood using the approximate analytical form of the \wow{} for the SIS lens given by Eq.~\eqref{eq:go_sis_feature}, valid at large $w$. Here, the \wow{} arises from the lens center as $\delta F_c = i / (w y^3) e^{i w \phi_c}$. Since we are considering large $y$ values, this is a small correction to the amplification factor, and the critical curves can be derived by expanding the mismatch $\Cc M$ of Eq.~\eqref{eq:def_mismatch} in this quantity. Moreover, we assume the mismatch to be dominated by a single frequency, $f_\star$. This is either the BBH merger frequency at the detector $f_{\rm max}$ (that we take to be double the ISCO frequency, assuming negligible final spin, $f_{\rm max} \simeq 1 / ( 6 \sqrt{6} \pi G (1+z_S) M_{\rm BBH} )$, see e.g.~\cite{maggiore2007gravitational}), or the frequency at the detector's sensitivity curve minimum, $f_{\rm det}$. 

At sufficiently small $w_\star \equiv 8 \pi G \MLz f_\star $, the approximation used for $\delta F_c$ breaks down, and one has to resort to the full WO result, Eq.~\eqref{eq:wl_sis_feature}. 
However, in the regime of small $w$ we expect the lensing effect to be negligible, with $F(w)\simeq 1$, and the WL critical impact parameter value to drop down rapidly. This condition can be implemented, roughly, by cutting-off the curves at the onset of WO. By inspecting Eqs.~\eqref{eq:wl_sis_feature} and \eqref{eq:go_sis_feature}, this happens when $w_{\star} y^2 / 2 \simeq m \pi$, where $m$ is a small integer. This corresponds to a peak value of the curves $y_{\rm cr}^{\rm max} \simeq {\rm SNR} / (2^{3/2} m \pi)$, that depends only on the SNR of the signal.

Following these prescriptions, the critical curves can be approximated as follows:
\begin{equation}
    y_{\rm cr}\simeq
        \left(
        \frac{{\rm SNR}}
        {\sqrt{2} w_\star }
        \right)^{1/3}
         \Theta\left( 
        \frac{\MLz}{\MLz^{\rm max}}>1
        \right)
        \;,
        \end{equation}
where the curve is truncated at the maximum, ${\MLz^{\rm max}\simeq 32 \pi^3 m^3/(w_\star {\rm SNR}^2 )}$, and 
\begin{equation}\nonumber
        w_\star\simeq
        8 \pi G  \MLz \times
        \begin{cases}
        (6 \sqrt{6} \pi G  M^D_{\rm BBH})^{-1}
        & \text{if  ${f_{\rm max}}<{f_{\rm det}}$}\\
        f_{\rm det}
        & \text{if  ${f_{\rm max}}>{f_{\rm det}}$}
        \end{cases}\,,
\end{equation}
with $M_{\rm BBH}^D=(1+z_S)M_{\rm BBH}$ the redshifted detector-frame mass.
We find that setting $m=2$ when ${f_{\rm max}<f_{\rm det}}$ and $m=1$ otherwise returns the closest match with the numerical results for LISA. Moreover, the agreement is improved if the $f_{\rm max}$ is taken to be the actual merger frequency instead of using the ISCO approximation (for numerical fits cf.~Eq.~(29) in \cite{Pompili:2023tna}). 
For ET the noise curves are flatter, making the single-frequency approximation less accurate.
Notice that $\MLz$ depends on the virial mass and distance of the lens through Eq.~\eqref{eq:virial_mass}. 
We stress that different mismatch thresholds, i.e.~$\mathcal{M}(y_{\rm cr}) \times {\rm SNR}^2 = \Lambda^2$, lead to re-scalings and shifts of the critical curves. In particular, the peak's position is $\MLz^{\rm max}\propto 1/\Lambda^2$ and its value as $y_{\rm cr}\propto1/\Lambda$. These scalings can be used to extrapolate our results to more stringent detection criteria ($\Lambda>1$).

Let us now compare our results with previous analyses that addressed the detectability of WO effects by LISA by different methods.
Reference \cite{Caliskan:2022hbu, Caliskan:2023zqm} employed a Fisher-matrix analysis, including source parameters. The authors define the critical impact parameter in terms of $\Delta\theta/\theta$, the ratio of the marginalized posterior width to the fiducial value of the lens parameter $\theta$. Specifically, they require $M_{Lz}$ at the critical impact parameter to be measurable at 1-sigma (the relative error is order one). This criterion can account for potential degeneracy in the lens-source parameters.
Table II in Ref.~\cite{Caliskan:2022hbu} shows results of $y_{\rm cr}$ for an SIS based on $\theta = \MLz$ or $\theta = y$ in three cases: the two estimates differ by a factor $\sim 2$, with $\Delta y/y$ giving the larger $y_{\rm cr}$. 
Our analysis gives slightly larger values, with $y_{\rm cr} = 58$, $48$ and $26$, which are $28\%$, $29\%$ and $33\%$ larger than the results in Ref.~\cite{Caliskan:2022hbu} for the same source properties. 

\subsection{WO optical depth} \label{sec:observation_rates}

Here we forecast the probability for a GW signal to carry a detectable \wow{} signature. We will focus on isolated matter halos, described by the SIS profile and with a mass distribution characterized by a halo mass function. We will further assume that halos at any given redshift are distributed homogeneously in space. Lensing probabilities are then described by Poisson statistics
\begin{equation}\label{eq:poisson_dist}
    \mathcal{P}(k,\lambda) = \frac{\lambda^k}{k!}e^{-\lambda}\,.
\end{equation}
Here $\lambda$ is the optical depth, which we define below. $k$ is the number of lenses contributing a detectable signature to the signal: for \wows{}, cases with $k>1$ would produce signatures similar to those shown in Figs.~\ref{fig:comp_lens_N} and \ref{fig:comp_lens_line}. 
The probability of having any number of detectable imprints is simply 
$\mathcal{P}(k\geq 1) = 1-e^{-\lambda} \simeq \lambda$, where the approximation holds for $\lambda\ll1$.

The lensing probability is given by the optical depth $\lambda$ evaluated at the source's redshift. The total optical depth is given by an integral over halo masses $\lambda = \int \de \log(\Mvir)\frac{\de \lambda}{\de \log(\Mvir)}$, where the \textit{differential optical depth} per logarithmic virial mass is
\begin{equation}\label{eq:diff_optical_depth_0}
    \dfrac{\de \lambda}{\de \log \Mvir}
    =
    \int \de \chi_L \, 
    \chi_L^2\, 
    \pi\,\theta_{\rm cr}^2(z_L) \dfrac{\de n_L(\Mvir,z_L)}{\de \log \Mvir}
    \;.
\end{equation}
Here $n_L$ is the number density of lenses at fixed $M_{\rm vir}$, $\chi_L$ is the lens' comoving distance. The integrand consists of two factors. The first is the lensing angular cross section, namely $\pi \theta_{\rm cr}^2$, representing the projected area where \wow{} is detectable. For a symmetric lens $\theta_{\rm cr}= y_{\rm cr} \xi_0/D_L$, where the critical impact parameter is the one discussed in the previous section.
\begin{figure}[t!]
    \centering
    \includegraphics[width=\columnwidth]{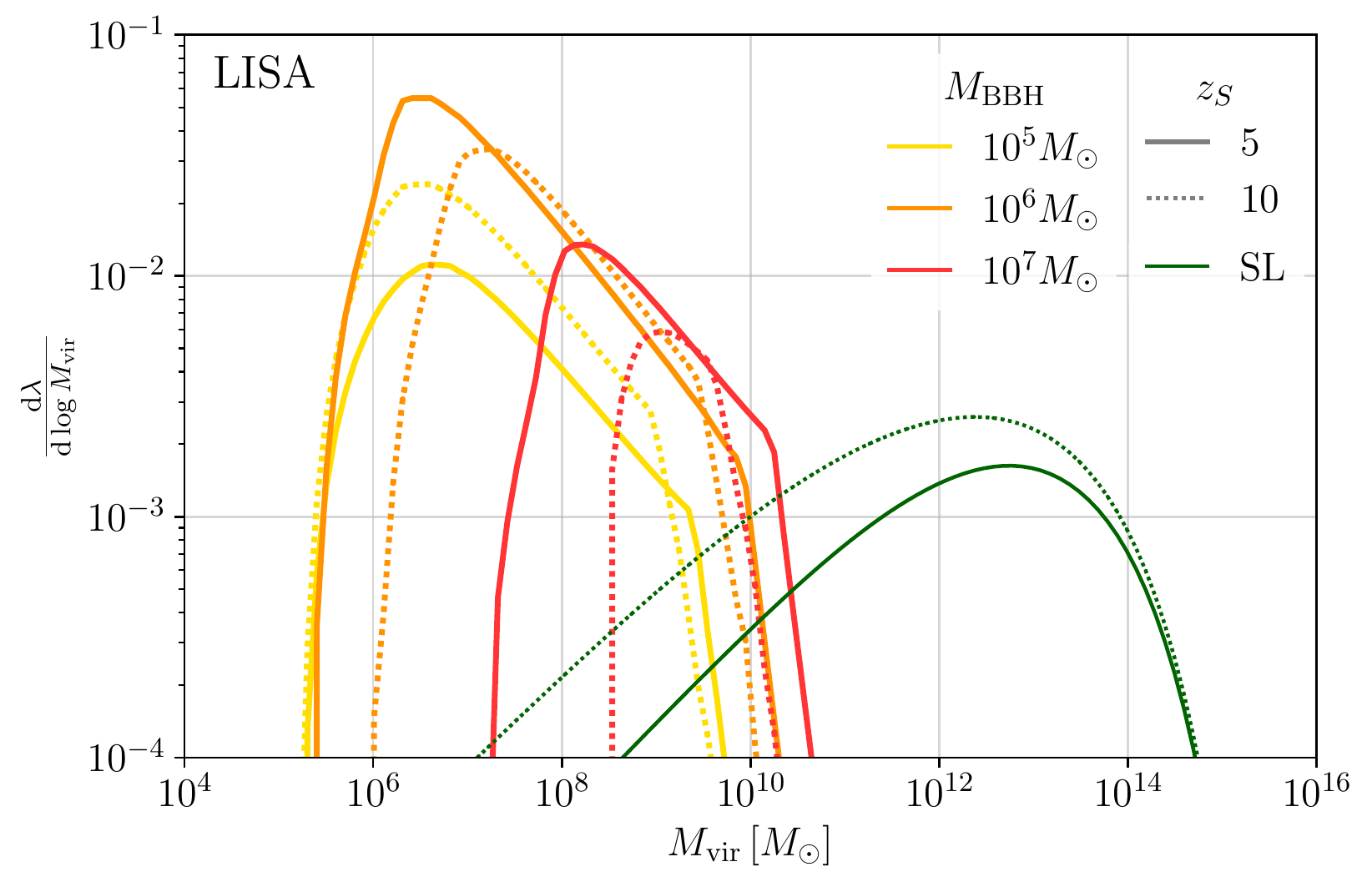}
    \caption{Prospects of \wowsfull{} observation by LISA. Lines show the differential optical depth per logarithmic virial mass. We considered equal-mass binaries at different total source-frame masses (colors) and different redshifts (line styles). The green lines show the strong-lensing (i.e.~multiple images) probability.}
    \label{fig:dNdlogMv_LISA}
\end{figure}
The second factor in Eq.~\eqref{eq:diff_optical_depth_0} is the \emph{halo mass function}, ${\rm d} n_L(\Mvir,z_L)/{\rm d}\log\Mvir$, i.e.~the number density of lenses per logarithmic virial mass unit. We adopt the Tinker halo mass function \cite{Tinker:2008ff} extrapolating it to low halo masses, i.e.~$\Mvir<10^{10} \, M_\odot$ (other halo mass functions produce similar results). We note that our $\de n/\de \log(\Mvir)$ only accounts for isolated halos: subhalos within the larger structure may further contribute to the probabilities presented here. We will comment on subhalo signatures in Sec.~\ref{sec:observation_subhalos}.

The optical depth depends on two definitions of the mass: $\Mvir$ in $\de n/\de \log(\Mvir)$ and $\MLz$, Eq.~\eqref{eq:def_MLz}, in $y_{\rm cr}$.
They are related by:
\begin{align}\label{eq:virial_mass}
    \MLz
    &=
    2.3
    \times 
    10^6 \, M_\odot (1 + z_L)^2
    \left(
    \frac{d_{\rm eff}}{1\,{\rm Gpc}}
    \right)
    \times
    \\ \nonumber
    &
    \hspace{3cm}
    \times
    \left(
    \frac{\Mvir}{10^{9} \, M_\odot} \frac{H(z_L)}{H_0}
    \right)^{4/3}
    \,,
\end{align}
(see Eq.~\eqref{eq:virial_mass_appendix} for further details).
Note that both quantities differ by several orders of magnitude for typical halos. The above relation depends on the source and lens' redshift. $z_S$ can be inferred by the amplitude of the signal, which scales as $\propto D_L(z_S)^{-1}$, assuming a cosmology. As we mentioned multiple times, while $z_L$ is in general unknown, it is possible to place some restrictions on $\Mvir$, given an observed $\MLz$, the quantity that lensing is sensitive to. Let us give more details about this procedure. Assuming a halo mass function, one can assign a probability distribution to $P(\Mvir|\MLz,z_S)$.
Because $d_{\rm eff}$ is bounded from above, it defines a minimum value $\Mvir^{\rm min}$ (at fixed $\MLz$).
The probability $P(\Mvir|\MLz,z_S)$ is sharply peaked around $\Mvir^{\rm min}$ although larger values are possible, see Sec.~V-A in \cite{Tambalo:2022wlm} for details.

\begin{figure}[t!]
    \centering
    \includegraphics[width=\columnwidth]{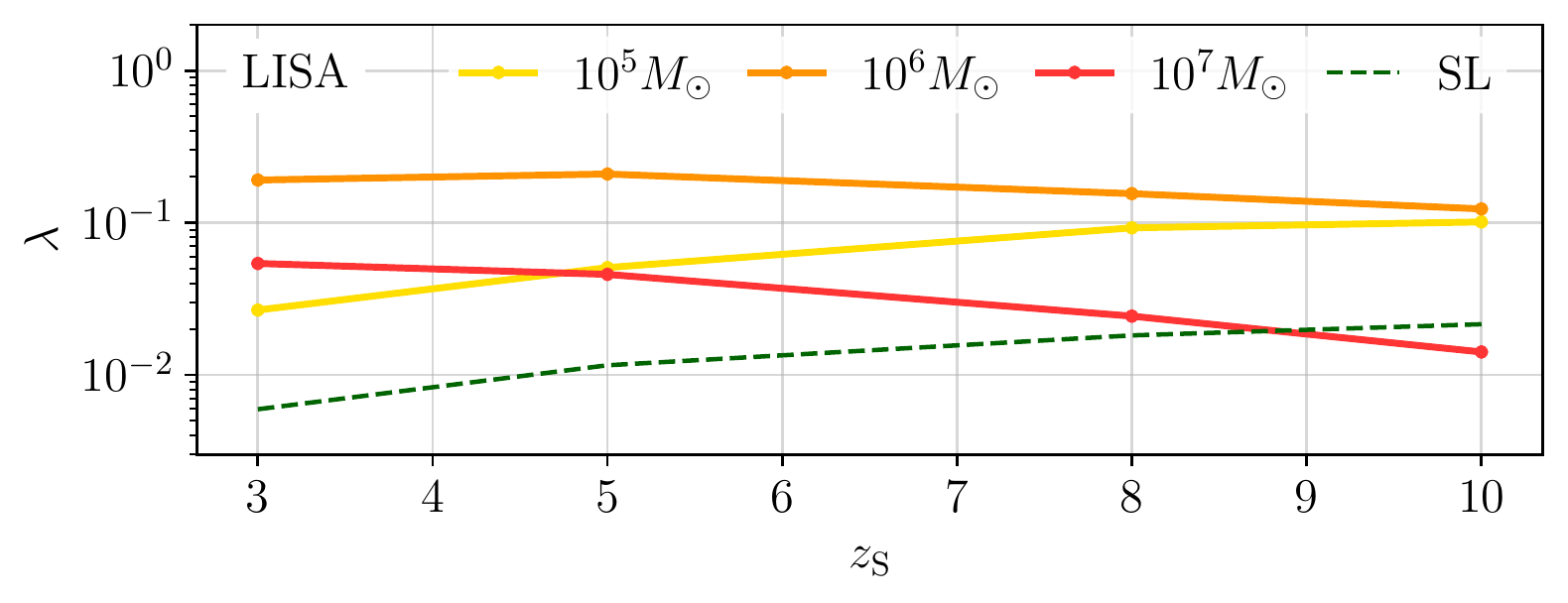}
        \includegraphics[width=\columnwidth]{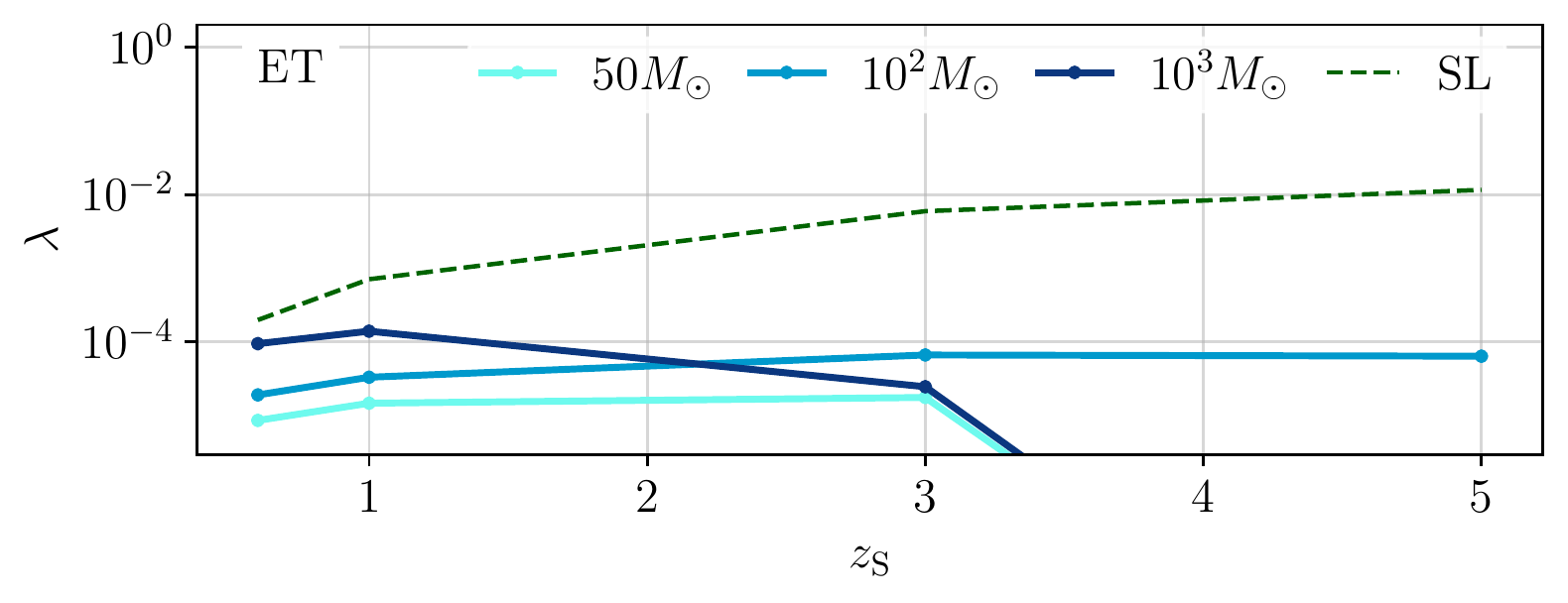}
    \caption{Lensing optical depth, $\lambda$, against the redshift of the source for LISA (top) and ET (bottom), for different source masses (colors). Dashed lines show the strong-lensing optical depth (independent of the source mass).}
    \label{fig:WO_rates}
\end{figure}

Using the definition of the effective lens mass, Eq.~\eqref{eq:def_MLz}, and $ {\rm d} \chi_L \chi_L^2 = {\rm d} z_L (1+z_L)^2 D_L^2/H(z_L) $, the differential optical depth can be recasted as:
\begin{equation}\label{eq:diff_optical_depth}
    \dfrac{\de \lambda}{\de \log \Mvir}
    =
    4\pi G
    \int_0^{z_S} \de z_L 
    y_{\rm cr}^2 \MLz 
    d_{\rm eff}
    \frac{(1+z_L)^3}{H(z_L)} 
    \dfrac{\de n_L}{\de \log \Mvir}
    \,,
\end{equation}
We will consider two types of lensing observation:
\begin{itemize}
    \item $\lambda_{\rm WO}$ for \wow{} detection, with $y_{\rm cr}$ computed as described in Sec.~\ref{sec:observation_ycrit}.
    \item $\lambda_{\rm SL}$ for strong lensing, computed with $y_{\rm cr}=1$. 
\end{itemize}
Note that $\lambda_{\rm WO}$ depends on $z_S$ as well as the source properties ($M_{\rm BBH}$), while $\lambda_{\rm SL}$ only depends on $z_S$. 
Note that the strong-lensing criterion is based solely on the formation of multiple images for the SIS model. Requiring that the secondary image is brighter than a detection threshold would reduce $\lambda_{\rm SL}$. However, lensing situations with $y\gtrsim 1$ could still produce strong-WO phenomena even in the single-image regime \cite{Dai:2018enj}. The probability of these situations is approximately represented by $\lambda_{\rm SL}$.

Figure \ref{fig:dNdlogMv_LISA} shows the differential optical depth as a function of the virial mass of the lens, for various redshifts and total BBH masses (source-frame), within LISA's reach. The higher peak is reached by the $10^6 \, M_\odot$ curve at $z_S=5$, which corresponds to the largest critical impact parameter and SNR among those considered. We observe that larger source redshifts do not necessarily reduce the differential optical depth. The SNR is reduced by a factor $(1+z_S)D_S$, while the number of intervening halos and their typical Einstein radii increase with $z_S$. In addition, as the signal gets redshifted, i.e.~$M^D_{\rm BBH}=(1+z_S)M_{\rm BBH}$, the merger can be displaced to a frequency band with lower/higher sensitivity.
The final behaviour depends on $\Mbbh$. For instance, for $\Mbbh=10^7 \, M_\odot$, the closer source curve ($z_S=5$) encompasses the further one ($z_S=10$), while the opposite is true for $\Mbbh=10^5 \, M_\odot$. The optical depth for strong lensing is independent on the BBH mass and increases with redshift as expected. 

We plot the total optical depth $\lambda$ in Fig.~\ref{fig:WO_rates} for typical sources detectable by LISA (upper panel) and ET (lower panel). Physically, this quantity represents the average number of lenses per source entering the Poisson distribution, Eq.~\eqref{eq:poisson_dist}.

We find that that $\lambda(z_S)$ displays a nontrivial dependence with $M_{\rm BBH}$: it can increase or decrease with $z_S$, as a result of the competing effects discussed above (SNR, lens projection and frequency shift). In the $10^2$ and $10^3 \, M_\odot$ curves for ET we can see the effect of the signal falling out of the detector's horizon at high $z_S$.

The prospect of observing \wows{} is optimistic: our mismatch analysis shows that $\sim 20$\% of LISA binaries with $M_{\rm BBH}\sim 10^6 \, M_\odot$ would carry observable \wow{} signatures. 
\wow-detection by LISA is substantially more likely than strong lensing, except for the heavier sources at very high redshift ($M_{\rm BBH}=10^7 \, M_\odot$, $z_S>7$).
While \wow{} detection will be dominated by events where a single lens contributes ($k=1$ in Eq.~\ref{eq:poisson_dist}), the probability of detecting \wow{} signatures from multiple lenses $\sim \lambda^2$ is non-negligible for $M_{\rm BBH}\sim 10^6\, M_\odot$, even when considering only isolated halos.
For ET, strong lensing is more likely. Still, \wows{} can plausibly be observed given the large number of expected sources, with rates $\sim 10^4-10^5\,{\rm yr}^{-1}$ \cite{Iacovelli:2022bbs,Borhanian:2022czq}.

We conclude by recalling that, while the flipped Lindblom  tends to overestimate the critical impact parameter, two assumptions make our estimate conservative. First, $y_{\rm cr}$ was computed from averaged noise curves. This accounts for typical sources, but we expect it to underestimate $\lambda_{\rm WO}$, as $\lambda \sim y_{\rm cr}^2\sim {\rm SNR}^2$ and $\langle{\rm SNR}^2\rangle > \langle {\rm SNR}\rangle^2$, i.e.~events with better source/detector alignment at fixed $z_S$ will compensate for the fainter ones. This information (e.g.~based on the posteriors for the sky localization, binary inclination, etc...) could be folded into the estimates discussed in Secs. \ref{sec:observation_constraints} and \ref{sec:applications_halo_dm}. 
Second, the halo mass function only accounts for isolated halos. Light subhalos ($M\lesssim 10^9 \, M_\odot$) are more abundant than their isolated counterpart, and will further contribute to the optical depth. They may also be distinguishable through characteristic \wow{} signatures, see Sec.~\ref{sec:observation_subhalos}. 

Now, we compare our estimates with previous results in the literature. The estimates in Ref. \cite{Caliskan:2023zqm} are less optimistic due to the differences in the critical impact parameter definition, as discussed at the end of the previous section.
Ref. \cite{Fairbairn:2022xln} obtained probabilities almost 4 orders of magnitude lower. In this case, the difference can be traced back to different modelling of the source and the lens. They use the lowest order post-Newtonian expansion of the waveform, truncated at the innermost stable circular orbit, which covers only the inspiral stage of the coalescence. As discussed above, the peak of the critical curves is associated with the large SNR contribution coming from the merger.  Neglecting the merger can severely underestimate the height of the peak in $y_{\rm cr}$ and its position. Finally, they work with a Navarro-Frenk-White lens profile, which is shallower than the SIS towards the center and consequently produces a less prominent WOF. Along these lines, Ref. \cite{Guo:2022dre} finds larger detection probabilities for cuspier profiles. This analysis is also based on inspiral-only waveforms and thus leads to low detection prospects. The impact of the lens profile on the detection prospects will be addressed in future work.

\subsection{Total rates: probing the halo mass function}\label{sec:observation_constraints}

We can now estimate the total number of events with observable \wow{} and use that information to constrain the halo mass function on the mass range where WO effects are observable. We will focus on LISA, for which the optical depth was found to be significant. 

\begin{table*}[t!]
\def\arraystretch{1.2}
\setlength\tabcolsep{5pt}
    \begin{tabular}{c | c c c | c c}
   \multicolumn{1}{c}{} & \multicolumn{3}{c}{Rates $[{\rm yr}^{-1}]$} & \multicolumn{2}{c}{$f_{\rm H}$ c.l.} \\[2pt]
    Limit &  $\dot N_{\rm det}$ &  $\dot N_{\rm WO}$ & $\dot N_{\rm SL}$ &   5\%  & 95\% \\
    \hline
    Agnostic 95\% & $1.3\times 10^{5}$ & $2.6\times 10^{4}$  & $1.5\times 10^{3}$ & $0.995$ & $1.005$ \\
    Agnostic 75\% & $5.2\times 10^{3}$ & $1.0\times 10^{3}$  & $60.2$ & $0.976$ & $1.025$ \\
    Agnostic 50\% & $66.3$ & $8.8$  & $0.77$ & $0.77$ & $1.26$ \\
    Agnostic 25\% & $0.076$ & $4.7\times 10^{-3}$  & $8.8\times 10^{-4}$ & $-$ & $-$ \\
    \hline
    Astro 95\% & $3.3\times 10^{2}$ & $35.2$  & $3.79$ & $0.88$ & $1.13$ \\
    Astro 75\% & $40.6$ & $3.26$  & $0.47$ & $0.66$ & $1.49$ \\
    Astro 50\% & $8.55$ & $0.48$  & $0.099$ & $0.35$ & $2.84$ \\
    Astro 25\% & $2.16$ & $0.11$  & $0.025$ & $-$ & $7.14$ \\
    Astro 5\% & $0.22$ & $6.8\times 10^{-3}$  & $2.5\times 10^{-3}$ & $-$ & $480$ \\
    \hline
    \end{tabular}
    \caption{LISA rates and prospective constraints on the halo mass function. The first three columns show rates of total mergers, WO detections and multiple-image events with $M_{\rm BBH}\geq 10^6\, M_\odot$, $z_S\leq 5$. Rows correspond t the agnostic and astro-informed models in Ref.~\cite{Steinle:2023vxs}, at different confidence levels (Agnostic 5\% c.l. is not shown, as it is several orders of magnitude lower). The last two columns show the lower and upper limits on the amplitude of the halo mass function, Eq.~\eqref{eq:rescaled_halo_mass_function}.
    \label{tab:combined}}
\end{table*}

The results depend on the total number of detectable signals, their redshifts and mass distributions.
For LISA, a recent analysis placed bounds on the detection rate compatible with results from pulsar timing arrays \cite{Steinle:2023vxs}. We will use the results for their ``agnostic'' \cite{Middleton:2015oda} and ``astro-informed'' \cite{Chen:2018znx} models (obtained from their Fig.~2) for sources with $z_S\leq 5$ and $M_{\rm BBH}\geq 10^6 \, M_\odot$.
We will estimate the rate of lensed events (WL, SL) as $\lambda_L(z_S^*, M_{\rm BBH})\cdot \dot N_{\rm det}(M_{\rm BBH}) T_{\rm obs}$, integrating over the binary masses. 
We evaluate the optical depth $z_S^*=5$, the upper end of their interval. This is conservative for weak-lensing, see Fig.~\ref{fig:WO_rates} ($\lambda_{\rm WO}$ increases with $z_S$ for lighter sources $M_{\rm BBH}=10^5 \, M_\odot$, which are not included). The choice is optimistic for strong lensing/multiple images, as the optical depth grows with $z_S$. 

Table \ref{tab:combined} shows the rates of detected events, WO distortion and multiple images under the above assumptions. Each row corresponds to a confidence interval within the agnostic/astro-motivated models. The detection rates vary several orders of magnitude, especially in the agnostic model. Note that the above numbers for \wow{} detection neglect sources at high redshift ($z_S>5$) and with light masses ($M_{\rm BHB}\leq 10^5  \, M_\odot$), whose inclusion would increase the detection rate. Even so, multiplying by the observation time $T_{\rm obs}\sim 5$ yr gives reasonable observation prospects in all but the most pessimistic cases. 

Detecting \wow{} enables novel probes of the halo mass function. In the simplest cases, constraints rely on the ratio of observed $\dot N_{\rm WO}/\dot N_{\rm det}$ to the values predicted in Eq.~\eqref{eq:diff_optical_depth}.
To perform a quantitative estimation of the sensitivity, let us consider a constant rescaling of the halo mass function
\begin{equation}\label{eq:rescaled_halo_mass_function}
    \frac{\de n}{\de \Mvir} 
    = 
    f_{\rm H} \left.\frac{\de n}{\de \Mvir}\right|_{\rm fid}\,,
\end{equation}
where the fiducial case $f_{\rm H}=1$ corresponds to the case discussed in Sec.\ref{sec:observation_rates}. This represents a phenomenological parametrization in the range of $\Mvir$ for which a detector is sensitive to \wow, cf.~Fig.~\ref{fig:dNdlogMv_LISA}.

Let us now derive limits on $f_{\rm H}$ given the observation of $k_{\rm det}$ events, of which $k_{\rm WO}$ carry \wow{} signatures. We will not consider the possibility of events with multiple \wows, with probability $\mathcal{O}(\lambda_{\rm WO}^2)$. Our theoretical model depends on $N_{\rm det} = r \dot N_{\rm det}^{(0)} T_{\rm obs}$ and $N_{\rm WO}= f_{\rm H} \dot N_{\rm WO}^{(0)} T_{\rm obs}$, where $T_{\rm obs}$ is the survey time and the free parameters $f_{\rm H}$, $r$ encode variation with respect to fiducial values $\dot N_X^{(0)}$. 
The likelihood of the data ($k_{\rm wo},k_{\rm det}$) given the model ($f_{H},r$) is 
\begin{equation}
L=\mathcal{P}(k_{\rm wo}| N_{\rm WO}(r,f))\mathcal{P}(k_{\rm det}| N_{\rm det}(f)) 
\;,
\end{equation}
where $\mathcal{P}(k|\lambda)$ is the Poisson distribution, Eq.~\eqref{eq:poisson_dist}.
We obtain the 1-dimensional confidence levels in $f$ by evaluating the posterior on a grid in $r,f$ and integrating $r$. This implicitly assumes a wide prior, i.e.~$\Pi(r,f)\simeq 1$ around the peak of the likelihood. 
For $k_{\rm WO}>0$, the limits are consistent with estimating the posterior by sampling $f_{\rm H} = \frac{1}{\lambda_{\rm WO}}\frac{k_{\rm WO}}{k_{\rm det}}$, where $k_{\rm WO},k_{\rm det}$ are taken from a Poisson distributions with rates $N_{\rm WO},N_{\rm det}$. 

Table \ref{tab:combined} shows the marginalized $90\%$ confidence interval on $f_{\rm H}$ that can be achieved by a $T_{\rm obs} = 5\,{\rm yr}$ LISA mission, given the rates shown in Table \ref{tab:combined}. We have assumed that the number of events $k_{\rm det}, k_{\rm WO}$ is $\lfloor N_{\rm det}^{(0)}\rfloor ,\lfloor N_{\rm WO}^{(0)}\rfloor$, where $\lfloor x\rfloor$ is the floor function. 
The sensitivity ranges from sub-percent in the most optimistic case to 1-2 orders of magnitude upper bounds in the most pessimistic cases. Note that cases with $k_{\rm WO} =0$ allow upper limits on $f_{\rm H}$, as long as $k_{\rm det}\leq 1$. In particular, the Astro 25\% c.l.~scenario has a single detection, leading to a $\mathcal{O}(10^2)$ limit.
This is a very simple estimate of prospective constraints on the halo mass function. In Sec.~\ref{sec:applications_halo_dm} we will discuss how more information on $\de n/\de \Mvir$ can be obtained from \wow{} observations.

\subsection{Subhalos vs isolated halos} \label{sec:observation_subhalos}

The rates presented above are conservative, as the halo mass function accounts only for isolated halos, excluding structures that have incorporated into more massive halos, i.e.~subhalos (see discussion in Sec.~4 of Ref.~\cite{Dai:2019lud}). While some subhalos will be disrupted, those that survive may contribute to the probability of detecting WO effects, increasing the rates presented in Sec.~\ref{sec:observation_rates}.

Besides increasing the probability of \wow{} detection, subhalos  differ from isolated halos in several regards. Even when comparing objects of equal mass, we expect the following effects in subhalos.
\begin{enumerate}
    \item Enhanced probability of detecting signatures from multiple subhalos, e.g.~similar to Figs.~\ref{fig:comp_lens_N},\ref{fig:comp_lens_line} but with a larger spread in $\tau$.
    \item Distortion of the lensing potential by the main halo through external convergence, shear and flexion terms in $\psi(\vect x)$.
    \item In some cases, multiple strong-lensing images are produced by the main halo. Then, each image may contain \wow{} of different subhalos.
\end{enumerate}
In addition, subhalos will differ from isolated halos (e.g.~in their shape) due to their assembly history and interactions with the main halo and other subhalos \cite{Zavala:2019gpq}. These evolutionary features will affect the probability of subhalo lensing and may have observational imprints.

Massive halos contain a large number of lighter subhalos \cite{Zavala:2019gpq}. 
A signal propagating through a galactic-scale halo can have an enhanced probability of encountering multiple low-mass subhalos that can produce a \wow, increasing the detection rates computed for isolated halos in Sec.~\ref{sec:observation_rates}.
The subhalo lensing probabilities can be modelled using a Poisson distribution with a rate $\bar N_{\rm WO,s}$, which depends on the point where the image forms in the lens plane and the details of the subhalo mass and spatial distribution \cite{Giocoli:2007uv,Jiang:2014nsa,Zavala:2019gpq}. Lens configurations for which $\bar N_{\rm WO,s} \gtrsim 1$ are likely to produce rich \wow{} with multiple peaks, as discussed in Sec.~\ref{sec:pheno_composite_signatures}.

Another effect of substructures is the distortion of the lensing potential by the main halo. The leading order corrections, convergence and shear, have been shown to enhance the diffraction pattern appreciably \cite{Dai:2018enj} and can allow lighter microlenses to contribute to diffraction effects \cite{Diego:2019lcd} (although see \cite{Meena:2023qdq}). 
They may thus increase the prospects of detecting the \wow, particularly sources whose image forms in the inner part of the halo. The imprint of convergence and shear may also serve to distinguish subhalos from isolated halos and place them within their host halo.

For closely aligned systems, the main halo will split the source into multiple images. The associated probability is given by the strong lensing rate ($\lambda_{\rm SL}$, for $y_{\rm cr}=1$) in Sec.~\ref{sec:observation_rates}. 
In this case, each image may contain information from nearby subhalos, which will also be affected by convergence and shear from the main halo. Identifying multiple GW events as strongly lensed images from the same source (e.g.~by overlapping sky-localization and intrinsic parameters) would provide a unique opportunity to constrain the properties of subhalos.

Finally, we note that \wows{} could also be imprinted by close-by systems, such as subhalos of our own galactic halo. The critical curves presented in Fig.~\ref{fig:ycrit} do not depend on the lens position and retain their validity when $z_L$, $d_{\rm eff}\rightarrow 0$. However, the \emph{physical} impact parameter shrinks -- the Einstein radius goes as $\sqrt{d_{\rm eff}}$ -- and the conversion between $\MLz$ and $\Mvir$ gets offset. For example, inverting Eq.~\eqref{eq:virial_mass} the interval $\MLz=10^{-1}-10^3  \, M_\odot$ is mapped into $\Mvir\simeq 10^6-10^{9}  \, M_\odot$ (at $z_S=1$ and $D_L=200$ kpc).
We stress that the geometry of the system does not suppress the lensing probability. Indeed, despite the contraction of the physical impact parameter, the quantity $\chi_L^2 \theta_{\rm cr}^2$ in Eq.~\eqref{eq:diff_optical_depth_0} is invariant. If a \wow{} is observed, sky localization information could be used to determine the probability of local versus cosmological origin.

\section{Potential applications} \label{sec:applications}

We will now discuss some possible uses of \wow{} to constrain properties of the large-scale structure. Our presentation will be qualitative and schematic. More detailed analyses are left for future work.

\subsection{Lens reconstruction} \label{sec:applications_reconstruction}

The simple expressions behind the perturbative weak-lensing framework (Sec.~\ref{sec:lensing_perturbative}) open the possibility of systematically reconstructing lens features. As already explained in the previous sections, a full reconstruction of the 2-dimensional lensing potential is impossible from 1-dimensional data from a single source ($\Cc G(\tau)$, $\Cc I(\tau)$ or $F(w)$).

Nonetheless, assuming that the lensing potential is symmetric, $\psi(\vect x)\to \psi(x)$, a formal relation between the time-domain integral and the lensing potential can be obtained from the leading-order term in Eq.~\eqref{eq:time_domain_perturbative_expansion}. First, we change integration variable from $\varphi$ to $x$ and obtain
\begin{equation}
\label{eq:lens_reconstruction}
    \Cc I(\tau) 
    = 
    \frac{\de}{\de \tau} 
    \int_0^{\infty} \de x \, 
    K(x,\tau, y) \psi(x)
    \;,
\end{equation}
with a kernel
\begin{equation}
    K(x,\tau,y) 
    \equiv
    \frac{2x}{\sqrt{2(\tau + \phi_m)}~y \sin \varphi}
    \Theta(x_{0} - x)
    \Theta(x-x_{\pi})
    \;,
\end{equation}
where $\varphi$ is a function of $x$, $y$ and $\tau$ (according to Eqs.~\eqref{eq:time_domain_perturbative_expansion_x1}, \eqref{eq:time_domain_perturbative_expansion_x2}), and $x_0$, $x_\pi$ are the values at which $\varphi(x, \tau, y) = 0$, $\pi$, respectively. 
By knowing $\Cc I(\tau)$, Eq.~\eqref{eq:lens_reconstruction} gives the lensing potential as the solution of an integral equation. One can then obtain the projected mass density $\Sigma(x)$ e.g.~Eqs.~(11) in Ref.~\cite{Tambalo:2022wlm}.

While potentially interesting, it is not clear how useful the above expression may be in practice. Besides assuming linearity and symmetry of $\psi$, Eq.~\eqref{eq:lens_reconstruction} requires knowing the impact parameter $y$. If this value is not constrained (e.g.~from the \wow{} peak), one possibility is to perform the reconstruction for different values of $y$ and consider the most plausible reconstructed lens, according to some prior, e.g.~from theory or simulations. Ultimately, the reconstruction will be limited by how well $\Cc I(\tau)$ can be inferred from real data.

\subsection{GW delensing}\label{sec:applications_delensing}

The correlation between the \wow{} and the GO magnification opens the possibility of inferring $\sqrt{|\mu|}$.
Eqs.~\eqref{eq:delensing_simple_lens_relations_peak} and \eqref{eq:delensing_simple_lens_relations_broadband} give approximate relationships for symmetric SIS models.

This information would enable a re-calibration of the intrinsic luminosity of the source, a major source of uncertainty for standard sirens at high redshift \cite{Holz:2005df}. Such procedures are known as ``delensing". Proposed methods for GW delensing usually rely on EM follow-up to characterize the lensing potential in the direction of the source \cite{Shapiro:2009sr,Wu:2022vrq}. Because GW sources are poorly localized, EM follow-up methods need to cover a large portion of the sky, which can become very costly.

The main limitation is that magnification is dominated by galactic scale lenses, too heavy to produce an observable \wow{} by themselves. Hence, WO-based delensing would rely on detecting substructure within the main lens (cf.~Sec.~\ref{sec:observation_subhalos}). This may limit the applicability to a subset of sources. 
However, delensing based on the \wow{} would not require additional observations and could be attempted without any costly follow-up. This is similar to delensing of the cosmic microwave background, which can be performed at the level of the observed maps \cite{Hotinli:2021umk}.

\subsection{LSS morphology} \label{sec:applications_LSS_morphology}

The large-scale structure (LSS) of the universe displays a rich pattern in the distribuiton of dark and baryonic matter. Different universe regions can be broadly classified by their morphology, determined by the number of independent directions that are expanding vs contracting. This gives rise to 4 categories: voids (3 expanding directions), sheets/walls (2 expanding, 1 contracting), filaments (1 expanding, 2 contracting) and halos (3 contracting) \cite{Sousbie:2006tg,Hahn:2006mk,Coil:2013}. 

In Fig.~\ref{fig:comp_lens_line} we explored an idealized representation of a filamentary structure. Our model, a 1-dimensional chain of equal-mass and equally-spaced lenses, left a characteristic imprint in Green's function, with a clear dependence on the angle between the chain and the location of the source. Weakly-lensed GWs have the potential to distinguish between these patterns, allowing not only to identify a filament but also to reconstruct its angle, and perhaps even other properties (e.g.~mass and relative spacing between sub-lenses).

Realistic realizations of the LSS are vastly more complex. Nonetheless, information about the lens morphology will be present in the \wow, e.g.~through the statistics of the peak distribution in $\Cc G(\tau)$. This information is projected from a 3D distribution into the 1D Green's function, and thus a complete reconstruction is not possible (see Sec.~\ref{sec:applications_reconstruction}). Nonetheless, it might be possible to obtain information on the lens morphology, e.g.~in high-SNR observations where multiple peaks can be clearly located. More likely, morphology reconstruction will, at best, assign a probability to each category given an observation.

\subsection{Probing low-mass halos and dark matter} \label{sec:applications_halo_dm}

Light halos ($\Mvir\lesssim 10^{10} \, M_\odot$) are pristine test-beds for structure formation and dark matter (DM) theories: they form at high redshift, their baryonic mass is subdominant, and many DM scenarios impact their abundances and profiles \cite{Bullock:2017xww,Buckley:2017ijx,Tulin:2017ara,Ferreira:2020fam,Hui:2021tkt}. 
However, such light halos are very difficult to observe, relying on the high-redshift observations \cite{Menci:2016eui,Irsic:2017yje,Rogers:2020ltq} or close-by systems in the Milky Way environment \cite{DES:2019ltu,Banik:2019cza,Bonaca:2020psc}.
Observations of strongly-lensed signals can also identify individual structures \cite{Vegetti:2012mc,Hezaveh:2016ltk,Sengul:2023olf} or their collective distortions \cite{Cyr-Racine:2015jwa,DiazRivero:2018oxk,Wagner-Carena:2022mrn}.
While these observations are promising, modelling a strongly-lensed system is challenging and computationally intensive. 

Lensed GWs may offer a complementary means to identify light halos and subhalos,  distinguish between both and constrain their properties \cite{Guo:2022dre,Fairbairn:2022xln,Tambalo:2022wlm} and abundances.
Isolated light halos at cosmological distances can be observed by future detectors thanks to the large critical impact parameter (Fig.~\ref{fig:ycrit}). 
In Sec.~\ref{sec:observation_constraints} we showed how LISA can obtain constraints on the halo mass function from detection of \wows{} or their absence. In the most optimistic case, a constant rescaling on the scale of interest $f_{\rm H}$, Eq.~\eqref{eq:rescaled_halo_mass_function} can be constrained to sub-percent level, while a single unlensed event yields an $\mathcal{O}(10^3)$ limit. As argued in Sec.~\ref{sec:observation_rates}, including information on the source's parameters that affect the SNR (sky localization, inclination) is likely to produce more robust limits. Information on the impact parameter posterior can also be incorporated; see Ref.~\cite{Gais:2022xir}.

A more detailed analysis can also constrain the virial mass dependence, i.e.~$f_{\rm H}(\Mvir)$, using the fact that the source mass $M_{\rm BBH}$ determines the range of $\Mvir$ that can be probed (cf.~Fig.~\ref{fig:dNdlogMv_LISA}). However, as the rates are dominated by the optimal mass ($\sim 10^6 \, M_\odot$ for LISA), constraints on the mass dependence will be far less stringent than the overall amplitude. It may also be possible to constrain the redshift dependence: the rates for different $M_{\rm BBH}$ evolve differently with $z_S$ (cf.~Fig.~\ref{fig:WO_rates}) and $z_S$ is known from the luminosity distance (assuming a cosmology). While important degeneracies are expected in a generic $f_{\rm H}(\Mvir,z_L)$, it may be possible to test models with suppression of light halos (e.g. warm dark matter \cite{Angulo:2013sza}) or a minimum low-mass cutoff (e.g. ultra-light dark matter \cite{Hui:2016ltb,Fairbairn:2017sil}).

Additional information on the lens may be obtained via the \wow: the relevant observable quantities are the source's redshift (from the signal's amplitude, assuming a cosmology) and the lens' effective mass $\MLz$ (see discussion in Sec.~\ref{sec:pheno_symmetric}). 
Because the lens redshift is unknown, $\MLz$ only provides a lower bound on the virial mass of the halo. However, assuming a halo mass function allows one to define a probability distribution for $z_L$ and $\Mvir$, which is peaked around the minimum $\Mvir$ (see Sec.~VA in Ref.~\cite{Tambalo:2022wlm} for details). 
The inferred values from all events with \wow{} would then improve on the constraints from detection counts. This method bears analogy to inferring the properties of galactic-scale lenses from strongly lensed signals using the distribution of time-delays between multiple images \cite{Xu:2021bfn}.

Lensed GWs can also probe subhalos and distinguish them from isolated halos. As argued in Sec.~\ref{sec:observation_subhalos}, the expected signature of substructures  is the detection of multiple peaks in the \wow{} (from nearby subhalos) and/or the presence of convergence/shear caused by the main halo. Subhalos may have different abundances and properties from isolated halos of the same mass (e.g.~due to tidal stripping, shock heating and other interactions \cite{Zavala:2019gpq}). Being able to distinguish both populations separately may offer valuable clues about the assembly history of the large-scale structure.

\section{Conclusions} \label{sec:conclusions}

We have investigated the phenomenology of gravitational lensing in the single-image, wave optics (WO) regime, with an outlook on their potential to probe cosmic structures and prospects for observation by future GW detectors. Our results converge to the well-known limits of geometric optics (GO) in the large source frequency/lens mass limit. Large angular separation between source and lens corresponds to weak-lensing (WL), where WO corrections are subtle but potentially observable.

We presented two methods to solve the diffraction integral in the time-domain, adapted to the single-image regime but accounting for WO effects (Sec.~\ref{sec:lensing}).
Both approaches yield accurate results in their domain of validity.
First, we present an algorithm able to explore any single-image configuration, its accuracy limited only by numerical errors. This method is valid even in strong-lensing: it can be used to explore the outer regions of a caustic, where the GO magnification diverges and WO effects persist at high frequency.
Second, we develop a perturbative expansion on the lens potential. This method is faster and converges very rapidly to the full solution in the WL limit, at large impact parameter $y$. The leading-order expansion is linear in the lensing potential/projected lens density, making the study of composite lenses straightforward.
Both methods are fast enough for applications such as parameter estimation.

Using these algorithms, we analyze the phenomenology of \wowsfull{} (\wows), the WO imprint on lensed GWs (Sec.~\ref{sec:phenomenology}). This discussion is particularly clear when using Green's function $\Cc G(\tau)$. The most salient aspects of the \wow{} are its peak and broadband distribution. The peak forms at the center of the lens: its associated time delay and height contain information about the lens location (relative to the GO image) and the lens mass (Fig.~\ref{fig:SIS_vary_y}), while its shape is related to the lens projected density (Fig.~\ref{fig:CIS_vary_xc}).
The broadband profile is related to the large-scale properties of the lens, such as its total mass and spatial extent. In the frequency domain, the broadband feature corresponds to the first maximum of $F(w)$, while the peak appears as a damped oscillatory pattern at higher frequencies.
Our analysis also applies to composite lenses: we study the superposition of $N_{\rm sub}$ equal-mass sublenses with an SIS profile. Each sublens produces a peak, whose associated time delay and amplitude are set by its distance to the GO image in the lens plane. We computed the average profile for this distribution and showed that $\Cc G(\tau)$ and $F(w)$ converge to it in the limit of large $N_{\rm sub}$, although finite $N_{\rm sub}$ is associated to stochasticity in Green's function.

We then address the prospects of detecting \wows{} (Sec.~\ref{sec:observation}). 
Future detectors can potentially observe WO effects at impact parameter one to two orders of magnitude larger than the Einstein angle. 
Assuming a halo mass function (i.e.~extrapolating to the mass-scales producing WO) and uncorrelated spatial distribution, our results can be directly translated into detection probabilities (via the optical depth). The prospect of detection is optimistic for LISA, with rates $\sim 20\%$ for mergers with the optimal mass range.
The rates suggest that detecting WO effects is plausible, although subject to uncertainties on the LISA detection rate for MBHB mergers: the number of detections ranges from large to unlikely in the scenarios we have considered.
Detection probabilities are lower for ground detectors due to the reduced SNR and higher frequency, as \wows{} require lighter halos with smaller Einstein radii.
Finally, we note that these rates refer only to isolated halos: subhalos assembled into larger structures will increase these rates, and might be distinguishable from isolated halos in some cases.

We can summarize our main findings as follows
\begin{itemize}
\item Gravitational lensing imprints \wow: frequency-dependent modulations that can be observed in the waveform without the need of a counterpart, higher-mode emission or association of multiple events. The Green's function offers a particularly transparent way to analyze these features.
\item Features of the lens (effective mass, spatial distribution, inner structure) translate cleanly into features of the Green's function (broadband shape, location and height of peaks), thanks to (approximate) linearity in the WL regime. Distinguishing these features would allow constraints on the properties of individual sublenses and their relative positions.
\item Our framework explains clearly how a macroscopic lens arises effectively from a superposition of smaller objects. In the frequency domain a large number of sub-lenses add-up incoherently, suppressing the signatures at high frequencies, as in the average smooth lens (Fig.~\ref{fig:comp_lens_N}).
\item Events with high SNR enable detections of \wows{} at impact parameters much larger than the Einstein radius. For low-frequency detectors such as LISA, sensitive to heavier halos, this translates into promising prospects for observations, with probabilities well beyond those of strong lensing.
\item Observation of \wows{} (or lack thereof) constrains the amplitude of the halo mass function in the range $10^5 \, M_\odot \lesssim \Mvir \lesssim 10^8 \, M_\odot$, with precision between percent-level to order-of-magnitude upper-limits (cf.~Table \ref{tab:combined}). This information will enable constraints on halos that are both poorly constrained (from simulations and observations) and sensitive to the properties of dark matter.
\end{itemize}

The transparency offered by WO effects to probe large-scale structures and the prospect for detection suggest several applications (Sec.~\ref{sec:applications}). GW data may allow a reconstruction of the lensing potential under the assumption of a symmetric lens. In some cases, identifying \wows{} on a GW signal might be used to infer the magnification of the signal, mitigating a major uncertainty for standard sirens. Novel probes of large-scale structure could be developed: for instance, identifying several peaks in the \wow{} may serve to constrain the morphology of the gravitational lens. Finally, constraints on the abundance of subgalactic halos could be improved significantly thanks to the information on the lens (projected mass, impact parameter, etc...) obtained from the \wow.
In addition, we envision future directions regarding the computational framework, lens modelling and data analysis.

Our computational frameworks are flexible, accurate and efficient. While we have focused mostly on weak lensing, our methods can be readily applied to single-image strong lensing, e.g.~a source very close to a caustic on the side in which a single GO image forms. This regime has WO features extending to very high frequencies, and could be used to probe the phenomenology of strongly lensed GWs without the challenges of including multiple GO images \cite{Ulmer:1994ij,Tambalo:2022plm}. Another interesting extension is microlensing by extended structures, i.e.~considering the effects of lens sub-structure on a macro-image. 
Ultimately, our algorithms will be integrated in the ``Gravitational Lensing of Waves" (GLoW) code for public use by the scientific community.%
\footnote{Expected release December 2023, \url{https://github.com/miguelzuma/GLoW_code}}

Another important extension is improved lens modelling. Our analysis largely relied on SIS. This choice was motivated by both simplicity and this lens' particular stance in the single-image regime: the SIS is the ``cuspiest'' profile that does not form multiple images for arbitrarily large $y$. It leads to the sharpest possible peak in the \wow, which becomes smoother in the presence of a central core.
Other, well-motivated, symmetric profiles such as NFW \cite{Navarro:1996gj} typically have shallower inner cusps. It is interesting to consider these well-motivated distributions, as well as profiles motivated by dark matter theories \cite{Tambalo:2022wlm}. Beyond symmetric lenses, future models should include elliptic matter distributions, external convergence and shear and realistic realizations of substructure. 
Besides more complex lens models, addressing observational prospects will require reliable halo mass functions for $\Mvir\lesssim 10^8 \, M_\odot$ halos and detailed predictions for source rates (e.g.~Refs.~\cite{Erickcek:2006xc,Sesana:2007sh,Sesana:2010wy,Klein:2015hvg,Katz:2019qlu,Barausse:2020mdt,Toubiana:2021iuw}).

Important open questions remain on the signal analysis. A limitation of our analysis is the use of mismatch as a detectability criterion, which is optimistic (e.g.~ignoring parameter degeneracies). Future analyses should rely on Fisher matrix or Bayesian sampling. In addition, it is necessary to address the issue of false alarm triggers, i.e.~detector noise mimicking a \wow{} signature. These mundane effects will affect the detection prospects, although less significantly than uncertainties on the source rate, cf.~Table \ref{tab:combined}.
A standing open question is how to optimally identify and analyze \wows{} from GW signals. A non-parametric reconstruction of the Green's function from strain data could provide a partial reconstruction of the lensing potential and hence the projected mass (Eq.~\eqref{eq:lens_reconstruction}). These insights could be combined with methods for lens-agnostic analyses \cite{Liu:2023ikc}. Given the chance of signals being affected by multiple \wows{} (but with an unknown number), it will be necessary to devise a framework for data analysis that does not assume a fixed number of lenses, e.g.~along the lines of reversible-jump samplers \cite{Karnesis:2023ras}.
Regardless of identifying \wow, it is important to prevent unaccounted-for signatures to bias the remaining parameter estimation, e.g.~in the LISA global fit \cite{Littenberg:2023xpl}, or misinterpreting those residuals as new physics (e.g.~violation of Einstein's GR).

There is a promising future for GW lensing at the intersection between the WO and WL regimes. While subtle, \wow{} may be common enough to offer a window into cosmic structures and their properties, and will likely lead to applications beyond the ones outlined here. LISA stands out as a particularly promising probe, thanks to the combination of large SNR and low-frequency sensitivity. If deployed, other proposed space detectors will yield even more impressive results, thanks to larger SNR \cite{Baibhav:2019rsa,Sedda:2019uro}, a lower frequency band \cite{Sesana:2019vho} or both. These observations will provide novel means to probe low-mass halos that are both notoriously elusive and a prime test-bed for dark matter models. \wow{} may thus become a powerful probe of large-scale structure and fundamental physics thanks to the next generation of GW detectors on the ground and in space.

\acknowledgements{We are very grateful to G. Brando, M. \c{C}al\i{}\c{s}kan, M. Cheung, L. Choi, L. Dai, S. Delos, J. Gair, M. Katz, M. Lagos, S. Singh and J. Streibert for discussions in connection to this project and comments on the manuscript.
HVR is supported by the Spanish Ministry of Universities through a Margarita Salas Fellowship, with funding from the European Union under the NextGenerationEU programme.}

\appendix

\section{Derivation of lensing results}\label{sec:appendix_ders}
In this appendix, we give details in the derivations of the time-domain signal $\Cc I(\tau)$ of Eq.~\eqref{eq:contour_decomposition}. Also, we explain how to obtain the coefficients of the GO expansion in Eqs.~\eqref{eq:bgo_time_dom}, \eqref{eq:bgo_freq_dom}.

\subsubsection*{Derivation of $\Cc I(\tau)$}
Let us review how to obtain $\Cc I(\tau)$. This discussion mainly follows Refs.~\cite{Ulmer:1994ij,Tambalo:2022wlm}.
We start from the definition of Eq.~\eqref{eq:lensing_contour_time_integral}.
To simplify this expression we choose coordinates in the lens plane adapted to the Fermat potential $\phi(\vect x, \vect y)$. In particular, we introduce a variable $t \equiv \phi(\vect x, \vect y)$, while as a second coordinate we use the ``proper time'' $u$ along the curve of constant $t$ (later we will change it to the arc-length distance $s$ along constant-$t$ lines.
From this choice, it follows that the tangent vectors to the $t = {\rm const.}$ lines are orthogonal to the gradient of $\phi(\vect x, \vect y)$:
\begin{equation}\label{eq:contours_orth}
        \frac{\partial x^{i}}{\partial u} \partial_i \phi(\vect x, \vect y) = 0\;,
\end{equation}
where $x^i$ indicates the component $i$ of the vector $\vect x$, with $i, j, \hdots = 1,2$ and the summation is implicit. Due to Eq.~\eqref{eq:contours_orth} we can construct $\partial x^{i}/\partial u$ as
\begin{equation}
        \dot x^i 
        \equiv 
        \frac{\partial x^{i}}{\partial u} 
        =
        f \epsilon^{ij}\partial_j \phi(\vect x, \vect y)\;,
\end{equation}
where $\epsilon^{ij}$ is the Levi-Civita pseudo tensor, with $\epsilon^{12} = 1$, and $f$ is a normalization function to be chosen. The dot stands for the derivative with respect to $u$.

The dimensionless time-delay $t$ changes perpendicularly to the contours i.e.~$\partial_i t \, \dot x^i = 0$. Also, $u$ at constant $t$ changes orthogonally to $\partial_i t$:
\begin{equation}
        \partial_i u 
        = 
        g \, \epsilon_i^{\;j} \partial_j \phi(\vect x, \vect y)
        \;.
\end{equation}
Here $g$ is a function to be fixed.
We choose $u$ is such a way that the measure $\de ^2 \vect x$ becomes simple in the new coordinates. Let us call $X^a = \{t, u\}$. Then the measure changes by a Jacobian
\begin{align}
        {\rm det}\, \partial_i X^a 
        =
        \epsilon^{ij} \partial_i t\partial_j u
        =
        g \, \epsilon^{ij} \partial_i \phi\, \epsilon_j^{\;k}\partial_k \phi
        = g\, |\vect \nabla \phi|^2
        \;.
\label{eq:det_tu}
\end{align}
Here we used $\epsilon^{ij}\epsilon_j^{\;k} = \delta^{ik}$. Now we set this determinant to $1$ by choosing $g = 1/|\vect \nabla \phi|^2$. With this choice, we see that $\de^2 \vect x = \de t \, \de u$. Moreover, since the determinant in Eq.~\eqref{eq:det_tu} is equal to one, we also have $|{\rm det}\, \partial_a x^{i}| = 1$: it is easy to check this then fixes the function $f = 1$.

Now we introduce the arc-length distance $s$, which is related to $u$ by the differential relation $\de s^2 = \dot{\vect{x}}^2 \de u^2 = |\vect \nabla \phi|^2\de u^2$.
Expressed in terms of the variables $u$ and $s$, we have that Eq.~\eqref{eq:lensing_contour_time_integral} becomes respectively
\begin{align}
        \Cc I(\tau) 
        &= 
        \sum_k \int \de t \, \de u
        \,
        \delta\left(t-\tau\right)
        \\ \nonumber
        &= 
        \sum_k \int \frac{\de t \, \de s}{|\vect \nabla \phi|}
        \delta\left(t-\tau\right)        
        \;,
\end{align}
where $\sum_k$ is the sum over distinct contours with same time delay.
Since the Fermat potential $\phi(\vect x, \vect y)$ is positive by construction, we then have $t > 0$ in the integration above. 

\subsubsection*{Beyond Geometric Optics coefficients}
In the limit of large $w$, or for small time delays, the lensing signal can be expressed as a GO expansion. 
Indeed, at high frequencies the diffraction integral \eqref{eq:amplification_F} can be obtained using a stationary-phase approximation. This involves solving Gaussian integrals weighted by powers of derivatives of $\psi$ at the location of the image (see Ref.~\cite{Takahashi:2004mc,Tambalo:2022plm}).
Here we provide the expression for the GO coefficients $\mu$, $\Delta_{(1)}$ and  $\Delta_{(2)}$ appearing in Eqs.~\eqref{eq:bgo_freq_dom}, \eqref{eq:bgo_time_dom} for axi-symemetric lenses.
These quantities are given by
\begin{align}
    |\mu|^{-1}
    &\equiv
    4 a b
    \;,
    \\
    \Delta_{(1)} 
    &\equiv
    \frac{1}{16}\left[
    \frac{\psi^{(4)}}{2 a^{2}}
    +\frac{5}{12 a^{3}} (\psi^{(3)})^{2}
    +\frac{\psi^{(3)}}{a^{2} x}
    +\frac{a-b}{a b x^{2}}
    \right]
    \;,
    \\
    \Delta_{(2)}
    & \equiv
    \frac{1}{512}
    \bigg[
    \frac{3 \left(3 a^2+2 a b-5 b^2\right)}{a^2 b^2 x^4}
    +\frac{385 (\psi^{(3)})^4}{144 a^6}
    \nn
    \\
    &  
    +\frac{35 (\psi^{(3)})^3}{6 a^5 x}
    +\frac{35 (\psi^{(4)})^2}{12 a^4}
    +\frac{(a-5 b) \psi^{(4)}}{a^3 b x^2}
    \nn
    \\
    & 
    -2\psi^{(3)} \left(-\frac{7 \psi^{(5)}}{3 a^4}
    -\frac{35 \psi^{(4)}}{3 a^4 x}
    +\frac{2(a-5 b)}{ a^3 b x^3}\right)
    \nn
    \\
    & 
    -(\psi^{(3)})^2 \left(-\frac{35 \psi^{(4)}}{4 a^5}
    -\frac{5 (a-7 b)}{6 a^4 b x^2}\right)
    \nn
    \\
    & 
    +\frac{4\psi^{(6)}}{3a^3}
    +\frac{4 \psi^{(5)}}{a^3 x}
    \bigg]\;.
\end{align}
Here we defined $a\equiv (1 - \psi'')/2 $, $b \equiv (1 - \psi' / x) / 2$, $\psi^{(n)} \equiv \de^n \psi / \de x^n$ and all quantities are evaluated at the location of the image, $x = x_m$.
As far as we know, the coefficient $\Delta_{(2)}$ was not given in the literature before.

\section{Symmetric lens models}\label{sec:lens_models}

Here we provide some details of the symmetric lens models considered in the text (SIS, CIS) and the relation between the effective lens mass and the virial mass of the halo.

\subsubsection*{Singular Isothermal Sphere}
A commonly used approximation to the density of a halo is given by the SIS profile
\begin{equation}
  \rho(r)  
  =  
  \frac{\sigma_v^2}{2\pi G r^2}  
  \;,
\end{equation}
where $\sigma_v$ is the velocity dispersion of the halo.
For this lens, a convenient choice for the arbitrary scale $\xi_0$ in Eq.~\eqref{eq:def_MLz} is $\xi_0 = \sigma_v^2 / (G \, \Sigma_{\rm cr})$, with $\Sigma_{\rm cr}=(4 \pi G(1+z_L) d_{\rm eff})^{-1}$.
The lensing potential $\psi(x)$ associated to $\rho(r)$ then becomes very simple $\psi(x) = x$.

In the GO limit, the SIS can have one or two images depending on whether the impact parameter is outside or inside the caustic $y_{\rm cr} = 1$, respectively.
For $y < y_{\rm cr}$, two images form (a minimum and a saddle, labelled respectively by $(+)$ and $(-)$) with magnifications $\mu_{\pm} = 1/y\pm 1$, time delays $\phi_{\pm} = \mp y -1/2$ and Morse phases $n_+ = 0$ (minimum), $n_- = 1/2$ (saddle). Only the image corresponding to the minimum survives for $y > y_{\rm cr}$. The minimum time delay is given by $\phi_m(y)  = -y - 1/2$.

\subsubsection*{Cored Isothermal Sphere}
We will also consider a CIS, a variant of the SIS in which the presence of a central core smoothes the density profile \cite{1987ApJ...320..468H,Flores:1995dc}:
\begin{equation}\label{eq:rho_cored}
    \rho 
    = 
    \rho_0\frac{r_c^2}{r^2+r_c^2}
    \;,
\end{equation}
where $\rho_0$ is the central density and $r_c$ is the core radius.
The surface density is $\Sigma(\xi) =  \pi \rho_0 r_c^2 / \sqrt{\xi^2+r_c^2}$. Choosing a normalization scale $\xi_0  = 2\pi \rho_0 r_c^2 / \Sigma_{\rm cr}$, gives the lensing potential 
\begin{equation}
\label{eq:cis_lensing_pot}
    \psi  = \sqrt{x_c^2+x^2} +x_c \log \left(2 x_c / (\sqrt{x_c^2+x^2}+x_c)\right) \ ,
\end{equation}
where $x_c \equiv r_c/\xi_0$. 

Similarly to the SIS, multiple images form for sources within the caustic $y_{rc}(x_c)\leq 1$, which is smaller than for SIS. An additional requirement for multiple images is $x_c < 1/2$, so the lens' central density $\Sigma_{\rm cr}$.
The eventual additional GO image is associated to the maximum of the Fermat potential, and forms close to the center of the lens. In the SIS limit $x_c\to 0$, the GO magnification vanishes and this image is replaced by the cusp feature.
This lens and the properties of the central image are discussed in detail in Ref.~\cite{Tambalo:2022wlm} Sec.~IIIC. 

\subsubsection*{Relation between $M_{Lz}$ and $M_{\rm vir}$}

For extended lenses, the redshifted lens mass defined in Eq.~\eqref{eq:def_MLz} does not typically coincide with the physical mass of the halo. The two quantities can differ even by a few orders of magnitude. 
As a definition for the physical mass we consider the \emph{virial mass} $\Mvir$, defined as the mass up to the virial radius $\rvir$, i.e.~$\Mvir \equiv 4 \pi \int_0^{\rvir}\de r  r^2 \rho(r)$ (see Ref.~\cite{Tambalo:2022wlm} for the full expressions for SIS and CIS lenses).
On the other hand, the virial radius is defined as $\rho(r_{\rm vir }(z_L)) \equiv \Delta_c \rho_c(z_L)$, with $\Delta_c(z_L>1)\simeq 18 \pi^2$ and $\rho_c$ being the critical density at redshift $z_L$, $\rho_c= 3 H(z_L)/(8\pi G)$. Due to these relations, the virial mass is a function of the lens redshift \cite{Bryan:1997dn}.

 In the case of the SIS, with $\xi_0$ as in the previous subsections, $\MLz$ and the virial mass are related by:
\begin{align}\label{eq:virial_mass_appendix}
    \MLz
    &=
    \frac{4 \pi^2}{G}
    (1+z_L)^2 d_{\rm eff}
    \left(
    \frac{5\sqrt{6}}{2}G H(z_L) \Mvir
    \right)^{4/3}
    \\ \nonumber
    &=
    2.3
    \times 
    10^6 \, M_\odot (1 + z_L)^2
    \left(
    \frac{d_{\rm eff}}{1\,{\rm Gpc}}
    \right)
    \times
    \\ \nonumber
    &
    \hspace{3cm}
    \times
    \left(
    \frac{\Mvir}{10^{9} \, M_\odot} \frac{H(z_L)}{H_0}
    \right)^{4/3}
    \,.
\end{align}
Similar expressions for the CIS profile can be adapted from Sec.~III-C-1 in Ref.~\cite{Tambalo:2022wlm}. 

\section{Non-perturbative results for the SIS}
The amplification factor for the SIS can be computed analytically both in the time and frequency domain. These expressions are valid both in the weak and strong lensing regimes. 
In this appendix we derive the explicit expression for $\Cc I(\tau)$, without assuming single-image or weak-lensing limits.

Here, we exceptionally write the lensing potential for the SIS as $\psi(x) = \psi_0 x$, where $\psi_0$ is a constant.\footnote{In general, $\psi_0$ will be related to the choice for $\xi_0$, and will be equal to one only for the specific choice used in the main text.} The full expression for $\Cc I(\tau)$, in the single-image regime and expanded in powers of $\psi_0$, needs to reduce to the perturbative calculation outlined in Sec.~\ref{sec:lensing_perturbative}. This will be a useful check of our results.

\subsubsection*{Time domain result}
  
Starting with the time-domain version of the amplification factor 
in Eq.~\eqref{eq:lensing_contour_time_integral}, we can write it as
\begin{align}
    \mathcal{I}_\text{\tiny SIS}(\tau) 
    = 
    \int\de^2 \vect x \, 
        \delta\Big(\frac{1}{2}x_1^2 
    &- x_1y +\frac{1}{2}x_2^2 + \frac{1}{2}y^2\nonumber\\ 
    &- \psi_0x - \tau - \phi_m\Big)\ ,
\end{align}
where, in this case, the minimum time delay is
\begin{equation}
    \phi_m = -\frac{1}{2}\psi_0(\psi_0+2y)\ .
\end{equation}
After changing to polar coordinates, the integral over the $\delta$ function can be solved analytically, both in the radial and in the angular coordinate, since the Fermat potential is quadratic both in the radius and in the cosine of the angle. 
Finally, owing to the simplicity of the SIS, the second integral can also be performed analytically. 
We will express the integral as a function of two new variables, $u$ and $R$:
\begin{equation}\label{eq:def_u_R}
    u \equiv \frac{\sqrt{2\tau}}{\psi_0 + y}\ ,\hspace{1cm}
    R \equiv \frac{\psi_0-y}{\psi_0 + y}\ .
\end{equation}
The variable $u$ is a redefined time parameter while $R$ is a constant, ranging between $-1$ and $1$.

The final result can be compactly expressed as 
\begin{align}\label{eq:I_sis_analytic}
    \mathcal{I}_\text{\tiny SIS}(\tau) 
    &= 
    \frac{8(b-c)}{\sqrt{(a-c)(b-d)}}
    \left[
            \Pi\left(\frac{a-b}{a-c}, r\right) + \frac{c\,K(r)}{(b-c)}
    \right]
    \ ,
\end{align}
with
\begin{equation}
    r \equiv \sqrt{\frac{(a-b)(c-d)}{(a-c)(b-d)}}\ ,
\end{equation}
and where $\Pi$ and $K$ are, respectively, the complete elliptic integrals of the third and first kind, see e.g.~Ref.\cite{Gradshtein2007-ov}. 

The coefficients $a$, $b$, $c$ and $d$ are functions of the variables $u$ and $R$ defined in Eq.~\eqref{eq:def_u_R} above.
We must however distinguish between three regions:
\begin{itemize}
    \item \emph{Region 1}: $(u>1)$
        \begin{alignat*}{3}
            a &= 1+u \ ,                   &\qquad  c&= 1-u \ , \\
            b &= R+\sqrt{u^2 + R^2 -1} \ , &\qquad  d&= R-\sqrt{u^2 + R^2 -1} \ .
        \end{alignat*}
    \item \emph{Region 2}: $(\sqrt{1-R^2}<u<1)$
        \begin{itemize}
            \item \emph{Case A:} $(R>0)$
                \begin{alignat*}{3}
                    a &= 1 \ , &\qquad  c&= \sqrt{1-u^2} \ , \\
                    b &= R \ , &\qquad  d&= -\sqrt{1-u^2} \ .
                \end{alignat*}
            \item \emph{Case B:} $(R<0)$
                \begin{alignat*}{3}
                    a &= 1            \ , &\qquad c&= -\sqrt{1-u^2} \ , \\
                    b &= \sqrt{1-u^2} \ , &\qquad d&= R \ .
                \end{alignat*}
        \end{itemize}
    \item \emph{Region 3}: $(0<u<\sqrt{1-R^2})$
        \begin{alignat*}{3}
            a &= 1            \ , &\qquad c&= R \ , \\
            b &= \sqrt{1-u^2} \ , &\qquad d&= -\sqrt{1-u^2} \ .
        \end{alignat*}
\end{itemize}

  \subsubsection*{Frequency domain result}

At this point, one can move to the Fourier transform of $\Cc I(\tau)$, so to obtain the amplification factor $F(w)$.
Starting from Eq.~\eqref{eq:amplification_F}, we could not find a closed-form expression for $F(w)$, but we were able to reduce it to a single angular integral. 
Using polar coordinates again, we can solve the radial integral, obtaining
\begin{align}
    F(w) 
    = 
    e^{iw(y^2/2-\phi_m)}
    \Big[
        1
        &+\int^\pi_0\de\theta\,\alpha f(-\alpha)
        \nonumber\\
        &-i\int^\pi_0\de\theta\,\alpha g(-\alpha)
    \Big]
    \ ,
\end{align}
with
\begin{equation}
    \alpha(\theta) 
    \equiv
    \sqrt{\frac{w}{\pi}}
    \left(\psi_0 + y\cos\theta\right) 
    \ ,
\end{equation}
and where $f$ and $g$ are the auxiliary functions for the Fresnel integrals.
Using the conventions of \cite{NIST:DLMF}, they can be written in terms of the Fresnel sine $S$ and cosine $C$ as
\begin{subequations}
\begin{align}
    f(z) &\equiv \left(\frac{1}{2} - S(z)\right)
            \cos\left(\frac{\pi}{2}z^2\right)\nonumber\\
        &\quad - \left(\frac{1}{2} - C(z)\right)\sin\left(\frac{\pi}{2}z^2\right)\ ,\\
    g(z) &\equiv \left(\frac{1}{2} - C(z)\right)
            \cos\left(\frac{\pi}{2}z^2\right)\nonumber\\
        &\quad + \left(\frac{1}{2} - S(z)\right)\sin\left(\frac{\pi}{2}z^2\right)\ .
\end{align}
\end{subequations}
Other representations for $F(w)$ are also available. See, for instance, \cite{Tambalo:2022plm} for a series representation.

\section{Weak lensing expansion for the SIS}\label{sec:appendix_an_res}

In this appendix, we will apply the weak-lensing expansion of Sec.~\ref{sec:lensing_perturbative} to the case of the SIS profile. Thanks to the simplicity of this model, we can obtain explicit expressions and characterize their \wows{}.

\subsubsection*{Time domain result} 

Let us apply the time-domain approximation to the SIS, $\psi = x$. Let us focus on the single-image region, $y > 1$.
Starting from Eq.~\eqref{eq:time_domain_perturbative_expansion}, it is possible to have a closed-form expression for $\Cc I_{(1)}$:
\begin{align}
	\Cc I_{(1)}(\tau)
	&=
	4\frac{\de }{\de\tau}
	\left[
		(a+y)
		\int_{0}^{\pi/2}\de \varphi\,\sqrt{1-q \sin^2 \varphi}
	\right]
	\nn \\
	&=
	4\frac{\de }{\de\tau} 
	\left[
	(a+y) E\left(q\right)
	\right]
	\nn \\
	&=
	2\frac{(a+y)}{a^{2}}E\left(q\right)
	+
	2\frac{(a-y)}{a^{2}}K\left(q\right)
	\;.
	\label{eq:I1_sis}
\end{align}
Here we defined $q \equiv 4 ay / (a+y)^2$, $a \equiv \sqrt{2 t}$, and $t = \tau + \phi_m$ while the functions $K(q)$ and $E(q)$ are the complete elliptic integrals of the first and second kind, respectively. 

The functions $K(q)$ and $E(q)$ are real for $q \leq 1$. Notice also that the parameter $q$ as a function of $t$ spans from $0$ to $1$. The latter value is attained at $t = y^2/2$, while zero is reached asymptotically for large $t$.
We notice that the function \eqref{eq:I1_sis} is always positive. 
More specifically, it decays for large $\tau$ and has a peak for time delays corresponding to the center of the lens (i.e.~around $t = y^2/2$).

Although there is no analytic expression for the location of the maximum, we can see that it is located around $q = 1$ (here $E(q)$ has a maximum). This is physically reasonable since it corresponds to the features due to the center of the lens. 
We can actually find a good approximation for the maximum by expanding $\Cc I_{(1)}$ around $t = y^2/2$. To do so, we first write $q = 1 - \epsilon$ and expand for small $\epsilon$:
\begin{align}
	\Cc I_{(1)}(\tau)
	&\simeq 
	\frac{4}{y}
	-\frac{2\sqrt{\epsilon}}{y}\left[6+\log(\epsilon/16) \right]
     \nn \\
	&
	+\frac{\epsilon}{y}\left[19+5\log(\epsilon/16)\right]
    +\Cc O(\epsilon^{3/2})
	\;.
\end{align}
Setting the derivative with respect to $\epsilon$ of this expression to zero yields an equation for the maximum, that can be solved. Such value $\bar{\epsilon}$ does not depend on $y$, and is approximately $\bar{\epsilon} \simeq 2.1 \times 10^{-3}$.
Having found the approximate maximum $\bar{q} = 1-\bar{\epsilon}$, we can translate to $t$. 
For small $\epsilon$, $t \simeq y^2/2(1+4 \sqrt{\epsilon} + 8 \epsilon)$, so that $\bar{t} \simeq 0.6 y^2$.
With these values we can evaluate $\Cc I_{(1)}(\tau)$ at the peak, $\Cc I_{(1)}(\bar{\tau}) \simeq 4.22/y$.
We can actually have an analytic form for the peak, with a very good approximation of the tail at large $\tau$. To get it, we expand the elliptic integrals around $q = 1$, and obtain
\begin{align}
	\Cc I_{(1)}(\tau)
	&\simeq
	\log (1-q) 
	\left[
	\frac{(q+3) y}{8 t}
	+\frac{3q-7}{4 \sqrt{2t} }
	\right]
     \nn \\
	&
	+\frac{1}{\sqrt{2t}}[1+q-q \log 8+\log 128]
     \nn \\
	&
	-\frac{y}{2 t}[q \log 2-2+\log 8]
	\;,
\label{eq:approx_I1_sis}
\end{align}
This expression well approximates the full $\Cc I_{(1)}$ but becomes unreliable at small $t$, where it diverges. In the latter region, one can use the GO expansion instead.
It is also interesting to evaluate Green's function at the peak. From the definition of $G(\tau)$, Eq.~\eqref{eq:green_function_def} and by expanding Eq.~\eqref{eq:approx_I1_sis} around the peak, we obtain
\begin{equation}
\label{eq:G_sis_singularity}
    G(\tau) 
    \simeq 
    - \frac{1}{\pi y^3} \log\left(\left| t - y^2 / 2 \right| \right)
    \;.
\end{equation}

We can also analyse the asymptotic behaviour for large $\tau$. This limit maps to the small-frequency limit since we are considering large time delays from the image.
At leading order for large $\tau$, the radius is approximated by $x \simeq \sqrt{2t}$ and is independent on the angle $\varphi$: see Eqs.~\eqref{eq:time_domain_perturbative_expansion_x1} and \eqref{eq:time_domain_perturbative_expansion_x2}. Then, taking the $\tau$ derivative in Eq.~\eqref{eq:time_domain_perturbative_expansion} inside the integral we obtain
\begin{equation}
    \Cc I_{(1)}(\tau)
    \simeq
    \frac{2 \pi}{\sqrt{2 t }}\psi'(\sqrt{2t})
    \;.
\end{equation}
Here $'$ stands for the derivative with respect to $x$. 
The result above applies to all symmetric lenses, particularly the SIS, where $\psi'(\sqrt{2t}) = 1$. We conclude that the falloff of $\Cc I_{(1)}(\tau)$ as a function of $\tau$ is related to the asymptotic properties of the lensing potential at large radii. In particular, more compact lenses have faster falloffs (e.g.~for a point lens $\psi(x) \propto \log x$, the decay is $\Cc I_{(1)}(\tau) \propto \tau^{-1}$).

\subsubsection*{Frequency domain result}

As we discussed in the previous paragraphs, the center of the lens can be responsible for a peak in Green's function in the time domain, even in the absence of an image. This is indeed the case for the SIS. In the section, using the frequency-domain approximation derived in Eq.~\eqref{eq:F_wl}, we obtain the analogous feature in the amplification factor.
For the SIS, it is straightforward to integrate Eq.~\eqref{eq:F_wl} directly:
\begin{equation}
\label{eq:wl_sis_feature}
   F(w)
   \simeq 
   1 
   + 
   i w y
   -\sqrt{\frac{\pi w}{2}} 
   e^{i z -i \frac{\pi}{4}} 
   \Bigl[
   2 z J_1(z) + (2 i z - 1) J_0(z)
   \Bigr]
   \;,
\end{equation}
where $z \equiv w y^2 / 4$ and $J_\nu(z)$ are the Bessel functions of the first kind. 
We can better appreciate the effect of the center by taking the high-$w$ limit of the expression just obtained:
\begin{equation}
\label{eq:go_sis_feature}
    F(w) 
    \simeq 
    \sqrt{|\mu|}
    + 
    \frac{i}{8 w y^3}
    + 
    \frac{e^{i w \phi_c}}{w y^3}
    \;,
\end{equation}
where $\sqrt{|\mu|} \simeq 1 + 2 / y$ at large $y$ and the time delay of the center is $\phi_c \equiv y^2 / 2$.
We can recognize the second term, going as $\propto w^{-1}$, as the approximate bGO correction from the image (it does not contain phases/time delays with respect to the magnification term). On the other hand, the third contribution containing $\phi_c$ originates from the center of the lens. In the SIS case, the dependence on $w$ of these two contributions scales in the same way, both in $w$ and $y$. 
Moreover, we notice that the bGO contribution is suppressed by a factor of $8$.

\subsubsection*{Application to cored lens}\label{sec:time_domain_pert_exp_cored}
In this subsection, we discuss the weak-lensing regime for a cored lens, focussing for simplicity on a variation of the CIS lens used in the main text.
In general, it is difficult to obtain analytic expressions for lenses more complicated than the SIS. For instance, the CIS's $\Cc I_{(1)}(\tau)$ is hard to evaluate analytically (the problematic term is the $\log$ term in Eq.~\eqref{eq:cis_lensing_pot}).
To have a sense of the effect of a core, we can use a similar cored profile with density
\begin{equation}
	\rho(r)
	= 
	\rho_0 \frac{r_c^2}{r^2+r_c^2}
	\left[
	1+ \frac{2 r_c^2}{r^2+r_c^2}
	\right]
	\;.
\end{equation}
Notice that the central density of this profile is $3 \rho_0$. This is the main difference from the CIS used in the main text, while this profile also approaches asymptotically the SIS.
It is immediate to see that this leads, after a proper choice for $\xi_0$, to the following lensing potential
\begin{equation}
	\psi(x) = \sqrt{x^2+x_c^2}\;,
\end{equation}
where $x_c = r_c/\xi_0$. 
We can obtain $\Cc I_{(1)}(\tau)$ by following the same steps as for SIS, with some slight modifications.
The final result is
\begin{equation}
  \Cc I_{(1)} 
  =
  \frac{2}{a^2}
  \left[
    \sqrt{(a+y)^2 +x_c^2} \, E(\tilde q)
    +
    \frac{a^2-y^2-x_c^2}{\sqrt{(a+y)^2 +x_c^2}} \, K(\tilde q)
  \right]
  \;.
  \label{eq:I1_cored}
\end{equation}
Here we introduced the new variable $\tilde q \equiv 4 a y/ [(a+y)^2 + x_c^2]$, while $a = \sqrt{2t}$ as for the SIS. 
In comparing this expression with Eq.~\eqref{eq:I1_sis}, the main difference resides in $\tilde q$: in the cored case, $\tilde q$ is strictly smaller than 1 while for SIS $q$ can reach 1. Since the peak in $G(\tau)$ is roughly given by the maximum value of $\tilde q$ or $q$, this implies a smoother feature in the cored case. Recall that for the SIS Green's function develops a log-divergence at the peak, whereas in the presence of a core the feature becomes regular. This is also seen for the CIS lens, in Fig.~\ref{fig:CIS_vary_xc}.
The location of the peak is instead mildly affected by $x_c$.
    
\section{Accuracy and tests of the weak lensing expansion}\label{app:Comparisons}

\begin{figure}
        \centering
        \includegraphics[width=\columnwidth]{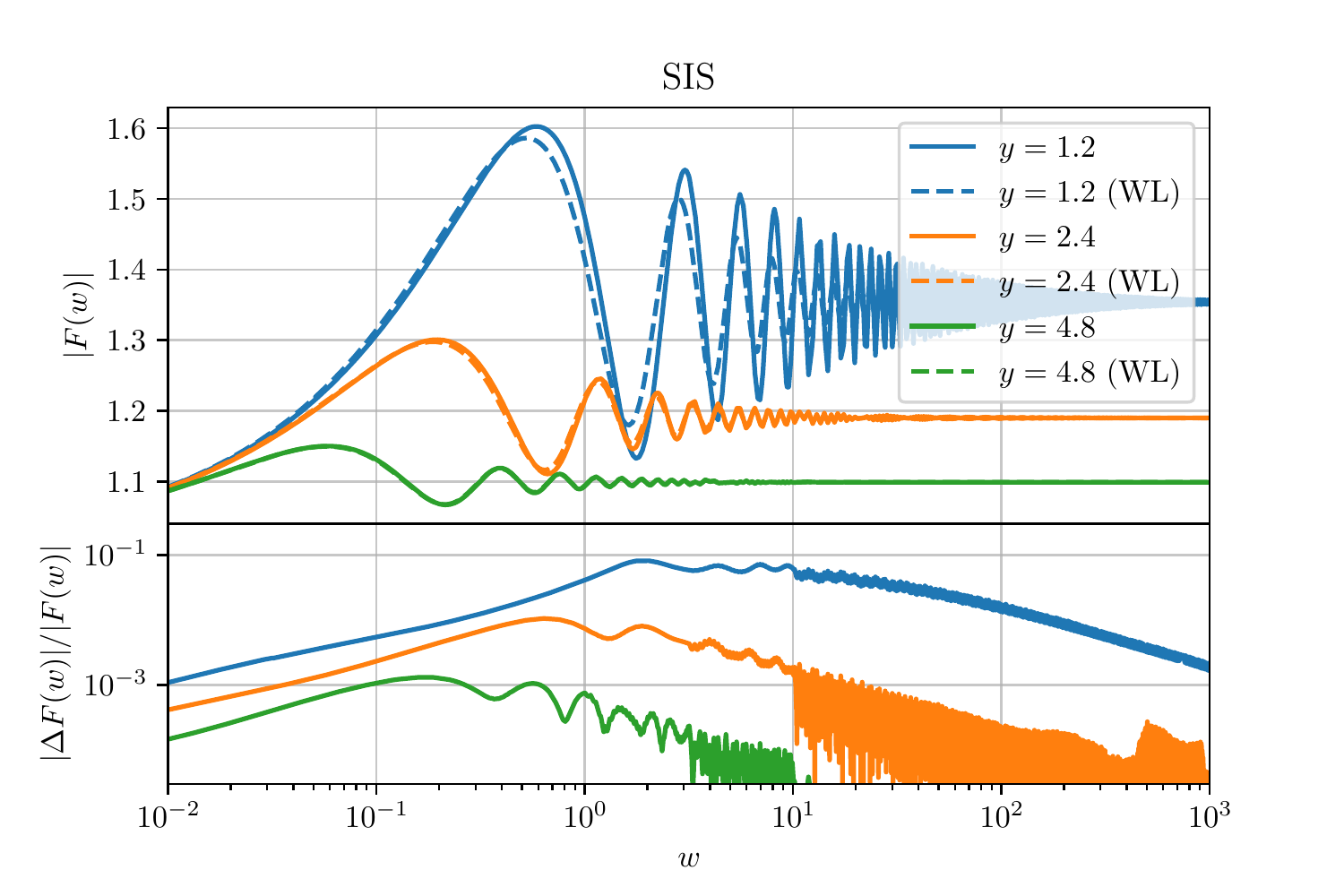}
        \caption{Difference between the non-perturbative computation using 
            \eqref{eq:It_non_pert} and the WL linear approximation, described in Section 
            \ref{sec:lensing_perturbative}, for an SIS.}
        \label{fig:comparison_Fw_SIS}
\end{figure}
\begin{figure}
        \centering
        \includegraphics[width=\columnwidth]{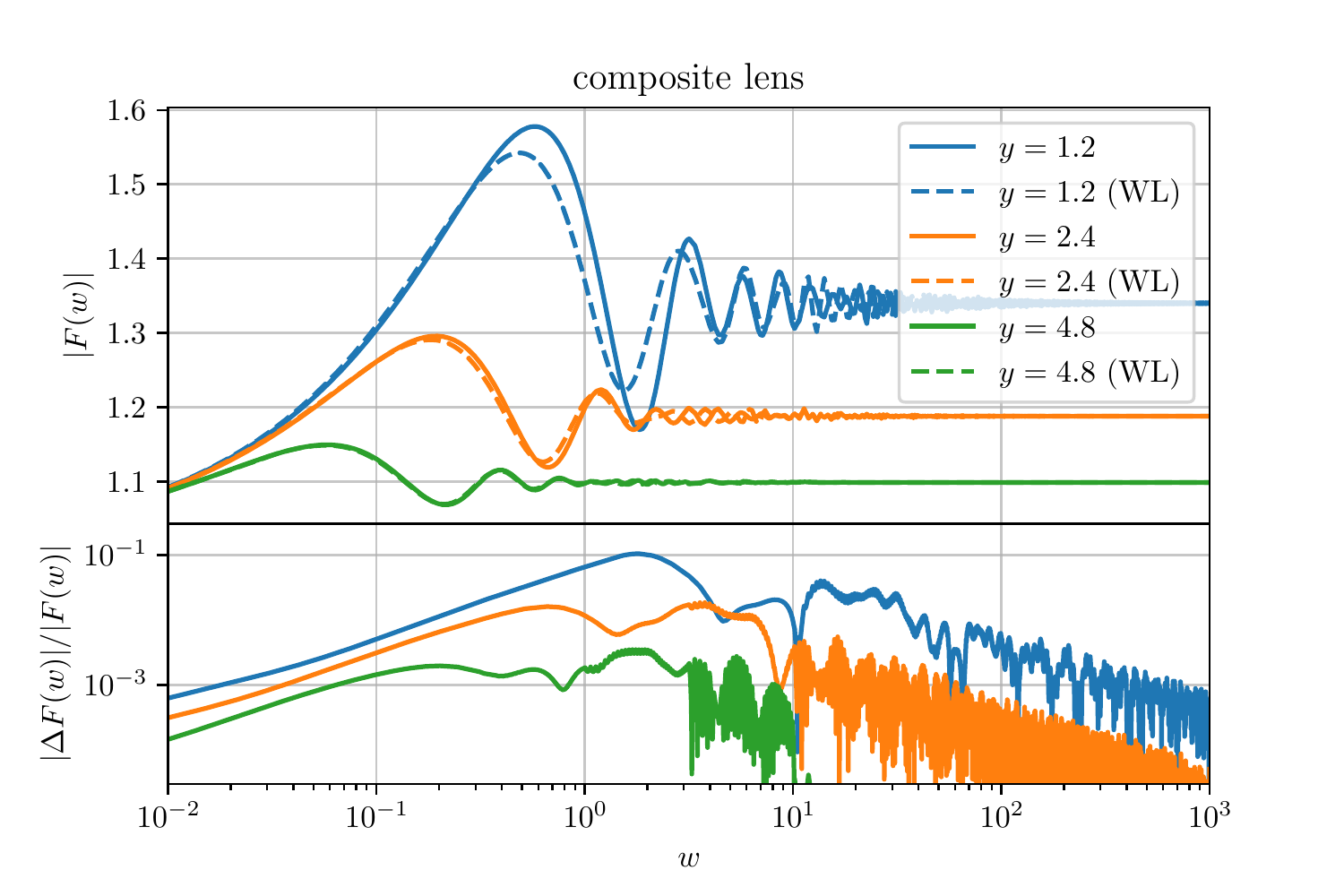}
        \caption{Difference between the non-perturbative computation using 
            \eqref{eq:It_non_pert} and the WL linear approximation, described in Section 
            \ref{sec:lensing_perturbative}, for a lens composed of three equal-mass SISs located at $(0.4, -0.5)$, 
            $(-0.1, -0.3)$, and $(-0.3, 0.8)$. The total mass of the lens is the same as the single
            SIS in Fig.~\ref{fig:comparison_Fw_SIS}.}
        \label{fig:comparison_Fw_composite}
\end{figure}

The results of the weak-lensing linear approximation introduced in Sec.~\ref{sec:lensing_perturbative} 
can be compared with the full non-perturbative computation, described in Sec.~\ref{sec:lensing_nonperturbative}. 
The performance of the WL approximation is excellent for large $y$, with relative differences falling
below the 1$\%$ level for $y\geq 3$ and remaining quite good, around $10\%$, even for impact parameters
very close to the strong-lensing limit $y\sim 1$. These tests have been performed for a variety of
lenses, axisymmetric and non-axisymmetric. Some results for the amplification factor are shown in 
Figs.~\ref{fig:comparison_Fw_SIS} and \ref{fig:comparison_Fw_composite}.
The time-domain results, i.e.~$\mathcal{I}(\tau)$, show a similar level of agreement. 
The inclusion of higher-order terms in the expansion would further increase the agreement with the non-perturbative method.
    
\newpage
\bibliographystyle{apsrev4-1}
\bibliography{gw_lensing}

\begin{thebibliography}{100}%
    \makeatletter
    \providecommand \@ifxundefined [1]{%
     \@ifx{#1\undefined}
    }%
    \providecommand \@ifnum [1]{%
     \ifnum #1\expandafter \@firstoftwo
     \else \expandafter \@secondoftwo
     \fi
    }%
    \providecommand \@ifx [1]{%
     \ifx #1\expandafter \@firstoftwo
     \else \expandafter \@secondoftwo
     \fi
    }%
    \providecommand \natexlab [1]{#1}%
    \providecommand \enquote  [1]{``#1''}%
    \providecommand \bibnamefont  [1]{#1}%
    \providecommand \bibfnamefont [1]{#1}%
    \providecommand \citenamefont [1]{#1}%
    \providecommand \href@noop [0]{\@secondoftwo}%
    \providecommand \href [0]{\begingroup \@sanitize@url \@href}%
    \providecommand \@href[1]{\@@startlink{#1}\@@href}%
    \providecommand \@@href[1]{\endgroup#1\@@endlink}%
    \providecommand \@sanitize@url [0]{\catcode `\\12\catcode `\$12\catcode
      `\&12\catcode `\#12\catcode `\^12\catcode `\_12\catcode `\%12\relax}%
    \providecommand \@@startlink[1]{}%
    \providecommand \@@endlink[0]{}%
    \providecommand \url  [0]{\begingroup\@sanitize@url \@url }%
    \providecommand \@url [1]{\endgroup\@href {#1}{\urlprefix }}%
    \providecommand \urlprefix  [0]{URL }%
    \providecommand \Eprint [0]{\href }%
    \providecommand \doibase [0]{http://dx.doi.org/}%
    \providecommand \selectlanguage [0]{\@gobble}%
    \providecommand \bibinfo  [0]{\@secondoftwo}%
    \providecommand \bibfield  [0]{\@secondoftwo}%
    \providecommand \translation [1]{[#1]}%
    \providecommand \BibitemOpen [0]{}%
    \providecommand \bibitemStop [0]{}%
    \providecommand \bibitemNoStop [0]{.\EOS\space}%
    \providecommand \EOS [0]{\spacefactor3000\relax}%
    \providecommand \BibitemShut  [1]{\csname bibitem#1\endcsname}%
    \let\auto@bib@innerbib\@empty
    \bibitem [{\citenamefont {{Schneider}}\ \emph {et~al.}(1992)\citenamefont
      {{Schneider}}, \citenamefont {{Ehlers}},\ and\ \citenamefont
      {{Falco}}}]{Schneider:1992}%
      \BibitemOpen
      \bibfield  {author} {\bibinfo {author} {\bibfnamefont {P.}~\bibnamefont
      {{Schneider}}}, \bibinfo {author} {\bibfnamefont {J.}~\bibnamefont
      {{Ehlers}}}, \ and\ \bibinfo {author} {\bibfnamefont {E.~E.}\ \bibnamefont
      {{Falco}}},\ }\href {\doibase 10.1007/978-3-662-03758-4} {\emph {\bibinfo
      {title} {{Gravitational Lenses}}}}\ (\bibinfo {year} {1992})\BibitemShut
      {NoStop}%
    \bibitem [{\citenamefont {Bartelmann}(2010)}]{Bartelmann:2010fz}%
      \BibitemOpen
      \bibfield  {author} {\bibinfo {author} {\bibfnamefont {M.}~\bibnamefont
      {Bartelmann}},\ }\href {\doibase 10.1088/0264-9381/27/23/233001} {\bibfield
      {journal} {\bibinfo  {journal} {Class. Quant. Grav.}\ }\textbf {\bibinfo
      {volume} {27}},\ \bibinfo {pages} {233001} (\bibinfo {year} {2010})},\
      \Eprint {http://arxiv.org/abs/1010.3829} {arXiv:1010.3829 [astro-ph.CO]}
      \BibitemShut {NoStop}%
    \bibitem [{\citenamefont {Mukherjee}\ \emph
      {et~al.}(2020{\natexlab{a}})\citenamefont {Mukherjee}, \citenamefont
      {Wandelt},\ and\ \citenamefont {Silk}}]{Mukherjee:2019wfw}%
      \BibitemOpen
      \bibfield  {author} {\bibinfo {author} {\bibfnamefont {S.}~\bibnamefont
      {Mukherjee}}, \bibinfo {author} {\bibfnamefont {B.~D.}\ \bibnamefont
      {Wandelt}}, \ and\ \bibinfo {author} {\bibfnamefont {J.}~\bibnamefont
      {Silk}},\ }\href {\doibase 10.1103/PhysRevD.101.103509} {\bibfield  {journal}
      {\bibinfo  {journal} {Phys. Rev. D}\ }\textbf {\bibinfo {volume} {101}},\
      \bibinfo {pages} {103509} (\bibinfo {year} {2020}{\natexlab{a}})},\ \Eprint
      {http://arxiv.org/abs/1908.08950} {arXiv:1908.08950 [astro-ph.CO]}
      \BibitemShut {NoStop}%
    \bibitem [{\citenamefont {Mukherjee}\ \emph
      {et~al.}(2020{\natexlab{b}})\citenamefont {Mukherjee}, \citenamefont
      {Wandelt},\ and\ \citenamefont {Silk}}]{Mukherjee:2019wcg}%
      \BibitemOpen
      \bibfield  {author} {\bibinfo {author} {\bibfnamefont {S.}~\bibnamefont
      {Mukherjee}}, \bibinfo {author} {\bibfnamefont {B.~D.}\ \bibnamefont
      {Wandelt}}, \ and\ \bibinfo {author} {\bibfnamefont {J.}~\bibnamefont
      {Silk}},\ }\href {\doibase 10.1093/mnras/staa827} {\bibfield  {journal}
      {\bibinfo  {journal} {Mon. Not. Roy. Astron. Soc.}\ }\textbf {\bibinfo
      {volume} {494}},\ \bibinfo {pages} {1956} (\bibinfo {year}
      {2020}{\natexlab{b}})},\ \Eprint {http://arxiv.org/abs/1908.08951}
      {arXiv:1908.08951 [astro-ph.CO]} \BibitemShut {NoStop}%
    \bibitem [{\citenamefont {Balaudo}\ \emph {et~al.}(2022)\citenamefont
      {Balaudo}, \citenamefont {Garoffolo}, \citenamefont {Martinelli},
      \citenamefont {Mukherjee},\ and\ \citenamefont
      {Silvestri}}]{Balaudo:2022znx}%
      \BibitemOpen
      \bibfield  {author} {\bibinfo {author} {\bibfnamefont {A.}~\bibnamefont
      {Balaudo}}, \bibinfo {author} {\bibfnamefont {A.}~\bibnamefont {Garoffolo}},
      \bibinfo {author} {\bibfnamefont {M.}~\bibnamefont {Martinelli}}, \bibinfo
      {author} {\bibfnamefont {S.}~\bibnamefont {Mukherjee}}, \ and\ \bibinfo
      {author} {\bibfnamefont {A.}~\bibnamefont {Silvestri}},\ }\href@noop {} {\
      (\bibinfo {year} {2022})},\ \Eprint {http://arxiv.org/abs/2210.06398}
      {arXiv:2210.06398 [astro-ph.CO]} \BibitemShut {NoStop}%
    \bibitem [{\citenamefont {Leung}\ \emph {et~al.}(2023)\citenamefont {Leung},
      \citenamefont {Jow}, \citenamefont {Saha}, \citenamefont {Dai}, \citenamefont
      {Oguri},\ and\ \citenamefont {Koopmans}}]{Leung:2023lmq}%
      \BibitemOpen
      \bibfield  {author} {\bibinfo {author} {\bibfnamefont {C.}~\bibnamefont
      {Leung}}, \bibinfo {author} {\bibfnamefont {D.}~\bibnamefont {Jow}}, \bibinfo
      {author} {\bibfnamefont {P.}~\bibnamefont {Saha}}, \bibinfo {author}
      {\bibfnamefont {L.}~\bibnamefont {Dai}}, \bibinfo {author} {\bibfnamefont
      {M.}~\bibnamefont {Oguri}}, \ and\ \bibinfo {author} {\bibfnamefont
      {L.~V.~E.}\ \bibnamefont {Koopmans}},\ }\href@noop {} {\  (\bibinfo {year}
      {2023})},\ \Eprint {http://arxiv.org/abs/2304.01202} {arXiv:2304.01202
      [astro-ph.HE]} \BibitemShut {NoStop}%
    \bibitem [{\citenamefont {Dai}\ \emph {et~al.}(2018)\citenamefont {Dai},
      \citenamefont {Li}, \citenamefont {Zackay}, \citenamefont {Mao},\ and\
      \citenamefont {Lu}}]{Dai:2018enj}%
      \BibitemOpen
      \bibfield  {author} {\bibinfo {author} {\bibfnamefont {L.}~\bibnamefont
      {Dai}}, \bibinfo {author} {\bibfnamefont {S.-S.}\ \bibnamefont {Li}},
      \bibinfo {author} {\bibfnamefont {B.}~\bibnamefont {Zackay}}, \bibinfo
      {author} {\bibfnamefont {S.}~\bibnamefont {Mao}}, \ and\ \bibinfo {author}
      {\bibfnamefont {Y.}~\bibnamefont {Lu}},\ }\href {\doibase
      10.1103/PhysRevD.98.104029} {\bibfield  {journal} {\bibinfo  {journal} {Phys.
      Rev. D}\ }\textbf {\bibinfo {volume} {98}},\ \bibinfo {pages} {104029}
      (\bibinfo {year} {2018})},\ \Eprint {http://arxiv.org/abs/1810.00003}
      {arXiv:1810.00003 [gr-qc]} \BibitemShut {NoStop}%
    \bibitem [{\citenamefont {Diego}\ \emph {et~al.}(2019)\citenamefont {Diego},
      \citenamefont {Hannuksela}, \citenamefont {Kelly}, \citenamefont
      {Broadhurst}, \citenamefont {Kim}, \citenamefont {Li}, \citenamefont
      {Smoot},\ and\ \citenamefont {Pagano}}]{Diego:2019lcd}%
      \BibitemOpen
      \bibfield  {author} {\bibinfo {author} {\bibfnamefont {J.~M.}\ \bibnamefont
      {Diego}}, \bibinfo {author} {\bibfnamefont {O.~A.}\ \bibnamefont
      {Hannuksela}}, \bibinfo {author} {\bibfnamefont {P.~L.}\ \bibnamefont
      {Kelly}}, \bibinfo {author} {\bibfnamefont {T.}~\bibnamefont {Broadhurst}},
      \bibinfo {author} {\bibfnamefont {K.}~\bibnamefont {Kim}}, \bibinfo {author}
      {\bibfnamefont {T.~G.~F.}\ \bibnamefont {Li}}, \bibinfo {author}
      {\bibfnamefont {G.~F.}\ \bibnamefont {Smoot}}, \ and\ \bibinfo {author}
      {\bibfnamefont {G.}~\bibnamefont {Pagano}},\ }\href {\doibase
      10.1051/0004-6361/201935490} {\bibfield  {journal} {\bibinfo  {journal}
      {Astron. Astrophys.}\ }\textbf {\bibinfo {volume} {627}},\ \bibinfo {pages}
      {A130} (\bibinfo {year} {2019})},\ \Eprint {http://arxiv.org/abs/1903.04513}
      {arXiv:1903.04513 [astro-ph.CO]} \BibitemShut {NoStop}%
    \bibitem [{\citenamefont {Yeung}\ \emph {et~al.}(2021)\citenamefont {Yeung},
      \citenamefont {Cheung}, \citenamefont {Gais}, \citenamefont {Hannuksela},\
      and\ \citenamefont {Li}}]{Yeung:2021roe}%
      \BibitemOpen
      \bibfield  {author} {\bibinfo {author} {\bibfnamefont {S.~M.~C.}\
      \bibnamefont {Yeung}}, \bibinfo {author} {\bibfnamefont {M.~H.~Y.}\
      \bibnamefont {Cheung}}, \bibinfo {author} {\bibfnamefont {J.~A.~J.}\
      \bibnamefont {Gais}}, \bibinfo {author} {\bibfnamefont {O.~A.}\ \bibnamefont
      {Hannuksela}}, \ and\ \bibinfo {author} {\bibfnamefont {T.~G.~F.}\
      \bibnamefont {Li}},\ }\href@noop {} {\  (\bibinfo {year} {2021})},\ \Eprint
      {http://arxiv.org/abs/2112.07635} {arXiv:2112.07635 [gr-qc]} \BibitemShut
      {NoStop}%
    \bibitem [{\citenamefont {Meena}\ \emph {et~al.}(2022)\citenamefont {Meena},
      \citenamefont {Mishra}, \citenamefont {More}, \citenamefont {Bose},\ and\
      \citenamefont {Bagla}}]{Meena:2022unp}%
      \BibitemOpen
      \bibfield  {author} {\bibinfo {author} {\bibfnamefont {A.~K.}\ \bibnamefont
      {Meena}}, \bibinfo {author} {\bibfnamefont {A.}~\bibnamefont {Mishra}},
      \bibinfo {author} {\bibfnamefont {A.}~\bibnamefont {More}}, \bibinfo {author}
      {\bibfnamefont {S.}~\bibnamefont {Bose}}, \ and\ \bibinfo {author}
      {\bibfnamefont {J.~S.}\ \bibnamefont {Bagla}},\ }\href {\doibase
      10.1093/mnras/stac2721} {\bibfield  {journal} {\bibinfo  {journal} {Mon. Not.
      Roy. Astron. Soc.}\ }\textbf {\bibinfo {volume} {517}},\ \bibinfo {pages}
      {872} (\bibinfo {year} {2022})},\ \Eprint {http://arxiv.org/abs/2205.05409}
      {arXiv:2205.05409 [astro-ph.GA]} \BibitemShut {NoStop}%
    \bibitem [{\citenamefont {Shan}\ \emph {et~al.}(2023)\citenamefont {Shan},
      \citenamefont {Chen}, \citenamefont {Hu},\ and\ \citenamefont
      {Cai}}]{Shan:2023ngi}%
      \BibitemOpen
      \bibfield  {author} {\bibinfo {author} {\bibfnamefont {X.}~\bibnamefont
      {Shan}}, \bibinfo {author} {\bibfnamefont {X.}~\bibnamefont {Chen}}, \bibinfo
      {author} {\bibfnamefont {B.}~\bibnamefont {Hu}}, \ and\ \bibinfo {author}
      {\bibfnamefont {R.-G.}\ \bibnamefont {Cai}},\ }\href@noop {} {\  (\bibinfo
      {year} {2023})},\ \Eprint {http://arxiv.org/abs/2301.06117} {arXiv:2301.06117
      [astro-ph.IM]} \BibitemShut {NoStop}%
    \bibitem [{\citenamefont {Meena}(2023)}]{Meena:2023qdq}%
      \BibitemOpen
      \bibfield  {author} {\bibinfo {author} {\bibfnamefont {A.~K.}\ \bibnamefont
      {Meena}},\ }\href@noop {} {\  (\bibinfo {year} {2023})},\ \Eprint
      {http://arxiv.org/abs/2305.02880} {arXiv:2305.02880 [astro-ph.CO]}
      \BibitemShut {NoStop}%
    \bibitem [{\citenamefont {Takahashi}\ and\ \citenamefont
      {Nakamura}(2003)}]{Takahashi:2003ix}%
      \BibitemOpen
      \bibfield  {author} {\bibinfo {author} {\bibfnamefont {R.}~\bibnamefont
      {Takahashi}}\ and\ \bibinfo {author} {\bibfnamefont {T.}~\bibnamefont
      {Nakamura}},\ }\href {\doibase 10.1086/377430} {\bibfield  {journal}
      {\bibinfo  {journal} {Astrophys. J.}\ }\textbf {\bibinfo {volume} {595}},\
      \bibinfo {pages} {1039} (\bibinfo {year} {2003})},\ \Eprint
      {http://arxiv.org/abs/astro-ph/0305055} {arXiv:astro-ph/0305055} \BibitemShut
      {NoStop}%
    \bibitem [{\citenamefont {\c{C}al\i{}\c{s}kan}\ \emph
      {et~al.}(2022{\natexlab{a}})\citenamefont {\c{C}al\i{}\c{s}kan},
      \citenamefont {Ji}, \citenamefont {Cotesta}, \citenamefont {Berti},
      \citenamefont {Kamionkowski},\ and\ \citenamefont
      {Marsat}}]{Caliskan:2022hbu}%
      \BibitemOpen
      \bibfield  {author} {\bibinfo {author} {\bibfnamefont {M.}~\bibnamefont
      {\c{C}al\i{}\c{s}kan}}, \bibinfo {author} {\bibfnamefont {L.}~\bibnamefont
      {Ji}}, \bibinfo {author} {\bibfnamefont {R.}~\bibnamefont {Cotesta}},
      \bibinfo {author} {\bibfnamefont {E.}~\bibnamefont {Berti}}, \bibinfo
      {author} {\bibfnamefont {M.}~\bibnamefont {Kamionkowski}}, \ and\ \bibinfo
      {author} {\bibfnamefont {S.}~\bibnamefont {Marsat}},\ }\href@noop {} {\
      (\bibinfo {year} {2022}{\natexlab{a}})},\ \Eprint
      {http://arxiv.org/abs/2206.02803} {arXiv:2206.02803 [astro-ph.CO]}
      \BibitemShut {NoStop}%
    \bibitem [{\citenamefont {Tambalo}\ \emph
      {et~al.}(2022{\natexlab{a}})\citenamefont {Tambalo}, \citenamefont
      {Zumalac\'arregui}, \citenamefont {Dai},\ and\ \citenamefont
      {Cheung}}]{Tambalo:2022wlm}%
      \BibitemOpen
      \bibfield  {author} {\bibinfo {author} {\bibfnamefont {G.}~\bibnamefont
      {Tambalo}}, \bibinfo {author} {\bibfnamefont {M.}~\bibnamefont
      {Zumalac\'arregui}}, \bibinfo {author} {\bibfnamefont {L.}~\bibnamefont
      {Dai}}, \ and\ \bibinfo {author} {\bibfnamefont {M.~H.-Y.}\ \bibnamefont
      {Cheung}},\ }\href@noop {} {\  (\bibinfo {year} {2022}{\natexlab{a}})},\
      \Eprint {http://arxiv.org/abs/2212.11960} {arXiv:2212.11960 [astro-ph.CO]}
      \BibitemShut {NoStop}%
    \bibitem [{\citenamefont {Lin}\ \emph {et~al.}(2023)\citenamefont {Lin},
      \citenamefont {Zhang}, \citenamefont {Dai}, \citenamefont {Huang},\ and\
      \citenamefont {Mei}}]{Lin:2023ccz}%
      \BibitemOpen
      \bibfield  {author} {\bibinfo {author} {\bibfnamefont {X.-y.}\ \bibnamefont
      {Lin}}, \bibinfo {author} {\bibfnamefont {J.-d.}\ \bibnamefont {Zhang}},
      \bibinfo {author} {\bibfnamefont {L.}~\bibnamefont {Dai}}, \bibinfo {author}
      {\bibfnamefont {S.-J.}\ \bibnamefont {Huang}}, \ and\ \bibinfo {author}
      {\bibfnamefont {J.}~\bibnamefont {Mei}},\ }\href@noop {} {\  (\bibinfo {year}
      {2023})},\ \Eprint {http://arxiv.org/abs/2304.04800} {arXiv:2304.04800
      [gr-qc]} \BibitemShut {NoStop}%
    \bibitem [{\citenamefont {\c{C}al\i{}\c{s}kan}\ \emph
      {et~al.}(2022{\natexlab{b}})\citenamefont {\c{C}al\i{}\c{s}kan},
      \citenamefont {Ezquiaga}, \citenamefont {Hannuksela},\ and\ \citenamefont
      {Holz}}]{Caliskan:2022wbh}%
      \BibitemOpen
      \bibfield  {author} {\bibinfo {author} {\bibfnamefont {M.}~\bibnamefont
      {\c{C}al\i{}\c{s}kan}}, \bibinfo {author} {\bibfnamefont {J.~M.}\
      \bibnamefont {Ezquiaga}}, \bibinfo {author} {\bibfnamefont {O.~A.}\
      \bibnamefont {Hannuksela}}, \ and\ \bibinfo {author} {\bibfnamefont {D.~E.}\
      \bibnamefont {Holz}},\ }\href@noop {} {\  (\bibinfo {year}
      {2022}{\natexlab{b}})},\ \Eprint {http://arxiv.org/abs/2201.04619}
      {arXiv:2201.04619 [astro-ph.CO]} \BibitemShut {NoStop}%
    \bibitem [{\citenamefont {Dai}\ and\ \citenamefont
      {Venumadhav}(2017)}]{Dai:2017huk}%
      \BibitemOpen
      \bibfield  {author} {\bibinfo {author} {\bibfnamefont {L.}~\bibnamefont
      {Dai}}\ and\ \bibinfo {author} {\bibfnamefont {T.}~\bibnamefont
      {Venumadhav}},\ }\href@noop {} {\  (\bibinfo {year} {2017})},\ \Eprint
      {http://arxiv.org/abs/1702.04724} {arXiv:1702.04724 [gr-qc]} \BibitemShut
      {NoStop}%
    \bibitem [{\citenamefont {Ezquiaga}\ \emph {et~al.}(2021)\citenamefont
      {Ezquiaga}, \citenamefont {Holz}, \citenamefont {Hu}, \citenamefont {Lagos},\
      and\ \citenamefont {Wald}}]{Ezquiaga:2020gdt}%
      \BibitemOpen
      \bibfield  {author} {\bibinfo {author} {\bibfnamefont {J.~M.}\ \bibnamefont
      {Ezquiaga}}, \bibinfo {author} {\bibfnamefont {D.~E.}\ \bibnamefont {Holz}},
      \bibinfo {author} {\bibfnamefont {W.}~\bibnamefont {Hu}}, \bibinfo {author}
      {\bibfnamefont {M.}~\bibnamefont {Lagos}}, \ and\ \bibinfo {author}
      {\bibfnamefont {R.~M.}\ \bibnamefont {Wald}},\ }\href {\doibase
      10.1103/PhysRevD.103.064047} {\bibfield  {journal} {\bibinfo  {journal}
      {Phys. Rev. D}\ }\textbf {\bibinfo {volume} {103}},\ \bibinfo {pages}
      {064047} (\bibinfo {year} {2021})},\ \Eprint
      {http://arxiv.org/abs/2008.12814} {arXiv:2008.12814 [gr-qc]} \BibitemShut
      {NoStop}%
    \bibitem [{\citenamefont {Vijaykumar}\ \emph {et~al.}(2022)\citenamefont
      {Vijaykumar}, \citenamefont {Mehta},\ and\ \citenamefont
      {Ganguly}}]{Vijaykumar:2022dlp}%
      \BibitemOpen
      \bibfield  {author} {\bibinfo {author} {\bibfnamefont {A.}~\bibnamefont
      {Vijaykumar}}, \bibinfo {author} {\bibfnamefont {A.~K.}\ \bibnamefont
      {Mehta}}, \ and\ \bibinfo {author} {\bibfnamefont {A.}~\bibnamefont
      {Ganguly}},\ }\href@noop {} {\  (\bibinfo {year} {2022})},\ \Eprint
      {http://arxiv.org/abs/2202.06334} {arXiv:2202.06334 [gr-qc]} \BibitemShut
      {NoStop}%
    \bibitem [{\citenamefont {Shajib}\ \emph {et~al.}(2022)\citenamefont {Shajib},
      \citenamefont {Vernardos}, \citenamefont {Collett}, \citenamefont {Motta},
      \citenamefont {Sluse}, \citenamefont {Williams}, \citenamefont {Saha},
      \citenamefont {Birrer}, \citenamefont {Spiniello},\ and\ \citenamefont
      {Treu}}]{Shajib:2022con}%
      \BibitemOpen
      \bibfield  {author} {\bibinfo {author} {\bibfnamefont {A.~J.}\ \bibnamefont
      {Shajib}}, \bibinfo {author} {\bibfnamefont {G.}~\bibnamefont {Vernardos}},
      \bibinfo {author} {\bibfnamefont {T.~E.}\ \bibnamefont {Collett}}, \bibinfo
      {author} {\bibfnamefont {V.}~\bibnamefont {Motta}}, \bibinfo {author}
      {\bibfnamefont {D.}~\bibnamefont {Sluse}}, \bibinfo {author} {\bibfnamefont
      {L.~L.~R.}\ \bibnamefont {Williams}}, \bibinfo {author} {\bibfnamefont
      {P.}~\bibnamefont {Saha}}, \bibinfo {author} {\bibfnamefont {S.}~\bibnamefont
      {Birrer}}, \bibinfo {author} {\bibfnamefont {C.}~\bibnamefont {Spiniello}}, \
      and\ \bibinfo {author} {\bibfnamefont {T.}~\bibnamefont {Treu}},\ }\href@noop
      {} {\  (\bibinfo {year} {2022})},\ \Eprint {http://arxiv.org/abs/2210.10790}
      {arXiv:2210.10790 [astro-ph.GA]} \BibitemShut {NoStop}%
    \bibitem [{\citenamefont {Sereno}\ \emph {et~al.}(2010)\citenamefont {Sereno},
      \citenamefont {Sesana}, \citenamefont {Bleuler}, \citenamefont {Jetzer},
      \citenamefont {Volonteri},\ and\ \citenamefont {Begelman}}]{Sereno:2010dr}%
      \BibitemOpen
      \bibfield  {author} {\bibinfo {author} {\bibfnamefont {M.}~\bibnamefont
      {Sereno}}, \bibinfo {author} {\bibfnamefont {A.}~\bibnamefont {Sesana}},
      \bibinfo {author} {\bibfnamefont {A.}~\bibnamefont {Bleuler}}, \bibinfo
      {author} {\bibfnamefont {P.}~\bibnamefont {Jetzer}}, \bibinfo {author}
      {\bibfnamefont {M.}~\bibnamefont {Volonteri}}, \ and\ \bibinfo {author}
      {\bibfnamefont {M.~C.}\ \bibnamefont {Begelman}},\ }\href {\doibase
      10.1103/PhysRevLett.105.251101} {\bibfield  {journal} {\bibinfo  {journal}
      {Phys. Rev. Lett.}\ }\textbf {\bibinfo {volume} {105}},\ \bibinfo {pages}
      {251101} (\bibinfo {year} {2010})},\ \Eprint {http://arxiv.org/abs/1011.5238}
      {arXiv:1011.5238 [astro-ph.CO]} \BibitemShut {NoStop}%
    \bibitem [{\citenamefont {Gao}\ \emph {et~al.}(2022)\citenamefont {Gao},
      \citenamefont {Chen}, \citenamefont {Hu}, \citenamefont {Zhang},\ and\
      \citenamefont {Huang}}]{Gao:2021sxw}%
      \BibitemOpen
      \bibfield  {author} {\bibinfo {author} {\bibfnamefont {Z.}~\bibnamefont
      {Gao}}, \bibinfo {author} {\bibfnamefont {X.}~\bibnamefont {Chen}}, \bibinfo
      {author} {\bibfnamefont {Y.-M.}\ \bibnamefont {Hu}}, \bibinfo {author}
      {\bibfnamefont {J.-D.}\ \bibnamefont {Zhang}}, \ and\ \bibinfo {author}
      {\bibfnamefont {S.-J.}\ \bibnamefont {Huang}},\ }\href {\doibase
      10.1093/mnras/stac365} {\bibfield  {journal} {\bibinfo  {journal} {Mon. Not.
      Roy. Astron. Soc.}\ }\textbf {\bibinfo {volume} {512}},\ \bibinfo {pages} {1}
      (\bibinfo {year} {2022})},\ \Eprint {http://arxiv.org/abs/2102.10295}
      {arXiv:2102.10295 [astro-ph.CO]} \BibitemShut {NoStop}%
    \bibitem [{\citenamefont {Gil~Choi}\ \emph {et~al.}(2021)\citenamefont
      {Gil~Choi}, \citenamefont {Park},\ and\ \citenamefont {Jung}}]{Choi:2021jqn}%
      \BibitemOpen
      \bibfield  {author} {\bibinfo {author} {\bibfnamefont {H.}~\bibnamefont
      {Gil~Choi}}, \bibinfo {author} {\bibfnamefont {C.}~\bibnamefont {Park}}, \
      and\ \bibinfo {author} {\bibfnamefont {S.}~\bibnamefont {Jung}},\ }\href@noop
      {} {\  (\bibinfo {year} {2021})},\ \Eprint {http://arxiv.org/abs/2103.08618}
      {arXiv:2103.08618 [astro-ph.CO]} \BibitemShut {NoStop}%
    \bibitem [{\citenamefont {Ulmer}\ and\ \citenamefont
      {Goodman}(1995)}]{Ulmer:1994ij}%
      \BibitemOpen
      \bibfield  {author} {\bibinfo {author} {\bibfnamefont {A.}~\bibnamefont
      {Ulmer}}\ and\ \bibinfo {author} {\bibfnamefont {J.}~\bibnamefont
      {Goodman}},\ }\href {\doibase 10.1086/175422} {\bibfield  {journal} {\bibinfo
       {journal} {Astrophys. J.}\ }\textbf {\bibinfo {volume} {442}},\ \bibinfo
      {pages} {67} (\bibinfo {year} {1995})},\ \Eprint
      {http://arxiv.org/abs/astro-ph/9406042} {arXiv:astro-ph/9406042} \BibitemShut
      {NoStop}%
    \bibitem [{\citenamefont {Tambalo}\ \emph
      {et~al.}(2022{\natexlab{b}})\citenamefont {Tambalo}, \citenamefont
      {Zumalac\'arregui}, \citenamefont {Dai},\ and\ \citenamefont
      {Cheung}}]{Tambalo:2022plm}%
      \BibitemOpen
      \bibfield  {author} {\bibinfo {author} {\bibfnamefont {G.}~\bibnamefont
      {Tambalo}}, \bibinfo {author} {\bibfnamefont {M.}~\bibnamefont
      {Zumalac\'arregui}}, \bibinfo {author} {\bibfnamefont {L.}~\bibnamefont
      {Dai}}, \ and\ \bibinfo {author} {\bibfnamefont {M.~H.-Y.}\ \bibnamefont
      {Cheung}},\ }\href@noop {} {\  (\bibinfo {year} {2022}{\natexlab{b}})},\
      \Eprint {http://arxiv.org/abs/2210.05658} {arXiv:2210.05658 [gr-qc]}
      \BibitemShut {NoStop}%
    \bibitem [{\citenamefont {Takahashi}(2004)}]{Takahashi:2004mc}%
      \BibitemOpen
      \bibfield  {author} {\bibinfo {author} {\bibfnamefont {R.}~\bibnamefont
      {Takahashi}},\ }\href {\doibase 10.1051/0004-6361:20040212} {\bibfield
      {journal} {\bibinfo  {journal} {Astron. Astrophys.}\ }\textbf {\bibinfo
      {volume} {423}},\ \bibinfo {pages} {787} (\bibinfo {year} {2004})},\ \Eprint
      {http://arxiv.org/abs/astro-ph/0402165} {arXiv:astro-ph/0402165} \BibitemShut
      {NoStop}%
    \bibitem [{\citenamefont {Husa}\ \emph {et~al.}(2016)\citenamefont {Husa},
      \citenamefont {Khan}, \citenamefont {Hannam}, \citenamefont {P\"urrer},
      \citenamefont {Ohme}, \citenamefont {Jim\'enez~Forteza},\ and\ \citenamefont
      {Boh\'e}}]{Husa:2015iqa}%
      \BibitemOpen
      \bibfield  {author} {\bibinfo {author} {\bibfnamefont {S.}~\bibnamefont
      {Husa}}, \bibinfo {author} {\bibfnamefont {S.}~\bibnamefont {Khan}}, \bibinfo
      {author} {\bibfnamefont {M.}~\bibnamefont {Hannam}}, \bibinfo {author}
      {\bibfnamefont {M.}~\bibnamefont {P\"urrer}}, \bibinfo {author}
      {\bibfnamefont {F.}~\bibnamefont {Ohme}}, \bibinfo {author} {\bibfnamefont
      {X.}~\bibnamefont {Jim\'enez~Forteza}}, \ and\ \bibinfo {author}
      {\bibfnamefont {A.}~\bibnamefont {Boh\'e}},\ }\href {\doibase
      10.1103/PhysRevD.93.044006} {\bibfield  {journal} {\bibinfo  {journal} {Phys.
      Rev. D}\ }\textbf {\bibinfo {volume} {93}},\ \bibinfo {pages} {044006}
      (\bibinfo {year} {2016})},\ \Eprint {http://arxiv.org/abs/1508.07250}
      {arXiv:1508.07250 [gr-qc]} \BibitemShut {NoStop}%
    \bibitem [{\citenamefont {Nitz}\ \emph {et~al.}(2023)\citenamefont {Nitz},
      \citenamefont {Harry}, \citenamefont {Brown}, \citenamefont {Biwer},
      \citenamefont {Willis}, \citenamefont {Canton}, \citenamefont {Capano},
      \citenamefont {Dent}, \citenamefont {Pekowsky}, \citenamefont {De},
      \citenamefont {Cabero}, \citenamefont {Davies}, \citenamefont {Williamson},
      \citenamefont {Macleod}, \citenamefont {Machenschalk}, \citenamefont
      {Pannarale}, \citenamefont {Kumar}, \citenamefont {Reyes}, \citenamefont
      {dfinstad}, \citenamefont {Kumar}, \citenamefont {Wu}, \citenamefont
      {Tápai}, \citenamefont {Singer}, \citenamefont {veronica villa},
      \citenamefont {Khan}, \citenamefont {Fairhurst}, \citenamefont {Chandra},
      \citenamefont {Nielsen}, \citenamefont {Singh},\ and\ \citenamefont
      {Massinger}}]{alex_nitz_2023_7885796}%
      \BibitemOpen
      \bibfield  {author} {\bibinfo {author} {\bibfnamefont {A.}~\bibnamefont
      {Nitz}}, \bibinfo {author} {\bibfnamefont {I.}~\bibnamefont {Harry}},
      \bibinfo {author} {\bibfnamefont {D.}~\bibnamefont {Brown}}, \bibinfo
      {author} {\bibfnamefont {C.~M.}\ \bibnamefont {Biwer}}, \bibinfo {author}
      {\bibfnamefont {J.}~\bibnamefont {Willis}}, \bibinfo {author} {\bibfnamefont
      {T.~D.}\ \bibnamefont {Canton}}, \bibinfo {author} {\bibfnamefont
      {C.}~\bibnamefont {Capano}}, \bibinfo {author} {\bibfnamefont
      {T.}~\bibnamefont {Dent}}, \bibinfo {author} {\bibfnamefont {L.}~\bibnamefont
      {Pekowsky}}, \bibinfo {author} {\bibfnamefont {S.}~\bibnamefont {De}},
      \bibinfo {author} {\bibfnamefont {M.}~\bibnamefont {Cabero}}, \bibinfo
      {author} {\bibfnamefont {G.~S.~C.}\ \bibnamefont {Davies}}, \bibinfo {author}
      {\bibfnamefont {A.~R.}\ \bibnamefont {Williamson}}, \bibinfo {author}
      {\bibfnamefont {D.}~\bibnamefont {Macleod}}, \bibinfo {author} {\bibfnamefont
      {B.}~\bibnamefont {Machenschalk}}, \bibinfo {author} {\bibfnamefont
      {F.}~\bibnamefont {Pannarale}}, \bibinfo {author} {\bibfnamefont
      {P.}~\bibnamefont {Kumar}}, \bibinfo {author} {\bibfnamefont
      {S.}~\bibnamefont {Reyes}}, \bibinfo {author} {\bibnamefont {dfinstad}},
      \bibinfo {author} {\bibfnamefont {S.}~\bibnamefont {Kumar}}, \bibinfo
      {author} {\bibfnamefont {S.}~\bibnamefont {Wu}}, \bibinfo {author}
      {\bibfnamefont {M.}~\bibnamefont {Tápai}}, \bibinfo {author} {\bibfnamefont
      {L.}~\bibnamefont {Singer}}, \bibinfo {author} {\bibnamefont {veronica
      villa}}, \bibinfo {author} {\bibfnamefont {S.}~\bibnamefont {Khan}}, \bibinfo
      {author} {\bibfnamefont {S.}~\bibnamefont {Fairhurst}}, \bibinfo {author}
      {\bibfnamefont {K.}~\bibnamefont {Chandra}}, \bibinfo {author} {\bibfnamefont
      {A.}~\bibnamefont {Nielsen}}, \bibinfo {author} {\bibfnamefont
      {S.}~\bibnamefont {Singh}}, \ and\ \bibinfo {author} {\bibfnamefont
      {T.}~\bibnamefont {Massinger}},\ }\href {\doibase 10.5281/zenodo.7885796}
      {\enquote {\bibinfo {title} {gwastro/pycbc: v2.1.2 release of pycbc},}\ }
      (\bibinfo {year} {2023})\BibitemShut {NoStop}%
    \bibitem [{\citenamefont {Amaro-Seoane}\ \emph {et~al.}(2017)\citenamefont
      {Amaro-Seoane} \emph {et~al.}}]{LISA:2017pwj}%
      \BibitemOpen
      \bibfield  {author} {\bibinfo {author} {\bibfnamefont {P.}~\bibnamefont
      {Amaro-Seoane}} \emph {et~al.} (\bibinfo {collaboration} {LISA}),\
      }\href@noop {} {\  (\bibinfo {year} {2017})},\ \Eprint
      {http://arxiv.org/abs/1702.00786} {arXiv:1702.00786 [astro-ph.IM]}
      \BibitemShut {NoStop}%
    \bibitem [{\citenamefont {Auclair}\ \emph {et~al.}(2022)\citenamefont {Auclair}
      \emph {et~al.}}]{LISACosmologyWorkingGroup:2022jok}%
      \BibitemOpen
      \bibfield  {author} {\bibinfo {author} {\bibfnamefont {P.}~\bibnamefont
      {Auclair}} \emph {et~al.} (\bibinfo {collaboration} {LISA Cosmology Working
      Group}),\ }\href@noop {} {\  (\bibinfo {year} {2022})},\ \Eprint
      {http://arxiv.org/abs/2204.05434} {arXiv:2204.05434 [astro-ph.CO]}
      \BibitemShut {NoStop}%
    \bibitem [{\citenamefont {Maggiore}\ \emph {et~al.}(2020)\citenamefont
      {Maggiore} \emph {et~al.}}]{Maggiore:2019uih}%
      \BibitemOpen
      \bibfield  {author} {\bibinfo {author} {\bibfnamefont {M.}~\bibnamefont
      {Maggiore}} \emph {et~al.},\ }\href {\doibase 10.1088/1475-7516/2020/03/050}
      {\bibfield  {journal} {\bibinfo  {journal} {JCAP}\ }\textbf {\bibinfo
      {volume} {03}},\ \bibinfo {pages} {050} (\bibinfo {year} {2020})},\ \Eprint
      {http://arxiv.org/abs/1912.02622} {arXiv:1912.02622 [astro-ph.CO]}
      \BibitemShut {NoStop}%
    \bibitem [{\citenamefont {Kalogera}\ \emph {et~al.}(2021)\citenamefont
      {Kalogera} \emph {et~al.}}]{Kalogera:2021bya}%
      \BibitemOpen
      \bibfield  {author} {\bibinfo {author} {\bibfnamefont {V.}~\bibnamefont
      {Kalogera}} \emph {et~al.},\ }\href@noop {} {\  (\bibinfo {year} {2021})},\
      \Eprint {http://arxiv.org/abs/2111.06990} {arXiv:2111.06990 [gr-qc]}
      \BibitemShut {NoStop}%
    \bibitem [{\citenamefont {Robson}\ \emph {et~al.}(2019)\citenamefont {Robson},
      \citenamefont {Cornish},\ and\ \citenamefont {Liu}}]{Robson:2018ifk}%
      \BibitemOpen
      \bibfield  {author} {\bibinfo {author} {\bibfnamefont {T.}~\bibnamefont
      {Robson}}, \bibinfo {author} {\bibfnamefont {N.~J.}\ \bibnamefont {Cornish}},
      \ and\ \bibinfo {author} {\bibfnamefont {C.}~\bibnamefont {Liu}},\ }\href
      {\doibase 10.1088/1361-6382/ab1101} {\bibfield  {journal} {\bibinfo
      {journal} {Class. Quant. Grav.}\ }\textbf {\bibinfo {volume} {36}},\ \bibinfo
      {pages} {105011} (\bibinfo {year} {2019})},\ \Eprint
      {http://arxiv.org/abs/1803.01944} {arXiv:1803.01944 [astro-ph.HE]}
      \BibitemShut {NoStop}%
    \bibitem [{\citenamefont {Marsat}\ \emph {et~al.}(2021)\citenamefont {Marsat},
      \citenamefont {Baker},\ and\ \citenamefont {Dal~Canton}}]{Marsat:2020rtl}%
      \BibitemOpen
      \bibfield  {author} {\bibinfo {author} {\bibfnamefont {S.}~\bibnamefont
      {Marsat}}, \bibinfo {author} {\bibfnamefont {J.~G.}\ \bibnamefont {Baker}}, \
      and\ \bibinfo {author} {\bibfnamefont {T.}~\bibnamefont {Dal~Canton}},\
      }\href {\doibase 10.1103/PhysRevD.103.083011} {\bibfield  {journal} {\bibinfo
       {journal} {Phys. Rev. D}\ }\textbf {\bibinfo {volume} {103}},\ \bibinfo
      {pages} {083011} (\bibinfo {year} {2021})},\ \Eprint
      {http://arxiv.org/abs/2003.00357} {arXiv:2003.00357 [gr-qc]} \BibitemShut
      {NoStop}%
    \bibitem [{\citenamefont {Lindblom}\ \emph {et~al.}(2008)\citenamefont
      {Lindblom}, \citenamefont {Owen},\ and\ \citenamefont {Brown}}]{Lindblom}%
      \BibitemOpen
      \bibfield  {author} {\bibinfo {author} {\bibfnamefont {L.}~\bibnamefont
      {Lindblom}}, \bibinfo {author} {\bibfnamefont {B.~J.}\ \bibnamefont {Owen}},
      \ and\ \bibinfo {author} {\bibfnamefont {D.~A.}\ \bibnamefont {Brown}},\
      }\href {\doibase 10.1103/PhysRevD.78.124020} {\bibfield  {journal} {\bibinfo
      {journal} {Phys. Rev. D}\ }\textbf {\bibinfo {volume} {78}},\ \bibinfo
      {pages} {124020} (\bibinfo {year} {2008})},\ \Eprint
      {http://arxiv.org/abs/0809.3844} {arXiv:0809.3844 [gr-qc]} \BibitemShut
      {NoStop}%
    \bibitem [{\citenamefont {Vallisneri}(2008)}]{Vallisneri:2007ev}%
      \BibitemOpen
      \bibfield  {author} {\bibinfo {author} {\bibfnamefont {M.}~\bibnamefont
      {Vallisneri}},\ }\href {\doibase 10.1103/PhysRevD.77.042001} {\bibfield
      {journal} {\bibinfo  {journal} {Phys. Rev. D}\ }\textbf {\bibinfo {volume}
      {77}},\ \bibinfo {pages} {042001} (\bibinfo {year} {2008})},\ \Eprint
      {http://arxiv.org/abs/gr-qc/0703086} {arXiv:gr-qc/0703086} \BibitemShut
      {NoStop}%
    \bibitem [{\citenamefont {Maggiore}(2007)}]{maggiore2007gravitational}%
      \BibitemOpen
      \bibfield  {author} {\bibinfo {author} {\bibfnamefont {M.}~\bibnamefont
      {Maggiore}},\ }\href@noop {} {\emph {\bibinfo {title} {Gravitational waves:
      Volume 1: Theory and experiments}}}\ (\bibinfo  {publisher} {OUP Oxford},\
      \bibinfo {year} {2007})\BibitemShut {NoStop}%
    \bibitem [{\citenamefont {Pompili}\ \emph {et~al.}(2023)\citenamefont {Pompili}
      \emph {et~al.}}]{Pompili:2023tna}%
      \BibitemOpen
      \bibfield  {author} {\bibinfo {author} {\bibfnamefont {L.}~\bibnamefont
      {Pompili}} \emph {et~al.},\ }\href@noop {} {\  (\bibinfo {year} {2023})},\
      \Eprint {http://arxiv.org/abs/2303.18039} {arXiv:2303.18039 [gr-qc]}
      \BibitemShut {NoStop}%
    \bibitem [{\citenamefont {\c{C}al\i{}\c{s}kan}\ \emph
      {et~al.}(2023)\citenamefont {\c{C}al\i{}\c{s}kan}, \citenamefont
      {Anil~Kumar}, \citenamefont {Ji}, \citenamefont {Ezquiaga}, \citenamefont
      {Cotesta}, \citenamefont {Berti},\ and\ \citenamefont
      {Kamionkowski}}]{Caliskan:2023zqm}%
      \BibitemOpen
      \bibfield  {author} {\bibinfo {author} {\bibfnamefont {M.}~\bibnamefont
      {\c{C}al\i{}\c{s}kan}}, \bibinfo {author} {\bibfnamefont {N.}~\bibnamefont
      {Anil~Kumar}}, \bibinfo {author} {\bibfnamefont {L.}~\bibnamefont {Ji}},
      \bibinfo {author} {\bibfnamefont {J.~M.}\ \bibnamefont {Ezquiaga}}, \bibinfo
      {author} {\bibfnamefont {R.}~\bibnamefont {Cotesta}}, \bibinfo {author}
      {\bibfnamefont {E.}~\bibnamefont {Berti}}, \ and\ \bibinfo {author}
      {\bibfnamefont {M.}~\bibnamefont {Kamionkowski}},\ }\href@noop {} {\
      (\bibinfo {year} {2023})},\ \Eprint {http://arxiv.org/abs/2307.06990}
      {arXiv:2307.06990 [astro-ph.CO]} \BibitemShut {NoStop}%
    \bibitem [{\citenamefont {Tinker}\ \emph {et~al.}(2008)\citenamefont {Tinker},
      \citenamefont {Kravtsov}, \citenamefont {Klypin}, \citenamefont {Abazajian},
      \citenamefont {Warren}, \citenamefont {Yepes}, \citenamefont {Gottlober},\
      and\ \citenamefont {Holz}}]{Tinker:2008ff}%
      \BibitemOpen
      \bibfield  {author} {\bibinfo {author} {\bibfnamefont {J.~L.}\ \bibnamefont
      {Tinker}}, \bibinfo {author} {\bibfnamefont {A.~V.}\ \bibnamefont
      {Kravtsov}}, \bibinfo {author} {\bibfnamefont {A.}~\bibnamefont {Klypin}},
      \bibinfo {author} {\bibfnamefont {K.}~\bibnamefont {Abazajian}}, \bibinfo
      {author} {\bibfnamefont {M.~S.}\ \bibnamefont {Warren}}, \bibinfo {author}
      {\bibfnamefont {G.}~\bibnamefont {Yepes}}, \bibinfo {author} {\bibfnamefont
      {S.}~\bibnamefont {Gottlober}}, \ and\ \bibinfo {author} {\bibfnamefont
      {D.~E.}\ \bibnamefont {Holz}},\ }\href {\doibase 10.1086/591439} {\bibfield
      {journal} {\bibinfo  {journal} {Astrophys. J.}\ }\textbf {\bibinfo {volume}
      {688}},\ \bibinfo {pages} {709} (\bibinfo {year} {2008})},\ \Eprint
      {http://arxiv.org/abs/0803.2706} {arXiv:0803.2706 [astro-ph]} \BibitemShut
      {NoStop}%
    \bibitem [{\citenamefont {Iacovelli}\ \emph {et~al.}(2022)\citenamefont
      {Iacovelli}, \citenamefont {Mancarella}, \citenamefont {Foffa},\ and\
      \citenamefont {Maggiore}}]{Iacovelli:2022bbs}%
      \BibitemOpen
      \bibfield  {author} {\bibinfo {author} {\bibfnamefont {F.}~\bibnamefont
      {Iacovelli}}, \bibinfo {author} {\bibfnamefont {M.}~\bibnamefont
      {Mancarella}}, \bibinfo {author} {\bibfnamefont {S.}~\bibnamefont {Foffa}}, \
      and\ \bibinfo {author} {\bibfnamefont {M.}~\bibnamefont {Maggiore}},\ }\href
      {\doibase 10.3847/1538-4357/ac9cd4} {\bibfield  {journal} {\bibinfo
      {journal} {Astrophys. J.}\ }\textbf {\bibinfo {volume} {941}},\ \bibinfo
      {pages} {208} (\bibinfo {year} {2022})},\ \Eprint
      {http://arxiv.org/abs/2207.02771} {arXiv:2207.02771 [gr-qc]} \BibitemShut
      {NoStop}%
    \bibitem [{\citenamefont {Borhanian}\ and\ \citenamefont
      {Sathyaprakash}(2022)}]{Borhanian:2022czq}%
      \BibitemOpen
      \bibfield  {author} {\bibinfo {author} {\bibfnamefont {S.}~\bibnamefont
      {Borhanian}}\ and\ \bibinfo {author} {\bibfnamefont {B.~S.}\ \bibnamefont
      {Sathyaprakash}},\ }\href@noop {} {\  (\bibinfo {year} {2022})},\ \Eprint
      {http://arxiv.org/abs/2202.11048} {arXiv:2202.11048 [gr-qc]} \BibitemShut
      {NoStop}%
    \bibitem [{\citenamefont {Fairbairn}\ \emph {et~al.}(2022)\citenamefont
      {Fairbairn}, \citenamefont {Urrutia},\ and\ \citenamefont
      {Vaskonen}}]{Fairbairn:2022xln}%
      \BibitemOpen
      \bibfield  {author} {\bibinfo {author} {\bibfnamefont {M.}~\bibnamefont
      {Fairbairn}}, \bibinfo {author} {\bibfnamefont {J.}~\bibnamefont {Urrutia}},
      \ and\ \bibinfo {author} {\bibfnamefont {V.}~\bibnamefont {Vaskonen}},\
      }\href@noop {} {\  (\bibinfo {year} {2022})},\ \Eprint
      {http://arxiv.org/abs/2210.13436} {arXiv:2210.13436 [astro-ph.CO]}
      \BibitemShut {NoStop}%
    \bibitem [{\citenamefont {Guo}\ and\ \citenamefont {Lu}(2022)}]{Guo:2022dre}%
      \BibitemOpen
      \bibfield  {author} {\bibinfo {author} {\bibfnamefont {X.}~\bibnamefont
      {Guo}}\ and\ \bibinfo {author} {\bibfnamefont {Y.}~\bibnamefont {Lu}},\
      }\href {\doibase 10.1103/PhysRevD.106.023018} {\bibfield  {journal} {\bibinfo
       {journal} {Phys. Rev. D}\ }\textbf {\bibinfo {volume} {106}},\ \bibinfo
      {pages} {023018} (\bibinfo {year} {2022})},\ \Eprint
      {http://arxiv.org/abs/2207.00325} {arXiv:2207.00325 [astro-ph.CO]}
      \BibitemShut {NoStop}%
    \bibitem [{\citenamefont {Steinle}\ \emph {et~al.}(2023)\citenamefont
      {Steinle}, \citenamefont {Middleton}, \citenamefont {Moore}, \citenamefont
      {Chen}, \citenamefont {Klein}, \citenamefont {Pratten}, \citenamefont
      {Buscicchio}, \citenamefont {Finch},\ and\ \citenamefont
      {Vecchio}}]{Steinle:2023vxs}%
      \BibitemOpen
      \bibfield  {author} {\bibinfo {author} {\bibfnamefont {N.}~\bibnamefont
      {Steinle}}, \bibinfo {author} {\bibfnamefont {H.}~\bibnamefont {Middleton}},
      \bibinfo {author} {\bibfnamefont {C.~J.}\ \bibnamefont {Moore}}, \bibinfo
      {author} {\bibfnamefont {S.}~\bibnamefont {Chen}}, \bibinfo {author}
      {\bibfnamefont {A.}~\bibnamefont {Klein}}, \bibinfo {author} {\bibfnamefont
      {G.}~\bibnamefont {Pratten}}, \bibinfo {author} {\bibfnamefont
      {R.}~\bibnamefont {Buscicchio}}, \bibinfo {author} {\bibfnamefont
      {E.}~\bibnamefont {Finch}}, \ and\ \bibinfo {author} {\bibfnamefont
      {A.}~\bibnamefont {Vecchio}},\ }\href@noop {} {\  (\bibinfo {year} {2023})},\
      \Eprint {http://arxiv.org/abs/2305.05955} {arXiv:2305.05955 [astro-ph.HE]}
      \BibitemShut {NoStop}%
    \bibitem [{\citenamefont {Middleton}\ \emph {et~al.}(2016)\citenamefont
      {Middleton}, \citenamefont {Del~Pozzo}, \citenamefont {Farr}, \citenamefont
      {Sesana},\ and\ \citenamefont {Vecchio}}]{Middleton:2015oda}%
      \BibitemOpen
      \bibfield  {author} {\bibinfo {author} {\bibfnamefont {H.}~\bibnamefont
      {Middleton}}, \bibinfo {author} {\bibfnamefont {W.}~\bibnamefont
      {Del~Pozzo}}, \bibinfo {author} {\bibfnamefont {W.~M.}\ \bibnamefont {Farr}},
      \bibinfo {author} {\bibfnamefont {A.}~\bibnamefont {Sesana}}, \ and\ \bibinfo
      {author} {\bibfnamefont {A.}~\bibnamefont {Vecchio}},\ }\href {\doibase
      10.1093/mnrasl/slv150} {\bibfield  {journal} {\bibinfo  {journal} {Mon. Not.
      Roy. Astron. Soc.}\ }\textbf {\bibinfo {volume} {455}},\ \bibinfo {pages}
      {L72} (\bibinfo {year} {2016})},\ \Eprint {http://arxiv.org/abs/1507.00992}
      {arXiv:1507.00992 [astro-ph.CO]} \BibitemShut {NoStop}%
    \bibitem [{\citenamefont {Chen}\ \emph {et~al.}(2019)\citenamefont {Chen},
      \citenamefont {Sesana},\ and\ \citenamefont {Conselice}}]{Chen:2018znx}%
      \BibitemOpen
      \bibfield  {author} {\bibinfo {author} {\bibfnamefont {S.}~\bibnamefont
      {Chen}}, \bibinfo {author} {\bibfnamefont {A.}~\bibnamefont {Sesana}}, \ and\
      \bibinfo {author} {\bibfnamefont {C.~J.}\ \bibnamefont {Conselice}},\ }\href
      {\doibase 10.1093/mnras/stz1722} {\bibfield  {journal} {\bibinfo  {journal}
      {Mon. Not. Roy. Astron. Soc.}\ }\textbf {\bibinfo {volume} {488}},\ \bibinfo
      {pages} {401} (\bibinfo {year} {2019})},\ \Eprint
      {http://arxiv.org/abs/1810.04184} {arXiv:1810.04184 [astro-ph.GA]}
      \BibitemShut {NoStop}%
    \bibitem [{\citenamefont {Dai}\ and\ \citenamefont
      {Miralda-Escud\'e}(2020)}]{Dai:2019lud}%
      \BibitemOpen
      \bibfield  {author} {\bibinfo {author} {\bibfnamefont {L.}~\bibnamefont
      {Dai}}\ and\ \bibinfo {author} {\bibfnamefont {J.}~\bibnamefont
      {Miralda-Escud\'e}},\ }\href {\doibase 10.3847/1538-3881/ab5e83} {\bibfield
      {journal} {\bibinfo  {journal} {Astron. J.}\ }\textbf {\bibinfo {volume}
      {159}},\ \bibinfo {pages} {49} (\bibinfo {year} {2020})},\ \Eprint
      {http://arxiv.org/abs/1908.01773} {arXiv:1908.01773 [astro-ph.CO]}
      \BibitemShut {NoStop}%
    \bibitem [{\citenamefont {Zavala}\ and\ \citenamefont
      {Frenk}(2019)}]{Zavala:2019gpq}%
      \BibitemOpen
      \bibfield  {author} {\bibinfo {author} {\bibfnamefont {J.}~\bibnamefont
      {Zavala}}\ and\ \bibinfo {author} {\bibfnamefont {C.~S.}\ \bibnamefont
      {Frenk}},\ }\href {\doibase 10.3390/galaxies7040081} {\bibfield  {journal}
      {\bibinfo  {journal} {Galaxies}\ }\textbf {\bibinfo {volume} {7}},\ \bibinfo
      {pages} {81} (\bibinfo {year} {2019})},\ \Eprint
      {http://arxiv.org/abs/1907.11775} {arXiv:1907.11775 [astro-ph.CO]}
      \BibitemShut {NoStop}%
    \bibitem [{\citenamefont {Giocoli}\ \emph {et~al.}(2008)\citenamefont
      {Giocoli}, \citenamefont {Tormen},\ and\ \citenamefont
      {Bosch}}]{Giocoli:2007uv}%
      \BibitemOpen
      \bibfield  {author} {\bibinfo {author} {\bibfnamefont {C.}~\bibnamefont
      {Giocoli}}, \bibinfo {author} {\bibfnamefont {G.}~\bibnamefont {Tormen}}, \
      and\ \bibinfo {author} {\bibfnamefont {F.~C. v.~d.}\ \bibnamefont {Bosch}},\
      }\href {\doibase 10.1111/j.1365-2966.2008.13182.x} {\bibfield  {journal}
      {\bibinfo  {journal} {Mon. Not. Roy. Astron. Soc.}\ }\textbf {\bibinfo
      {volume} {386}},\ \bibinfo {pages} {2135} (\bibinfo {year} {2008})},\ \Eprint
      {http://arxiv.org/abs/0712.1563} {arXiv:0712.1563 [astro-ph]} \BibitemShut
      {NoStop}%
    \bibitem [{\citenamefont {Jiang}\ and\ \citenamefont {van~den
      Bosch}(2016)}]{Jiang:2014nsa}%
      \BibitemOpen
      \bibfield  {author} {\bibinfo {author} {\bibfnamefont {F.}~\bibnamefont
      {Jiang}}\ and\ \bibinfo {author} {\bibfnamefont {F.~C.}\ \bibnamefont
      {van~den Bosch}},\ }\href {\doibase 10.1093/mnras/stw439} {\bibfield
      {journal} {\bibinfo  {journal} {Mon. Not. Roy. Astron. Soc.}\ }\textbf
      {\bibinfo {volume} {458}},\ \bibinfo {pages} {2848} (\bibinfo {year}
      {2016})},\ \Eprint {http://arxiv.org/abs/1403.6827} {arXiv:1403.6827
      [astro-ph.CO]} \BibitemShut {NoStop}%
    \bibitem [{\citenamefont {Holz}\ and\ \citenamefont
      {Hughes}(2005)}]{Holz:2005df}%
      \BibitemOpen
      \bibfield  {author} {\bibinfo {author} {\bibfnamefont {D.~E.}\ \bibnamefont
      {Holz}}\ and\ \bibinfo {author} {\bibfnamefont {S.~A.}\ \bibnamefont
      {Hughes}},\ }\href {\doibase 10.1086/431341} {\bibfield  {journal} {\bibinfo
      {journal} {Astrophys. J.}\ }\textbf {\bibinfo {volume} {629}},\ \bibinfo
      {pages} {15} (\bibinfo {year} {2005})},\ \Eprint
      {http://arxiv.org/abs/astro-ph/0504616} {arXiv:astro-ph/0504616} \BibitemShut
      {NoStop}%
    \bibitem [{\citenamefont {Shapiro}\ \emph {et~al.}(2010)\citenamefont
      {Shapiro}, \citenamefont {Bacon}, \citenamefont {Hendry},\ and\ \citenamefont
      {Hoyle}}]{Shapiro:2009sr}%
      \BibitemOpen
      \bibfield  {author} {\bibinfo {author} {\bibfnamefont {C.}~\bibnamefont
      {Shapiro}}, \bibinfo {author} {\bibfnamefont {D.}~\bibnamefont {Bacon}},
      \bibinfo {author} {\bibfnamefont {M.}~\bibnamefont {Hendry}}, \ and\ \bibinfo
      {author} {\bibfnamefont {B.}~\bibnamefont {Hoyle}},\ }\href {\doibase
      10.1111/j.1365-2966.2010.16317.x} {\bibfield  {journal} {\bibinfo  {journal}
      {Mon. Not. Roy. Astron. Soc.}\ }\textbf {\bibinfo {volume} {404}},\ \bibinfo
      {pages} {858} (\bibinfo {year} {2010})},\ \Eprint
      {http://arxiv.org/abs/0907.3635} {arXiv:0907.3635 [astro-ph.CO]} \BibitemShut
      {NoStop}%
    \bibitem [{\citenamefont {Wu}\ \emph {et~al.}(2022)\citenamefont {Wu},
      \citenamefont {Chan}, \citenamefont {Hendry},\ and\ \citenamefont
      {Hannuksela}}]{Wu:2022vrq}%
      \BibitemOpen
      \bibfield  {author} {\bibinfo {author} {\bibfnamefont {Z.-F.}\ \bibnamefont
      {Wu}}, \bibinfo {author} {\bibfnamefont {L.~W.~L.}\ \bibnamefont {Chan}},
      \bibinfo {author} {\bibfnamefont {M.}~\bibnamefont {Hendry}}, \ and\ \bibinfo
      {author} {\bibfnamefont {O.~A.}\ \bibnamefont {Hannuksela}},\ }\href
      {\doibase 10.1093/mnras/stad1194} {\  (\bibinfo {year} {2022}),\
      10.1093/mnras/stad1194},\ \Eprint {http://arxiv.org/abs/2211.15160}
      {arXiv:2211.15160 [astro-ph.CO]} \BibitemShut {NoStop}%
    \bibitem [{\citenamefont {Hotinli}\ \emph {et~al.}(2022)\citenamefont
      {Hotinli}, \citenamefont {Meyers}, \citenamefont {Trendafilova},
      \citenamefont {Green},\ and\ \citenamefont {van Engelen}}]{Hotinli:2021umk}%
      \BibitemOpen
      \bibfield  {author} {\bibinfo {author} {\bibfnamefont {S.~C.}\ \bibnamefont
      {Hotinli}}, \bibinfo {author} {\bibfnamefont {J.}~\bibnamefont {Meyers}},
      \bibinfo {author} {\bibfnamefont {C.}~\bibnamefont {Trendafilova}}, \bibinfo
      {author} {\bibfnamefont {D.}~\bibnamefont {Green}}, \ and\ \bibinfo {author}
      {\bibfnamefont {A.}~\bibnamefont {van Engelen}},\ }\href {\doibase
      10.1088/1475-7516/2022/04/020} {\bibfield  {journal} {\bibinfo  {journal}
      {JCAP}\ }\textbf {\bibinfo {volume} {04}},\ \bibinfo {pages} {020} (\bibinfo
      {year} {2022})},\ \Eprint {http://arxiv.org/abs/2111.15036} {arXiv:2111.15036
      [astro-ph.CO]} \BibitemShut {NoStop}%
    \bibitem [{\citenamefont {Sousbie}\ \emph {et~al.}(2006)\citenamefont
      {Sousbie}, \citenamefont {Pichon}, \citenamefont {Courtois}, \citenamefont
      {Colombi},\ and\ \citenamefont {Novikov}}]{Sousbie:2006tg}%
      \BibitemOpen
      \bibfield  {author} {\bibinfo {author} {\bibfnamefont {T.}~\bibnamefont
      {Sousbie}}, \bibinfo {author} {\bibfnamefont {C.}~\bibnamefont {Pichon}},
      \bibinfo {author} {\bibfnamefont {H.~M.}\ \bibnamefont {Courtois}}, \bibinfo
      {author} {\bibfnamefont {S.}~\bibnamefont {Colombi}}, \ and\ \bibinfo
      {author} {\bibfnamefont {D.}~\bibnamefont {Novikov}},\ }\href@noop {} {\
      (\bibinfo {year} {2006})},\ \Eprint {http://arxiv.org/abs/astro-ph/0602628}
      {arXiv:astro-ph/0602628} \BibitemShut {NoStop}%
    \bibitem [{\citenamefont {Hahn}\ \emph {et~al.}(2007)\citenamefont {Hahn},
      \citenamefont {Porciani}, \citenamefont {Carollo},\ and\ \citenamefont
      {Dekel}}]{Hahn:2006mk}%
      \BibitemOpen
      \bibfield  {author} {\bibinfo {author} {\bibfnamefont {O.}~\bibnamefont
      {Hahn}}, \bibinfo {author} {\bibfnamefont {C.}~\bibnamefont {Porciani}},
      \bibinfo {author} {\bibfnamefont {C.~M.}\ \bibnamefont {Carollo}}, \ and\
      \bibinfo {author} {\bibfnamefont {A.}~\bibnamefont {Dekel}},\ }\href
      {\doibase 10.1111/j.1365-2966.2006.11318.x} {\bibfield  {journal} {\bibinfo
      {journal} {Mon. Not. Roy. Astron. Soc.}\ }\textbf {\bibinfo {volume} {375}},\
      \bibinfo {pages} {489} (\bibinfo {year} {2007})},\ \Eprint
      {http://arxiv.org/abs/astro-ph/0610280} {arXiv:astro-ph/0610280} \BibitemShut
      {NoStop}%
    \bibitem [{\citenamefont {{Coil}}(2013)}]{Coil:2013}%
      \BibitemOpen
      \bibfield  {author} {\bibinfo {author} {\bibfnamefont {A.~L.}\ \bibnamefont
      {{Coil}}},\ }in\ \href {\doibase 10.1007/978-94-007-5609-0_8} {\emph
      {\bibinfo {booktitle} {Planets, Stars and Stellar Systems. Volume 6:
      Extragalactic Astronomy and Cosmology}}},\ Vol.~\bibinfo {volume} {6},\
      \bibinfo {editor} {edited by\ \bibinfo {editor} {\bibfnamefont {T.~D.}\
      \bibnamefont {{Oswalt}}}\ and\ \bibinfo {editor} {\bibfnamefont {W.~C.}\
      \bibnamefont {{Keel}}}}\ (\bibinfo {year} {2013})\ p.\ \bibinfo {pages}
      {387}\BibitemShut {NoStop}%
    \bibitem [{\citenamefont {Bullock}\ and\ \citenamefont
      {Boylan-Kolchin}(2017)}]{Bullock:2017xww}%
      \BibitemOpen
      \bibfield  {author} {\bibinfo {author} {\bibfnamefont {J.~S.}\ \bibnamefont
      {Bullock}}\ and\ \bibinfo {author} {\bibfnamefont {M.}~\bibnamefont
      {Boylan-Kolchin}},\ }\href {\doibase 10.1146/annurev-astro-091916-055313}
      {\bibfield  {journal} {\bibinfo  {journal} {Ann. Rev. Astron. Astrophys.}\
      }\textbf {\bibinfo {volume} {55}},\ \bibinfo {pages} {343} (\bibinfo {year}
      {2017})},\ \Eprint {http://arxiv.org/abs/1707.04256} {arXiv:1707.04256
      [astro-ph.CO]} \BibitemShut {NoStop}%
    \bibitem [{\citenamefont {Buckley}\ and\ \citenamefont
      {Peter}(2018)}]{Buckley:2017ijx}%
      \BibitemOpen
      \bibfield  {author} {\bibinfo {author} {\bibfnamefont {M.~R.}\ \bibnamefont
      {Buckley}}\ and\ \bibinfo {author} {\bibfnamefont {A.~H.~G.}\ \bibnamefont
      {Peter}},\ }\href {\doibase 10.1016/j.physrep.2018.07.003} {\bibfield
      {journal} {\bibinfo  {journal} {Phys. Rept.}\ }\textbf {\bibinfo {volume}
      {761}},\ \bibinfo {pages} {1} (\bibinfo {year} {2018})},\ \Eprint
      {http://arxiv.org/abs/1712.06615} {arXiv:1712.06615 [astro-ph.CO]}
      \BibitemShut {NoStop}%
    \bibitem [{\citenamefont {Tulin}\ and\ \citenamefont
      {Yu}(2018)}]{Tulin:2017ara}%
      \BibitemOpen
      \bibfield  {author} {\bibinfo {author} {\bibfnamefont {S.}~\bibnamefont
      {Tulin}}\ and\ \bibinfo {author} {\bibfnamefont {H.-B.}\ \bibnamefont {Yu}},\
      }\href {\doibase 10.1016/j.physrep.2017.11.004} {\bibfield  {journal}
      {\bibinfo  {journal} {Phys. Rept.}\ }\textbf {\bibinfo {volume} {730}},\
      \bibinfo {pages} {1} (\bibinfo {year} {2018})},\ \Eprint
      {http://arxiv.org/abs/1705.02358} {arXiv:1705.02358 [hep-ph]} \BibitemShut
      {NoStop}%
    \bibitem [{\citenamefont {Ferreira}(2021)}]{Ferreira:2020fam}%
      \BibitemOpen
      \bibfield  {author} {\bibinfo {author} {\bibfnamefont {E.~G.~M.}\
      \bibnamefont {Ferreira}},\ }\href {\doibase 10.1007/s00159-021-00135-6}
      {\bibfield  {journal} {\bibinfo  {journal} {Astron. Astrophys. Rev.}\
      }\textbf {\bibinfo {volume} {29}},\ \bibinfo {pages} {7} (\bibinfo {year}
      {2021})},\ \Eprint {http://arxiv.org/abs/2005.03254} {arXiv:2005.03254
      [astro-ph.CO]} \BibitemShut {NoStop}%
    \bibitem [{\citenamefont {Hui}(2021)}]{Hui:2021tkt}%
      \BibitemOpen
      \bibfield  {author} {\bibinfo {author} {\bibfnamefont {L.}~\bibnamefont
      {Hui}},\ }\href {\doibase 10.1146/annurev-astro-120920-010024} {\bibfield
      {journal} {\bibinfo  {journal} {Ann. Rev. Astron. Astrophys.}\ }\textbf
      {\bibinfo {volume} {59}},\ \bibinfo {pages} {247} (\bibinfo {year} {2021})},\
      \Eprint {http://arxiv.org/abs/2101.11735} {arXiv:2101.11735 [astro-ph.CO]}
      \BibitemShut {NoStop}%
    \bibitem [{\citenamefont {Menci}\ \emph {et~al.}(2016)\citenamefont {Menci},
      \citenamefont {Grazian}, \citenamefont {Castellano},\ and\ \citenamefont
      {Sanchez}}]{Menci:2016eui}%
      \BibitemOpen
      \bibfield  {author} {\bibinfo {author} {\bibfnamefont {N.}~\bibnamefont
      {Menci}}, \bibinfo {author} {\bibfnamefont {A.}~\bibnamefont {Grazian}},
      \bibinfo {author} {\bibfnamefont {M.}~\bibnamefont {Castellano}}, \ and\
      \bibinfo {author} {\bibfnamefont {N.~G.}\ \bibnamefont {Sanchez}},\ }\href
      {\doibase 10.3847/2041-8205/825/1/L1} {\bibfield  {journal} {\bibinfo
      {journal} {Astrophys. J. Lett.}\ }\textbf {\bibinfo {volume} {825}},\
      \bibinfo {pages} {L1} (\bibinfo {year} {2016})},\ \Eprint
      {http://arxiv.org/abs/1606.02530} {arXiv:1606.02530 [astro-ph.CO]}
      \BibitemShut {NoStop}%
    \bibitem [{\citenamefont {Ir\v{s}i\v{c}}\ \emph {et~al.}(2017)\citenamefont
      {Ir\v{s}i\v{c}}, \citenamefont {Viel}, \citenamefont {Haehnelt},
      \citenamefont {Bolton},\ and\ \citenamefont {Becker}}]{Irsic:2017yje}%
      \BibitemOpen
      \bibfield  {author} {\bibinfo {author} {\bibfnamefont {V.}~\bibnamefont
      {Ir\v{s}i\v{c}}}, \bibinfo {author} {\bibfnamefont {M.}~\bibnamefont {Viel}},
      \bibinfo {author} {\bibfnamefont {M.~G.}\ \bibnamefont {Haehnelt}}, \bibinfo
      {author} {\bibfnamefont {J.~S.}\ \bibnamefont {Bolton}}, \ and\ \bibinfo
      {author} {\bibfnamefont {G.~D.}\ \bibnamefont {Becker}},\ }\href {\doibase
      10.1103/PhysRevLett.119.031302} {\bibfield  {journal} {\bibinfo  {journal}
      {Phys. Rev. Lett.}\ }\textbf {\bibinfo {volume} {119}},\ \bibinfo {pages}
      {031302} (\bibinfo {year} {2017})},\ \Eprint
      {http://arxiv.org/abs/1703.04683} {arXiv:1703.04683 [astro-ph.CO]}
      \BibitemShut {NoStop}%
    \bibitem [{\citenamefont {Rogers}\ and\ \citenamefont
      {Peiris}(2021)}]{Rogers:2020ltq}%
      \BibitemOpen
      \bibfield  {author} {\bibinfo {author} {\bibfnamefont {K.~K.}\ \bibnamefont
      {Rogers}}\ and\ \bibinfo {author} {\bibfnamefont {H.~V.}\ \bibnamefont
      {Peiris}},\ }\href {\doibase 10.1103/PhysRevLett.126.071302} {\bibfield
      {journal} {\bibinfo  {journal} {Phys. Rev. Lett.}\ }\textbf {\bibinfo
      {volume} {126}},\ \bibinfo {pages} {071302} (\bibinfo {year} {2021})},\
      \Eprint {http://arxiv.org/abs/2007.12705} {arXiv:2007.12705 [astro-ph.CO]}
      \BibitemShut {NoStop}%
    \bibitem [{\citenamefont {Nadler}\ \emph {et~al.}(2020)\citenamefont {Nadler}
      \emph {et~al.}}]{DES:2019ltu}%
      \BibitemOpen
      \bibfield  {author} {\bibinfo {author} {\bibfnamefont {E.~O.}\ \bibnamefont
      {Nadler}} \emph {et~al.} (\bibinfo {collaboration} {DES}),\ }\href {\doibase
      10.3847/1538-4357/ab846a} {\bibfield  {journal} {\bibinfo  {journal}
      {Astrophys. J.}\ }\textbf {\bibinfo {volume} {893}},\ \bibinfo {pages} {48}
      (\bibinfo {year} {2020})},\ \Eprint {http://arxiv.org/abs/1912.03303}
      {arXiv:1912.03303 [astro-ph.GA]} \BibitemShut {NoStop}%
    \bibitem [{\citenamefont {Banik}\ \emph {et~al.}(2021)\citenamefont {Banik},
      \citenamefont {Bovy}, \citenamefont {Bertone}, \citenamefont {Erkal},\ and\
      \citenamefont {de~Boer}}]{Banik:2019cza}%
      \BibitemOpen
      \bibfield  {author} {\bibinfo {author} {\bibfnamefont {N.}~\bibnamefont
      {Banik}}, \bibinfo {author} {\bibfnamefont {J.}~\bibnamefont {Bovy}},
      \bibinfo {author} {\bibfnamefont {G.}~\bibnamefont {Bertone}}, \bibinfo
      {author} {\bibfnamefont {D.}~\bibnamefont {Erkal}}, \ and\ \bibinfo {author}
      {\bibfnamefont {T.~J.~L.}\ \bibnamefont {de~Boer}},\ }\href {\doibase
      10.1093/mnras/stab210} {\bibfield  {journal} {\bibinfo  {journal} {Mon. Not.
      Roy. Astron. Soc.}\ }\textbf {\bibinfo {volume} {502}},\ \bibinfo {pages}
      {2364} (\bibinfo {year} {2021})},\ \Eprint {http://arxiv.org/abs/1911.02662}
      {arXiv:1911.02662 [astro-ph.GA]} \BibitemShut {NoStop}%
    \bibitem [{\citenamefont {Bonaca}\ \emph {et~al.}(2020)\citenamefont {Bonaca},
      \citenamefont {Conroy}, \citenamefont {Hogg}, \citenamefont {Cargile},
      \citenamefont {Caldwell}, \citenamefont {Naidu}, \citenamefont
      {Price-Whelan}, \citenamefont {Speagle},\ and\ \citenamefont
      {Johnson}}]{Bonaca:2020psc}%
      \BibitemOpen
      \bibfield  {author} {\bibinfo {author} {\bibfnamefont {A.}~\bibnamefont
      {Bonaca}}, \bibinfo {author} {\bibfnamefont {C.}~\bibnamefont {Conroy}},
      \bibinfo {author} {\bibfnamefont {D.~W.}\ \bibnamefont {Hogg}}, \bibinfo
      {author} {\bibfnamefont {P.~A.}\ \bibnamefont {Cargile}}, \bibinfo {author}
      {\bibfnamefont {N.}~\bibnamefont {Caldwell}}, \bibinfo {author}
      {\bibfnamefont {R.~P.}\ \bibnamefont {Naidu}}, \bibinfo {author}
      {\bibfnamefont {A.~M.}\ \bibnamefont {Price-Whelan}}, \bibinfo {author}
      {\bibfnamefont {J.~S.}\ \bibnamefont {Speagle}}, \ and\ \bibinfo {author}
      {\bibfnamefont {B.~D.}\ \bibnamefont {Johnson}},\ }\href {\doibase
      10.3847/2041-8213/ab800c} {\bibfield  {journal} {\bibinfo  {journal}
      {Astrophys. J. Lett.}\ }\textbf {\bibinfo {volume} {892}},\ \bibinfo {pages}
      {L37} (\bibinfo {year} {2020})},\ \Eprint {http://arxiv.org/abs/2001.07215}
      {arXiv:2001.07215 [astro-ph.GA]} \BibitemShut {NoStop}%
    \bibitem [{\citenamefont {Vegetti}\ \emph {et~al.}(2012)\citenamefont
      {Vegetti}, \citenamefont {Lagattuta}, \citenamefont {McKean}, \citenamefont
      {Auger}, \citenamefont {Fassnacht},\ and\ \citenamefont
      {Koopmans}}]{Vegetti:2012mc}%
      \BibitemOpen
      \bibfield  {author} {\bibinfo {author} {\bibfnamefont {S.}~\bibnamefont
      {Vegetti}}, \bibinfo {author} {\bibfnamefont {D.~J.}\ \bibnamefont
      {Lagattuta}}, \bibinfo {author} {\bibfnamefont {J.~P.}\ \bibnamefont
      {McKean}}, \bibinfo {author} {\bibfnamefont {M.~W.}\ \bibnamefont {Auger}},
      \bibinfo {author} {\bibfnamefont {C.~D.}\ \bibnamefont {Fassnacht}}, \ and\
      \bibinfo {author} {\bibfnamefont {L.~V.~E.}\ \bibnamefont {Koopmans}},\
      }\href {\doibase 10.1038/nature10669} {\bibfield  {journal} {\bibinfo
      {journal} {Nature}\ }\textbf {\bibinfo {volume} {481}},\ \bibinfo {pages}
      {341} (\bibinfo {year} {2012})},\ \Eprint {http://arxiv.org/abs/1201.3643}
      {arXiv:1201.3643 [astro-ph.CO]} \BibitemShut {NoStop}%
    \bibitem [{\citenamefont {Hezaveh}\ \emph {et~al.}(2016)\citenamefont {Hezaveh}
      \emph {et~al.}}]{Hezaveh:2016ltk}%
      \BibitemOpen
      \bibfield  {author} {\bibinfo {author} {\bibfnamefont {Y.~D.}\ \bibnamefont
      {Hezaveh}} \emph {et~al.},\ }\href {\doibase 10.3847/0004-637X/823/1/37}
      {\bibfield  {journal} {\bibinfo  {journal} {Astrophys. J.}\ }\textbf
      {\bibinfo {volume} {823}},\ \bibinfo {pages} {37} (\bibinfo {year} {2016})},\
      \Eprint {http://arxiv.org/abs/1601.01388} {arXiv:1601.01388 [astro-ph.CO]}
      \BibitemShut {NoStop}%
    \bibitem [{\citenamefont {\c{S}eng\"ul}\ \emph {et~al.}(2023)\citenamefont
      {\c{S}eng\"ul}, \citenamefont {Birrer}, \citenamefont {Natarajan},\ and\
      \citenamefont {Dvorkin}}]{Sengul:2023olf}%
      \BibitemOpen
      \bibfield  {author} {\bibinfo {author} {\bibfnamefont {A.~c.}\ \bibnamefont
      {\c{S}eng\"ul}}, \bibinfo {author} {\bibfnamefont {S.}~\bibnamefont
      {Birrer}}, \bibinfo {author} {\bibfnamefont {P.}~\bibnamefont {Natarajan}}, \
      and\ \bibinfo {author} {\bibfnamefont {C.}~\bibnamefont {Dvorkin}},\
      }\href@noop {} {\  (\bibinfo {year} {2023})},\ \Eprint
      {http://arxiv.org/abs/2303.14786} {arXiv:2303.14786 [astro-ph.GA]}
      \BibitemShut {NoStop}%
    \bibitem [{\citenamefont {Cyr-Racine}\ \emph {et~al.}(2016)\citenamefont
      {Cyr-Racine}, \citenamefont {Moustakas}, \citenamefont {Keeton},
      \citenamefont {Sigurdson},\ and\ \citenamefont
      {Gilman}}]{Cyr-Racine:2015jwa}%
      \BibitemOpen
      \bibfield  {author} {\bibinfo {author} {\bibfnamefont {F.-Y.}\ \bibnamefont
      {Cyr-Racine}}, \bibinfo {author} {\bibfnamefont {L.~A.}\ \bibnamefont
      {Moustakas}}, \bibinfo {author} {\bibfnamefont {C.~R.}\ \bibnamefont
      {Keeton}}, \bibinfo {author} {\bibfnamefont {K.}~\bibnamefont {Sigurdson}}, \
      and\ \bibinfo {author} {\bibfnamefont {D.~A.}\ \bibnamefont {Gilman}},\
      }\href {\doibase 10.1103/PhysRevD.94.043505} {\bibfield  {journal} {\bibinfo
      {journal} {Phys. Rev. D}\ }\textbf {\bibinfo {volume} {94}},\ \bibinfo
      {pages} {043505} (\bibinfo {year} {2016})},\ \Eprint
      {http://arxiv.org/abs/1506.01724} {arXiv:1506.01724 [astro-ph.CO]}
      \BibitemShut {NoStop}%
    \bibitem [{\citenamefont {D\'\i{}az~Rivero}\ \emph {et~al.}(2018)\citenamefont
      {D\'\i{}az~Rivero}, \citenamefont {Dvorkin}, \citenamefont {Cyr-Racine},
      \citenamefont {Zavala},\ and\ \citenamefont
      {Vogelsberger}}]{DiazRivero:2018oxk}%
      \BibitemOpen
      \bibfield  {author} {\bibinfo {author} {\bibfnamefont {A.}~\bibnamefont
      {D\'\i{}az~Rivero}}, \bibinfo {author} {\bibfnamefont {C.}~\bibnamefont
      {Dvorkin}}, \bibinfo {author} {\bibfnamefont {F.-Y.}\ \bibnamefont
      {Cyr-Racine}}, \bibinfo {author} {\bibfnamefont {J.}~\bibnamefont {Zavala}},
      \ and\ \bibinfo {author} {\bibfnamefont {M.}~\bibnamefont {Vogelsberger}},\
      }\href {\doibase 10.1103/PhysRevD.98.103517} {\bibfield  {journal} {\bibinfo
      {journal} {Phys. Rev. D}\ }\textbf {\bibinfo {volume} {98}},\ \bibinfo
      {pages} {103517} (\bibinfo {year} {2018})},\ \Eprint
      {http://arxiv.org/abs/1809.00004} {arXiv:1809.00004 [astro-ph.CO]}
      \BibitemShut {NoStop}%
    \bibitem [{\citenamefont {Wagner-Carena}\ \emph {et~al.}(2023)\citenamefont
      {Wagner-Carena}, \citenamefont {Aalbers}, \citenamefont {Birrer},
      \citenamefont {Nadler}, \citenamefont {Darragh-Ford}, \citenamefont
      {Marshall},\ and\ \citenamefont {Wechsler}}]{Wagner-Carena:2022mrn}%
      \BibitemOpen
      \bibfield  {author} {\bibinfo {author} {\bibfnamefont {S.}~\bibnamefont
      {Wagner-Carena}}, \bibinfo {author} {\bibfnamefont {J.}~\bibnamefont
      {Aalbers}}, \bibinfo {author} {\bibfnamefont {S.}~\bibnamefont {Birrer}},
      \bibinfo {author} {\bibfnamefont {E.~O.}\ \bibnamefont {Nadler}}, \bibinfo
      {author} {\bibfnamefont {E.}~\bibnamefont {Darragh-Ford}}, \bibinfo {author}
      {\bibfnamefont {P.~J.}\ \bibnamefont {Marshall}}, \ and\ \bibinfo {author}
      {\bibfnamefont {R.~H.}\ \bibnamefont {Wechsler}},\ }\href {\doibase
      10.3847/1538-4357/aca525} {\bibfield  {journal} {\bibinfo  {journal}
      {Astrophys. J.}\ }\textbf {\bibinfo {volume} {942}},\ \bibinfo {pages} {75}
      (\bibinfo {year} {2023})},\ \Eprint {http://arxiv.org/abs/2203.00690}
      {arXiv:2203.00690 [astro-ph.CO]} \BibitemShut {NoStop}%
    \bibitem [{\citenamefont {Gais}\ \emph {et~al.}(2022)\citenamefont {Gais},
      \citenamefont {Ng}, \citenamefont {Seo}, \citenamefont {Wong},\ and\
      \citenamefont {Li}}]{Gais:2022xir}%
      \BibitemOpen
      \bibfield  {author} {\bibinfo {author} {\bibfnamefont {J.}~\bibnamefont
      {Gais}}, \bibinfo {author} {\bibfnamefont {K.~K.~Y.}\ \bibnamefont {Ng}},
      \bibinfo {author} {\bibfnamefont {E.}~\bibnamefont {Seo}}, \bibinfo {author}
      {\bibfnamefont {K.~W.~K.}\ \bibnamefont {Wong}}, \ and\ \bibinfo {author}
      {\bibfnamefont {T.~G.~F.}\ \bibnamefont {Li}},\ }\href {\doibase
      10.3847/2041-8213/ac7052} {\bibfield  {journal} {\bibinfo  {journal}
      {Astrophys. J. Lett.}\ }\textbf {\bibinfo {volume} {932}},\ \bibinfo {pages}
      {L4} (\bibinfo {year} {2022})},\ \Eprint {http://arxiv.org/abs/2201.01817}
      {arXiv:2201.01817 [gr-qc]} \BibitemShut {NoStop}%
    \bibitem [{\citenamefont {Angulo}\ \emph {et~al.}(2013)\citenamefont {Angulo},
      \citenamefont {Hahn},\ and\ \citenamefont {Abel}}]{Angulo:2013sza}%
      \BibitemOpen
      \bibfield  {author} {\bibinfo {author} {\bibfnamefont {R.~E.}\ \bibnamefont
      {Angulo}}, \bibinfo {author} {\bibfnamefont {O.}~\bibnamefont {Hahn}}, \ and\
      \bibinfo {author} {\bibfnamefont {T.}~\bibnamefont {Abel}},\ }\href {\doibase
      10.1093/mnras/stt1246} {\bibfield  {journal} {\bibinfo  {journal} {Mon. Not.
      Roy. Astron. Soc.}\ }\textbf {\bibinfo {volume} {434}},\ \bibinfo {pages}
      {3337} (\bibinfo {year} {2013})},\ \Eprint {http://arxiv.org/abs/1304.2406}
      {arXiv:1304.2406 [astro-ph.CO]} \BibitemShut {NoStop}%
    \bibitem [{\citenamefont {Hui}\ \emph {et~al.}(2017)\citenamefont {Hui},
      \citenamefont {Ostriker}, \citenamefont {Tremaine},\ and\ \citenamefont
      {Witten}}]{Hui:2016ltb}%
      \BibitemOpen
      \bibfield  {author} {\bibinfo {author} {\bibfnamefont {L.}~\bibnamefont
      {Hui}}, \bibinfo {author} {\bibfnamefont {J.~P.}\ \bibnamefont {Ostriker}},
      \bibinfo {author} {\bibfnamefont {S.}~\bibnamefont {Tremaine}}, \ and\
      \bibinfo {author} {\bibfnamefont {E.}~\bibnamefont {Witten}},\ }\href
      {\doibase 10.1103/PhysRevD.95.043541} {\bibfield  {journal} {\bibinfo
      {journal} {Phys. Rev. D}\ }\textbf {\bibinfo {volume} {95}},\ \bibinfo
      {pages} {043541} (\bibinfo {year} {2017})},\ \Eprint
      {http://arxiv.org/abs/1610.08297} {arXiv:1610.08297 [astro-ph.CO]}
      \BibitemShut {NoStop}%
    \bibitem [{\citenamefont {Fairbairn}\ \emph {et~al.}(2018)\citenamefont
      {Fairbairn}, \citenamefont {Marsh}, \citenamefont {Quevillon},\ and\
      \citenamefont {Rozier}}]{Fairbairn:2017sil}%
      \BibitemOpen
      \bibfield  {author} {\bibinfo {author} {\bibfnamefont {M.}~\bibnamefont
      {Fairbairn}}, \bibinfo {author} {\bibfnamefont {D.~J.~E.}\ \bibnamefont
      {Marsh}}, \bibinfo {author} {\bibfnamefont {J.}~\bibnamefont {Quevillon}}, \
      and\ \bibinfo {author} {\bibfnamefont {S.}~\bibnamefont {Rozier}},\ }\href
      {\doibase 10.1103/PhysRevD.97.083502} {\bibfield  {journal} {\bibinfo
      {journal} {Phys. Rev. D}\ }\textbf {\bibinfo {volume} {97}},\ \bibinfo
      {pages} {083502} (\bibinfo {year} {2018})},\ \Eprint
      {http://arxiv.org/abs/1707.03310} {arXiv:1707.03310 [astro-ph.CO]}
      \BibitemShut {NoStop}%
    \bibitem [{\citenamefont {Xu}\ \emph {et~al.}(2022)\citenamefont {Xu},
      \citenamefont {Ezquiaga},\ and\ \citenamefont {Holz}}]{Xu:2021bfn}%
      \BibitemOpen
      \bibfield  {author} {\bibinfo {author} {\bibfnamefont {F.}~\bibnamefont
      {Xu}}, \bibinfo {author} {\bibfnamefont {J.~M.}\ \bibnamefont {Ezquiaga}}, \
      and\ \bibinfo {author} {\bibfnamefont {D.~E.}\ \bibnamefont {Holz}},\ }\href
      {\doibase 10.3847/1538-4357/ac58f8} {\bibfield  {journal} {\bibinfo
      {journal} {Astrophys. J.}\ }\textbf {\bibinfo {volume} {929}},\ \bibinfo
      {pages} {9} (\bibinfo {year} {2022})},\ \Eprint
      {http://arxiv.org/abs/2105.14390} {arXiv:2105.14390 [astro-ph.CO]}
      \BibitemShut {NoStop}%
    \bibitem [{\citenamefont {Navarro}\ \emph {et~al.}(1997)\citenamefont
      {Navarro}, \citenamefont {Frenk},\ and\ \citenamefont
      {White}}]{Navarro:1996gj}%
      \BibitemOpen
      \bibfield  {author} {\bibinfo {author} {\bibfnamefont {J.~F.}\ \bibnamefont
      {Navarro}}, \bibinfo {author} {\bibfnamefont {C.~S.}\ \bibnamefont {Frenk}},
      \ and\ \bibinfo {author} {\bibfnamefont {S.~D.~M.}\ \bibnamefont {White}},\
      }\href {\doibase 10.1086/304888} {\bibfield  {journal} {\bibinfo  {journal}
      {Astrophys. J.}\ }\textbf {\bibinfo {volume} {490}},\ \bibinfo {pages} {493}
      (\bibinfo {year} {1997})},\ \Eprint {http://arxiv.org/abs/astro-ph/9611107}
      {arXiv:astro-ph/9611107} \BibitemShut {NoStop}%
    \bibitem [{\citenamefont {Erickcek}\ \emph {et~al.}(2006)\citenamefont
      {Erickcek}, \citenamefont {Kamionkowski},\ and\ \citenamefont
      {Benson}}]{Erickcek:2006xc}%
      \BibitemOpen
      \bibfield  {author} {\bibinfo {author} {\bibfnamefont {A.~L.}\ \bibnamefont
      {Erickcek}}, \bibinfo {author} {\bibfnamefont {M.}~\bibnamefont
      {Kamionkowski}}, \ and\ \bibinfo {author} {\bibfnamefont {A.~J.}\
      \bibnamefont {Benson}},\ }\href {\doibase 10.1111/j.1365-2966.2006.10838.x}
      {\bibfield  {journal} {\bibinfo  {journal} {Mon. Not. Roy. Astron. Soc.}\
      }\textbf {\bibinfo {volume} {371}},\ \bibinfo {pages} {1992} (\bibinfo {year}
      {2006})},\ \Eprint {http://arxiv.org/abs/astro-ph/0604281}
      {arXiv:astro-ph/0604281} \BibitemShut {NoStop}%
    \bibitem [{\citenamefont {Sesana}\ \emph {et~al.}(2007)\citenamefont {Sesana},
      \citenamefont {Volonteri},\ and\ \citenamefont {Haardt}}]{Sesana:2007sh}%
      \BibitemOpen
      \bibfield  {author} {\bibinfo {author} {\bibfnamefont {A.}~\bibnamefont
      {Sesana}}, \bibinfo {author} {\bibfnamefont {M.}~\bibnamefont {Volonteri}}, \
      and\ \bibinfo {author} {\bibfnamefont {F.}~\bibnamefont {Haardt}},\ }\href
      {\doibase 10.1111/j.1365-2966.2007.11734.x} {\bibfield  {journal} {\bibinfo
      {journal} {Mon. Not. Roy. Astron. Soc.}\ }\textbf {\bibinfo {volume} {377}},\
      \bibinfo {pages} {1711} (\bibinfo {year} {2007})},\ \Eprint
      {http://arxiv.org/abs/astro-ph/0701556} {arXiv:astro-ph/0701556} \BibitemShut
      {NoStop}%
    \bibitem [{\citenamefont {Sesana}\ \emph {et~al.}(2011)\citenamefont {Sesana},
      \citenamefont {Gair}, \citenamefont {Berti},\ and\ \citenamefont
      {Volonteri}}]{Sesana:2010wy}%
      \BibitemOpen
      \bibfield  {author} {\bibinfo {author} {\bibfnamefont {A.}~\bibnamefont
      {Sesana}}, \bibinfo {author} {\bibfnamefont {J.}~\bibnamefont {Gair}},
      \bibinfo {author} {\bibfnamefont {E.}~\bibnamefont {Berti}}, \ and\ \bibinfo
      {author} {\bibfnamefont {M.}~\bibnamefont {Volonteri}},\ }\href {\doibase
      10.1103/PhysRevD.83.044036} {\bibfield  {journal} {\bibinfo  {journal} {Phys.
      Rev. D}\ }\textbf {\bibinfo {volume} {83}},\ \bibinfo {pages} {044036}
      (\bibinfo {year} {2011})},\ \Eprint {http://arxiv.org/abs/1011.5893}
      {arXiv:1011.5893 [astro-ph.CO]} \BibitemShut {NoStop}%
    \bibitem [{\citenamefont {Klein}\ \emph {et~al.}(2016)\citenamefont {Klein}
      \emph {et~al.}}]{Klein:2015hvg}%
      \BibitemOpen
      \bibfield  {author} {\bibinfo {author} {\bibfnamefont {A.}~\bibnamefont
      {Klein}} \emph {et~al.},\ }\href {\doibase 10.1103/PhysRevD.93.024003}
      {\bibfield  {journal} {\bibinfo  {journal} {Phys. Rev. D}\ }\textbf {\bibinfo
      {volume} {93}},\ \bibinfo {pages} {024003} (\bibinfo {year} {2016})},\
      \Eprint {http://arxiv.org/abs/1511.05581} {arXiv:1511.05581 [gr-qc]}
      \BibitemShut {NoStop}%
    \bibitem [{\citenamefont {Katz}\ \emph {et~al.}(2020)\citenamefont {Katz},
      \citenamefont {Kelley}, \citenamefont {Dosopoulou}, \citenamefont {Berry},
      \citenamefont {Blecha},\ and\ \citenamefont {Larson}}]{Katz:2019qlu}%
      \BibitemOpen
      \bibfield  {author} {\bibinfo {author} {\bibfnamefont {M.~L.}\ \bibnamefont
      {Katz}}, \bibinfo {author} {\bibfnamefont {L.~Z.}\ \bibnamefont {Kelley}},
      \bibinfo {author} {\bibfnamefont {F.}~\bibnamefont {Dosopoulou}}, \bibinfo
      {author} {\bibfnamefont {S.}~\bibnamefont {Berry}}, \bibinfo {author}
      {\bibfnamefont {L.}~\bibnamefont {Blecha}}, \ and\ \bibinfo {author}
      {\bibfnamefont {S.~L.}\ \bibnamefont {Larson}},\ }\href {\doibase
      10.1093/mnras/stz3102} {\bibfield  {journal} {\bibinfo  {journal} {Mon. Not.
      Roy. Astron. Soc.}\ }\textbf {\bibinfo {volume} {491}},\ \bibinfo {pages}
      {2301} (\bibinfo {year} {2020})},\ \Eprint {http://arxiv.org/abs/1908.05779}
      {arXiv:1908.05779 [astro-ph.HE]} \BibitemShut {NoStop}%
    \bibitem [{\citenamefont {Barausse}\ \emph {et~al.}(2020)\citenamefont
      {Barausse}, \citenamefont {Dvorkin}, \citenamefont {Tremmel}, \citenamefont
      {Volonteri},\ and\ \citenamefont {Bonetti}}]{Barausse:2020mdt}%
      \BibitemOpen
      \bibfield  {author} {\bibinfo {author} {\bibfnamefont {E.}~\bibnamefont
      {Barausse}}, \bibinfo {author} {\bibfnamefont {I.}~\bibnamefont {Dvorkin}},
      \bibinfo {author} {\bibfnamefont {M.}~\bibnamefont {Tremmel}}, \bibinfo
      {author} {\bibfnamefont {M.}~\bibnamefont {Volonteri}}, \ and\ \bibinfo
      {author} {\bibfnamefont {M.}~\bibnamefont {Bonetti}},\ }\href {\doibase
      10.3847/1538-4357/abba7f} {\bibfield  {journal} {\bibinfo  {journal}
      {Astrophys. J.}\ }\textbf {\bibinfo {volume} {904}},\ \bibinfo {pages} {16}
      (\bibinfo {year} {2020})},\ \Eprint {http://arxiv.org/abs/2006.03065}
      {arXiv:2006.03065 [astro-ph.GA]} \BibitemShut {NoStop}%
    \bibitem [{\citenamefont {Toubiana}\ \emph {et~al.}(2021)\citenamefont
      {Toubiana}, \citenamefont {Wong}, \citenamefont {Babak}, \citenamefont
      {Barausse}, \citenamefont {Berti}, \citenamefont {Gair}, \citenamefont
      {Marsat},\ and\ \citenamefont {Taylor}}]{Toubiana:2021iuw}%
      \BibitemOpen
      \bibfield  {author} {\bibinfo {author} {\bibfnamefont {A.}~\bibnamefont
      {Toubiana}}, \bibinfo {author} {\bibfnamefont {K.~W.~K.}\ \bibnamefont
      {Wong}}, \bibinfo {author} {\bibfnamefont {S.}~\bibnamefont {Babak}},
      \bibinfo {author} {\bibfnamefont {E.}~\bibnamefont {Barausse}}, \bibinfo
      {author} {\bibfnamefont {E.}~\bibnamefont {Berti}}, \bibinfo {author}
      {\bibfnamefont {J.~R.}\ \bibnamefont {Gair}}, \bibinfo {author}
      {\bibfnamefont {S.}~\bibnamefont {Marsat}}, \ and\ \bibinfo {author}
      {\bibfnamefont {S.~R.}\ \bibnamefont {Taylor}},\ }\href {\doibase
      10.1103/PhysRevD.104.083027} {\bibfield  {journal} {\bibinfo  {journal}
      {Phys. Rev. D}\ }\textbf {\bibinfo {volume} {104}},\ \bibinfo {pages}
      {083027} (\bibinfo {year} {2021})},\ \Eprint
      {http://arxiv.org/abs/2106.13819} {arXiv:2106.13819 [gr-qc]} \BibitemShut
      {NoStop}%
    \bibitem [{\citenamefont {Liu}\ \emph {et~al.}(2023)\citenamefont {Liu},
      \citenamefont {Wong}, \citenamefont {Leong}, \citenamefont {More},
      \citenamefont {Hannuksela},\ and\ \citenamefont {Li}}]{Liu:2023ikc}%
      \BibitemOpen
      \bibfield  {author} {\bibinfo {author} {\bibfnamefont {A.}~\bibnamefont
      {Liu}}, \bibinfo {author} {\bibfnamefont {I.~C.~F.}\ \bibnamefont {Wong}},
      \bibinfo {author} {\bibfnamefont {S.~H.~W.}\ \bibnamefont {Leong}}, \bibinfo
      {author} {\bibfnamefont {A.}~\bibnamefont {More}}, \bibinfo {author}
      {\bibfnamefont {O.~A.}\ \bibnamefont {Hannuksela}}, \ and\ \bibinfo {author}
      {\bibfnamefont {T.~G.~F.}\ \bibnamefont {Li}},\ }\href {\doibase
      10.1093/mnras/stad1302} {\  (\bibinfo {year} {2023}),\
      10.1093/mnras/stad1302},\ \Eprint {http://arxiv.org/abs/2302.09870}
      {arXiv:2302.09870 [gr-qc]} \BibitemShut {NoStop}%
    \bibitem [{\citenamefont {Karnesis}\ \emph {et~al.}(2023)\citenamefont
      {Karnesis}, \citenamefont {Katz}, \citenamefont {Korsakova}, \citenamefont
      {Gair},\ and\ \citenamefont {Stergioulas}}]{Karnesis:2023ras}%
      \BibitemOpen
      \bibfield  {author} {\bibinfo {author} {\bibfnamefont {N.}~\bibnamefont
      {Karnesis}}, \bibinfo {author} {\bibfnamefont {M.~L.}\ \bibnamefont {Katz}},
      \bibinfo {author} {\bibfnamefont {N.}~\bibnamefont {Korsakova}}, \bibinfo
      {author} {\bibfnamefont {J.~R.}\ \bibnamefont {Gair}}, \ and\ \bibinfo
      {author} {\bibfnamefont {N.}~\bibnamefont {Stergioulas}},\ }\href@noop {} {\
      (\bibinfo {year} {2023})},\ \Eprint {http://arxiv.org/abs/2303.02164}
      {arXiv:2303.02164 [astro-ph.IM]} \BibitemShut {NoStop}%
    \bibitem [{\citenamefont {Littenberg}\ and\ \citenamefont
      {Cornish}(2023)}]{Littenberg:2023xpl}%
      \BibitemOpen
      \bibfield  {author} {\bibinfo {author} {\bibfnamefont {T.~B.}\ \bibnamefont
      {Littenberg}}\ and\ \bibinfo {author} {\bibfnamefont {N.~J.}\ \bibnamefont
      {Cornish}},\ }\href {\doibase 10.1103/PhysRevD.107.063004} {\bibfield
      {journal} {\bibinfo  {journal} {Phys. Rev. D}\ }\textbf {\bibinfo {volume}
      {107}},\ \bibinfo {pages} {063004} (\bibinfo {year} {2023})},\ \Eprint
      {http://arxiv.org/abs/2301.03673} {arXiv:2301.03673 [gr-qc]} \BibitemShut
      {NoStop}%
    \bibitem [{\citenamefont {Baibhav}\ \emph {et~al.}(2021)\citenamefont {Baibhav}
      \emph {et~al.}}]{Baibhav:2019rsa}%
      \BibitemOpen
      \bibfield  {author} {\bibinfo {author} {\bibfnamefont {V.}~\bibnamefont
      {Baibhav}} \emph {et~al.},\ }\href {\doibase 10.1007/s10686-021-09741-9}
      {\bibfield  {journal} {\bibinfo  {journal} {Exper. Astron.}\ }\textbf
      {\bibinfo {volume} {51}},\ \bibinfo {pages} {1385} (\bibinfo {year}
      {2021})},\ \Eprint {http://arxiv.org/abs/1908.11390} {arXiv:1908.11390
      [astro-ph.HE]} \BibitemShut {NoStop}%
    \bibitem [{\citenamefont {Sedda}\ \emph {et~al.}(2020)\citenamefont {Sedda}
      \emph {et~al.}}]{Sedda:2019uro}%
      \BibitemOpen
      \bibfield  {author} {\bibinfo {author} {\bibfnamefont {M.~A.}\ \bibnamefont
      {Sedda}} \emph {et~al.},\ }\href {\doibase 10.1088/1361-6382/abb5c1}
      {\bibfield  {journal} {\bibinfo  {journal} {Class. Quant. Grav.}\ }\textbf
      {\bibinfo {volume} {37}},\ \bibinfo {pages} {215011} (\bibinfo {year}
      {2020})},\ \Eprint {http://arxiv.org/abs/1908.11375} {arXiv:1908.11375
      [gr-qc]} \BibitemShut {NoStop}%
    \bibitem [{\citenamefont {Sesana}\ \emph {et~al.}(2021)\citenamefont {Sesana}
      \emph {et~al.}}]{Sesana:2019vho}%
      \BibitemOpen
      \bibfield  {author} {\bibinfo {author} {\bibfnamefont {A.}~\bibnamefont
      {Sesana}} \emph {et~al.},\ }\href {\doibase 10.1007/s10686-021-09709-9}
      {\bibfield  {journal} {\bibinfo  {journal} {Exper. Astron.}\ }\textbf
      {\bibinfo {volume} {51}},\ \bibinfo {pages} {1333} (\bibinfo {year}
      {2021})},\ \Eprint {http://arxiv.org/abs/1908.11391} {arXiv:1908.11391
      [astro-ph.IM]} \BibitemShut {NoStop}%
    \bibitem [{\citenamefont {{Hinshaw}}\ and\ \citenamefont
      {{Krauss}}(1987)}]{1987ApJ...320..468H}%
      \BibitemOpen
      \bibfield  {author} {\bibinfo {author} {\bibfnamefont {G.}~\bibnamefont
      {{Hinshaw}}}\ and\ \bibinfo {author} {\bibfnamefont {L.~M.}\ \bibnamefont
      {{Krauss}}},\ }\href {\doibase 10.1086/165564} {\bibfield  {journal}
      {\bibinfo  {journal} {\apj}\ }\textbf {\bibinfo {volume} {320}},\ \bibinfo
      {pages} {468} (\bibinfo {year} {1987})}\BibitemShut {NoStop}%
    \bibitem [{\citenamefont {Flores}\ and\ \citenamefont
      {Primack}(1996)}]{Flores:1995dc}%
      \BibitemOpen
      \bibfield  {author} {\bibinfo {author} {\bibfnamefont {R.~A.}\ \bibnamefont
      {Flores}}\ and\ \bibinfo {author} {\bibfnamefont {J.~R.}\ \bibnamefont
      {Primack}},\ }\href {\doibase 10.1086/309879} {\bibfield  {journal} {\bibinfo
       {journal} {Astrophys. J. Lett.}\ }\textbf {\bibinfo {volume} {457}},\
      \bibinfo {pages} {L5} (\bibinfo {year} {1996})},\ \Eprint
      {http://arxiv.org/abs/astro-ph/9512063} {arXiv:astro-ph/9512063} \BibitemShut
      {NoStop}%
    \bibitem [{\citenamefont {Bryan}\ and\ \citenamefont
      {Norman}(1998)}]{Bryan:1997dn}%
      \BibitemOpen
      \bibfield  {author} {\bibinfo {author} {\bibfnamefont {G.~L.}\ \bibnamefont
      {Bryan}}\ and\ \bibinfo {author} {\bibfnamefont {M.~L.}\ \bibnamefont
      {Norman}},\ }\href {\doibase 10.1086/305262} {\bibfield  {journal} {\bibinfo
      {journal} {Astrophys. J.}\ }\textbf {\bibinfo {volume} {495}},\ \bibinfo
      {pages} {80} (\bibinfo {year} {1998})},\ \Eprint
      {http://arxiv.org/abs/astro-ph/9710107} {arXiv:astro-ph/9710107} \BibitemShut
      {NoStop}%
    \bibitem [{\citenamefont {Gradshteyn}\ and\ \citenamefont
      {Ryzhik}(2007)}]{Gradshtein2007-ov}%
      \BibitemOpen
      \bibfield  {author} {\bibinfo {author} {\bibfnamefont {I.~S.}\ \bibnamefont
      {Gradshteyn}}\ and\ \bibinfo {author} {\bibfnamefont {I.~M.}\ \bibnamefont
      {Ryzhik}},\ }\href@noop {} {\emph {\bibinfo {title} {Table of integrals,
      series, and products}}},\ \bibinfo {edition} {7th}\ ed.,\ edited by\ \bibinfo
      {editor} {\bibfnamefont {A.}~\bibnamefont {Jeffrey}}\ and\ \bibinfo {editor}
      {\bibfnamefont {D.}~\bibnamefont {Zwillinger}}\ (\bibinfo  {publisher}
      {Academic Press},\ \bibinfo {address} {San Diego, CA},\ \bibinfo {year}
      {2007})\BibitemShut {NoStop}%
    \bibitem [{{\relax DLMF}()}]{NIST:DLMF}%
      \BibitemOpen
      {\relax DLMF},\ \href {https://dlmf.nist.gov/} {\enquote {\bibinfo {title}
      {{\it NIST Digital Library of Mathematical Functions}},}\ }\bibinfo
      {howpublished} {\url{https://dlmf.nist.gov/}, Release 1.1.9 of 2023-03-15},\
      \bibinfo {note} {f.~W.~J. Olver, A.~B. {Olde Daalhuis}, D.~W. Lozier, B.~I.
      Schneider, R.~F. Boisvert, C.~W. Clark, B.~R. Miller, B.~V. Saunders, H.~S.
      Cohl, and M.~A. McClain, eds.}\BibitemShut {Stop}%
    \end{thebibliography}
\end{document}